\documentclass[12pt,twoside]{article}
\newcommand{\bq}{\begin{equation}}
\newcommand{\eq}{\end{equation}}
\newcommand{\ba}{\begin{eqnarray}}
\newcommand{\ea}{\end{eqnarray}}

\newcommand{\nl }{ \nonumber  }
\newcommand{\vf}{\varphi}
\newcommand{\ul}{\underline}
\newcommand{\p}{\partial}
\newcommand{\pu}{\p_\tau}

\newcommand{\h}{\hspace{1cm}}
\newcommand{\s}{\sigma}
\newcommand{\us}{\underline\sigma}
\newcommand{\da}{\delta^p(\us_1 - \us_2)}
\newcommand{\la}{\lambda}
\newcommand{\La}{\Lambda}
\newcommand{\pj}{\p_j}
\newcommand{\uz}{\underline{z}}
\newcommand{\pJ}{\p_J}
\newcommand{\pK}{\p_K}
\newcommand{\pk}{\p_k}
\newcommand{\Dj}{\bigl(\pu-\mu^{j}\pj\bigr)}
\newcommand{\Dk}{\bigl(\pu-\mu^{k}\pk\bigr)}

\def\be{\begin{eqnarray}}
\def\ee{\end{eqnarray}}

\def\half{{\textstyle{1 \over 2}}}

\begin{document}
\begin{titlepage}
\begin{center}
\LARGE{BULGARIAN ACADEMY OF SCIENCES \\ Institute for Nuclear
Research and Nuclear Energy}
\vspace*{4cm}
\\
\LARGE{Plamen Lubenov Bozhilov} \vspace*{1cm} \\

\Huge{NULL $P$-BRANES} \\ \vspace*{1cm} \LARGE{DISSERTATION \\
submitted for obtaining the educational and scientific degree \\
"doctor" \\  \vspace*{1cm} {\bf Scientific supervisor}: \\ Prof.
Dimitr Tz. Stoyanov \\ \vspace*{2cm} Sofia, 2 000}

\end{center}
\end{titlepage}

\begin{titlepage}

\end{titlepage}

\tableofcontents
\newpage
\vspace*{.5cm}
\section{\bf INTRODUCTION}
\hspace{1cm} In the attempt to unify all the forces present in
Nature, which entails having a consistent quantum theory of gravity,
superstring theories seem promising candidates. The evidence that
string theories could be unified theories is provided by the presence
in their massless spectrum of enough particles to account for those
present at low energies, including the graviton \cite{GSW87}.

String theory is by now a vast subject with almost three decades of
active research contributing to its development. During the last few
years, our understanding of string theory  has undergone a dramatic
change. The key to this development is the discovery of duality
symmetries, which relate the strong and weak coupling limits of
different string theories ($S$-duality). These symmetries not only
relate apparently different string theories, but give us a way to
compute certain strong coupling results in one string theory by
mapping it to a weak coupling result in a dual string theory (for
recent review on non-perturbative string theory see for instance
\cite{S98}).

Our main aim here is the description and further investigation of the
properties of the so called \ul{null} or \ul{tensionless} branes. To
explain how these extended objects appear and why it is worthwhile to
learn more about them, we first will give a brief introduction to the
related notions in string theory following mainly \cite{V97}.

\subsection{\bf Brief Overview of String Theory}
\hspace{1cm} String theory is a description of dynamics of objects
with one spatial direction, which we parameterize by $\sigma$,
propagating in a space parameterized by $x^\mu$. The world-sheet
of the string is parameterized by coordinates ($\tau,\sigma $)
where each $\tau ={\rm constant}$ denotes the string at a given
time. The amplitude for propagation of a string from an initial
configuration to a final one is given by sum over world-sheets
which interpolate between the two string configurations weighed by
${\exp} (i S)$, where \ba\label{a1} S\propto\int d\tau d\sigma \
\partial_J x^\mu \partial^J x^{\nu}g_{\mu \nu}(x) \ea where
$g_{\mu \nu}$ is the metric on space-time and $J$ runs over the
$\tau$ and $\sigma$ directions. Note that by slicing the
world-sheet we will get configurations where a single string
splits to a pair or vice versa, and combinations thereof.

If we consider propagation in flat space-time where $g_{\mu
\nu}=\eta_{\mu \nu}$ the fields $x^{\mu}$ on the world-sheet,
which describe the position in space-time of each bit of string,
are free fields and satisfy the 2 dimensional equation \ba\nl
\partial _J\partial^J x^\mu
=(\partial^2_\tau -\partial^2_\sigma )x^\mu =0 . \ea The solution of
which is given by \ba\nl x^\mu(\tau,\sigma)= x^\mu_L(\tau+\sigma
)+x^\mu_R(\tau -\sigma ). \ea In particular notice that the left- and
right-moving degrees of freedom are essentially independent. There
are two basic types of strings:  {\it Closed} strings and {\it Open}
strings depending on whether the string is a closed circle or an open
interval respectively.  If we are dealing with closed strings the
left- and right-moving degrees of freedom remain essentially
independent but if are dealing with open strings the left-moving
modes reflecting off the left boundary become the right-moving
modes--thus the left- and right-moving modes are essentially
identical in this case.  In this sense an open string has `half' the
degrees of freedom of a closed string and can be viewed as a
`folding' of a closed string so that it looks like an interval.

There are two basic types of string theories, bosonic and fermionic.
What distinguishes bosonic and fermionic strings is the existence of
supersymmetry on the world-sheet.  This means that in addition to the
coordinates $x^\mu$ we also have anti-commuting fermionic coordinates
$\psi^{\mu}_{L,R}$ which are space-time vectors but fermionic spinors
on the worldsheet whose chirality is denoted by subscript $L,R$.  The
action for superstrings takes the form \ba\nl S=\int
\partial_Lx^\mu
\partial_R x^\mu +\psi_R^\mu \partial_L \psi_R^\mu +\psi_L^\mu
\partial_R \psi _R^\mu.
\ea
There are two consistent boundary conditions on each of the
fermions, periodic ({\bf R}amond sector) or anti-periodic ({\bf
N}eveu-{\bf S}chwarz sector) (note that the coordinate $\sigma$ is
periodic).

A natural question arises as to what metric we should put on the
world-sheet.  In the above we have taken it to be flat.  However
in principle there is one degree of freedom that a metric can have
in two dimensions. This is because it is a $2\times 2$ symmetric
matrix (3 degrees of freedom) which is defined up to arbitrary
reparametrization of 2 dimensional space-time (2 degrees of
freedom) leaving us with one function. Locally we can take the 2
dimensional metric $g_{JK}$ to be conformally flat \ba\nl
g_{JK}={\rm exp}(\phi) \eta_{JK}. \ea Classically the action $S$
does not depend on $\phi$.  This is easily seen by noting that the
properly coordinate invariant action density goes as
 ${\sqrt{|g|}}g^{JK}\partial_J x^{\mu} \partial_K x^{\nu}\eta_{\mu\nu}$ and is
independent of $\phi$ only in $D=2$.  This is rather nice and
means that we can ignore all the local dynamics associated with
gravity on the world-sheet.  This case is what is known as the
critical string case which is the case of most interest.  It turns
out that this independence from the local dynamics of the
world-sheet metric survives quantum corrections only when the
dimension of space is 26 in the case of {\it bosonic strings} and
10 for {\it fermionic} or {\it superstrings}.  Each string can be
in a specific vibrational mode which gives rise to a particle.  To
describe the totality of such particles it is convenient to go to
`light-cone' gauge.  Roughly speaking this means that we take into
account that string vibration along their world-sheet is not
physical.  In particular for bosonic string the vibrational modes
exist only in 24 transverse directions and for superstrings they
exist in $8$ transverse directions.

Solving the free field equations for $x,\psi $ we have \ba\nl
\partial_L x^\mu = \sum_{n}\alpha_{-n}^\mu {\rm
e}^{-in(\tau+\sigma)}\\ \psi_L^\mu =\sum_n \psi_{-n}^\mu {\rm
e}^{-in(\tau+\sigma)} \ea and similarly for right-moving oscillator
modes $\tilde \alpha_{-n}^\mu$ and $\tilde \psi_{-n}^\mu$. The sum
over $n$ in the above runs over integers for the $\alpha_{-n}$.  For
fermions depending on whether we are in the {\bf{R}} sector or {\bf
{NS}} sector it runs over integers or integers shifted by ${1/2}$
respectively. Many things decouple between the left- and right-movers
in the construction of a single string Hilbert space and we sometimes
talk only about one of them.  For the open string Fock space the
left- and right-movers mix as mentioned before, and we simply get one
copy of the above oscillators.

A special role is played by the {\it zero modes} of the oscillators.
For the $x$-fields they correspond to the center of mass motion and
thus $\alpha_0$ gets identified with the left-moving momentum of the
center of mass. In particular we have for the center of mass \ba\nl
x=\alpha_0 (\tau+\sigma) +\tilde \alpha_0 (\tau -\sigma ), \ea where
we identify \ba\nl (\alpha_0,\tilde \alpha_0)=(P_L,P_R). \ea Note
that for closed string, periodicity of $x$ in $\sigma$ requires that
$P_L=P_R=P$ which we identify with the center of mass momentum of the
string.

In quantizing the fields on the strings we use the usual
(anti)commutation relations \ba\nl
[\alpha^{\mu}_{n},\alpha^{\nu}_{m}]=n\delta_{m+n,0}\eta^{\mu \nu}
\\ \nl
 \{\psi^{\mu}_{n},\psi^{\nu}_{m}\}=\eta^{\mu \nu}\delta_{m+n,0}.
\ea We choose the negative moded oscillators as creation operators.
In constructing the Fock space we have to pay special attention to
the zero modes. The zero modes of $\alpha$ should be diagonal in the
Fock space and we identify their eigenvalue with momentum.  For
$\psi$ in the $NS$ sector there is no zero mode so there is no
subtlety in construction of the Hilbert space. For the $R$ sector, we
have zero modes.  In this case the zero modes form a Clifford algebra
\ba\nl \{\psi^\mu_0,\psi^\nu_0\}=\eta^{\mu \nu}. \ea This implies
that in these cases the ground state is a spinor representation of
the Lorentz group.  Thus a typical element in the Fock space looks
like \ba\nl
\alpha_{-n_1}^{L\mu_1}....\psi_{-n_k}^{L\mu_k}...|P_L,a\rangle\otimes
\alpha_{-m_1}^{R\mu_1}....\psi_{-m_r}^{R\mu_k}...|P_R,b\rangle , \ea
where $a,b$ label spinor states  for R sectors and are absent in the
NS case; moreover for the bosonic string we only have the left and
right bosonic oscillators.

It is convenient to define the total oscillator number as sum of the
negative oscillator numbers, for left- and right- movers separately.
$N_L=n_1+...+n_k+...$, $N_R=m_1+...+m_r+...$. The condition that the
two dimensional gravity decouple implies that the energy momentum
tensor annihilate the physical states.  The trace of the energy
momentum tensor is zero here (and in all compactifications of string
theory) and so we have two independent components which can be
identified with the left- and right- moving hamiltonians $H_{L,R}$
and the physical states condition requires that \ba\label{onsh}
H_L=N_L+(1/2)P_L^2-\delta_L=0= H_R=N_R+(1/2)P_R^2-\delta_R ,\ea where
$\delta_{L,R}$ are normal ordering constants which depend on which
string theory and which sector we are dealing with.  For bosonic
string $\delta =1$, for superstrings we have two cases:  For $NS$
sector $\delta ={1/2}$ and for the $R$ sector $\delta =0$. The
equations (\ref{onsh}) give the spectrum of particles in the string
perturbation theory.  Note that $P_L^2=P_R^2=-m^2$ and so we see that
$m^2$ grows linearly with the oscillator number $N$, up to a shift:
\ba\label{spec} (1/2)m^2=N_L-\delta_L=N_R-\delta_R. \ea

\subsubsection{Massless States of Bosonic Strings}
\hspace{1cm} Let us consider the left-mover excitations.  Since
$\delta =1$ for bosonic string, (\ref{spec}) implies that if we do
not use any string oscillations, the ground state is tachyonic
${1/2}m^2=-1$.  This clearly implies that bosonic string by itself
is not a good starting point for perturbation theory.
Nevertheless in anticipation of a modified appearance of bosonic
strings in the context of heterotic strings, let us continue to
the next state.

If we consider oscillator number $N_L=1$, from (\ref{spec}) we learn
that excitation is massless.  Putting the right-movers together with
it, we find that it is given by \ba\nl \alpha^\mu_{-1}\tilde
\alpha^\nu_{-1}|P\rangle . \ea What is the physical interpretation of
these massless states? The most reliable method is to find how they
transform under the little group for massless states which in this
case is $SO(24)$.  If we go to the light cone gauge, and count the
physical states, which roughly speaking means taking the indices
$\mu$ to go over spatial directions transverse to a null vector, we
can easily deduce the content of states.   By decomposing the above
massless state under the little group of $SO(24)$, we find that we
have symmetric traceless tensor, anti-symmetric 2-tensor, and the
trace, which we identify as arising from 26 dimensional fields
\ba\label{grm} {g_{\mu \nu},B_{\mu\nu},\phi} \ea the metric, the
anti-symmetric field $B$ and the {\it dilaton}. This triple of fields
should be viewed as the stringy multiplet for gravity.  The quantity
${\rm \exp}[-\phi]$ is identified with the string coupling constant.
What this means is that a world-sheet configuration of a string which
sweeps a genus $g$ curve, which should be viewed as $g$-th loop
correction for string theory, will be weighed by ${\rm
exp}(-2(g-1)\phi)$. The existence of the field $B$ can also be
understood (and in some sense predicted) rather easily. If we have a
point particle it is natural to have it charged under a gauge field,
which introduces a term ${\rm exp}(i\int A)$ along the world-line.
For strings the natural generalization of this requires an
anti-symmetric 2-form to integrate over the world-sheet, and so we
say that the strings are {\it charged} under $B_{\mu \nu}$ and that
the amplitude for a world-sheet configuration will have an extra
factor of ${\rm exp}(i\int B)$.

Since bosonic string has tachyons we do not know how to make sense of
that theory by itself.

\subsubsection{Massless States of Type II Superstrings}
\hspace{1cm} Let us now consider the light particle states for
superstrings. We recall from the above discussion that there are
two sectors to consider, NS and R, separately for the left- and
the right-movers. As usual we will first treat the left- and
right-moving sectors separately and then combine them at the end.
Let us  consider the NS sector for left-movers. Then the formula
for masses (\ref{onsh}) implies that the ground state is tachyonic
with ${1/2}m^2={-1/2}$. The first excited states from the
left-movers are massless and corresponds to $\psi_{-1/2}^\mu
|0\rangle$, and so is a vector in space-time.  How do we deal with
the tachyons? It turns out that summing over the boundary
conditions of fermions on the world-sheet amounts to keeping the
states with a fixed fermion number $(-1)^F$ on the world-sheet.
Since in the NS sector the number of fermionic oscillator
correlates with the integrality/half-integrality of $N$, it turns
out that the consistent choice involves keeping only the
$N=$half-integral states. This is known as the GSO projection.
Thus the tachyon is projected out and the lightest left-moving
state is a massless vector.

For the R-sector using (\ref{onsh}) we see that the ground states are
massless.  As discussed above, quantizing the zero modes of fermions
implies that they are spinors.  Moreover GSO projection, which is
projection on a definite $(-1)^F$ state, amounts to projecting to
spinors of a given chirality. So after GSO projection we get a
massless spinor of a definite chirality. Let us denote the spinor of
one chirality by $s$ and the other one by $s'$.

Now let us combine the left- and right-moving sectors together. Here
we run into two distinct possibilities: A) The GSO projections on the
left- and right-movers are different and lead in the R sector to
ground states with different chirality. B) The GSO projections on the
left- and right-movers are the same and lead in the R sector to
ground states with the same chirality.  The first case is known as
type IIA superstring and the second one as type IIB.  Let us see what
kind of massless modes we get for either of them.  From NS$\otimes$NS
we find for both type IIA,B \ba\nl NS\otimes NS \rightarrow v\otimes
v \rightarrow (g_{\mu \nu},B_{\mu \nu},\phi ) \ea From the $NS
\otimes R$ and $R \otimes NS$ we get the fermions of the theory
(including the gravitinos).  However the IIA and IIB differ in that
the gravitinos of IIB are of the same chirality, whereas for IIA they
are of the opposite chirality.  This implies that IIB is a chiral
theory whereas IIA is non-chiral. Let us move to the R$\otimes$ R
sector.  We find \ba\nl IIA: R\otimes R =s\otimes s' \rightarrow
(A_\mu, C_{\mu \nu \rho})\\ \label{cont}IIB: R\otimes R =s\otimes
s\rightarrow (\chi,B'_{\mu \nu}, D_{\mu\nu\rho \lambda}), \ea where
all the tensors appearing above are fully antisymmetric. Moreover
$D_{\mu \nu \rho \lambda}$ has a self-dual field strength $F=dD=*F$.
It turns out that to write the equations of motion in a unified way
it is convenient to consider a generalized gauge fields ${\cal A}$
and ${\cal B}$ in the IIA and IIB case respectively by adding all the
fields in the RR sector together with the following properties: i)
${\cal A} ({\cal B})$ involve all the odd (even) dimensional
antisymmetric fields. ii) the equation of motion is $d{\cal A}=*
d{\cal A}$.  In the case of all fields (except $D_{\mu\nu\lambda
\rho}$) this equation allows us to solve for the forms with degrees
bigger than 4 in terms of the lower ones and moreover it implies the
field equation $d*dA=0$ which is the familiar field equation for the
gauge fields. In the case of the $D$-field it simply gives that its
field strength is self-dual.

\subsubsection{Open Superstring: Type I String}
\hspace{1cm} In the case of type IIB theory in 10 dimensions, we
note that the left- and right-moving degrees of freedom on the
worldsheet are the same.  In this case we can `mod out' by a
reflection symmetry on the string; this means keeping only the
states in the full Hilbert space which are invariant under the
left-/right-moving exchange of quantum numbers.  This is simply
projecting the Hilbert space onto the invariant subspace of the
projection operator $P={1\over 2}(1+\Omega)$ where $\Omega$
exchanges left- and right-movers. $\Omega$ is known as the
orientifold operation as it reverses the orientation on the
world-sheet.
 Note that this symmetry
only exists for IIB and not for IIA theory (unless we accompany it
with a parity reflection in spacetime).  Let us see which bosonic
states we will be left with after this projection. {}From the NS-NS
sector $B_{\mu\nu}$ is odd and projected out and thus we are left
with the symmetric parts of the tensor product \ba\nl
NS-NS\rightarrow (v\otimes v)_{symm.}=(g_{\mu \nu},\phi ). \ea From
the R-R sector since the degrees of freedom are fermionic from each
sector we get, when exchanging left- and right-movers an extra minus
sign which thus means we have to keep anti-symmetric parts of the
tensor product \ba\nl R-R\rightarrow (s\otimes
s)_{anti-symm.}={\tilde B}_{\mu \nu}. \ea
This is not the end of the
story, however.  In order to make the theory consistent we need to
introduce a new sector in this theory involving open strings.  This
comes about from the fact that in the R-R sector there actually is a
10 form gauge potential which has no propagating degree of freedom,
but acquires a tadpole.  Introduction of a suitable open string
sector cancels this tadpole.

As noted before the construction of open string sector Hilbert space
proceeds as in the closed string case, but now, the left-moving and
right-moving modes become indistinguishable due to reflection off the
boundaries of open string.  We thus get only one copy of the
oscillators. Moreover we can associate `Chan-Paton' factors to the
boundaries of open string . To cancel the tadpole it turns out that
we need 32 Chan-Paton labels on each end. We still have two sectors
corresponding to the NS and R sectors.  The NS sector gives a vector
field $A_\mu$ and the R sector gives the gaugino.  The gauge field
$A_{\mu}$ has two additional labels coming from the end points of the
open string and it turns out that the left-right exchange projection
of the type IIB theory translates to keeping the antisymmetric
component of $A_\mu =-A_\mu^T$, which means we have an adjoint of
$SO(32)$. Thus all put together, the bosonic degrees of freedom are
\ba\nl (g_{\mu \nu},{\tilde B}_{\mu\nu},\phi)+(A_\mu)_{SO(32)}. \ea
We should keep in mind here that $\tilde B$ came not from the NS-NS
sector, but from the R-R sector.

\subsubsection{Heterotic Strings}
\hspace{1cm} Heterotic string is a combination of bosonic string
and superstring, where roughly speaking the left-moving degrees of
freedom are as in the bosonic string and the right-moving degrees
of freedom are as in the superstring. It is clear that this makes
sense for the construction of the states because the left- and
right-moving sectors hardly talk with each other.  This is almost
true, however they are linked together by the zero modes of the
bosonic oscillators which give rise to momenta $(P_L,P_R)$.
Previously we had $P_L=P_R$ but now this cannot be the case
because $P_L$ is 26 dimensional but $P_R$ is 10 dimensional.  It
is natural to decompose $P_L$ to a 10+16 dimensional vectors,
where we identify the 10 dimensional part of it with $P_R$. It
turns out that for the consistency of the theory the extra 16
dimensional component should belong to the root lattice of
$E_8\times E_8$ or a $Z_2$ sublattice of $SO(32)$ weight lattice.
In either of these two cases the vectors in the lattice with
$(length)^2=2$ are in one to one correspondence with non-zero
weights in the adjoint of $E_8\times E_8$ and $SO(32)$
respectively.  These can also be conveniently represented (through
bosonization) by 32 fermions:  In the case of $E_8\times E_8$ we
group them to two groups of 16 and consider independent NS, R
sectors for each group.
 In the case of $SO(32)$ we only
have one group of $32$ fermions with either NS or R boundary
conditions.

Let us tabulate the massless modes using (\ref{spec}). The
right-movers can be either NS or R. The left-moving degrees of
freedom start out with a tachyonic mode.  But (\ref{spec}) implies
that this is not satisfying the level-matching condition because the
right-moving ground state is at zero energy.  Thus we should search
on the left-moving side for states with $L_0=0$ which means from
(\ref{spec}) that we have either $N_L=1$ or $(1/2)P_L^2=1$, where
$P_L$ is an internal 16 dimensional vector in one of the two lattices
noted above. The states with $N_L=1$ are \ba\nl 16\oplus v , \ea
where 16 corresponds to the oscillation direction in the extra 16
dimensions and $v$ corresponds to vector in 10 dimensional spacetime.
States with $(1/2)P_L^2=1$ correspond to the non-zero weights of the
adjoint of $E_8\times E_8$ or $SO(32)$ which altogether correspond to
$480$ states in both cases.  The extra 16 $N_L=1$ modes combine with
these $480$ states to form the adjoints of $E_8\times E_8$ or
$SO(32)$ respectively. The right-movers give, as before, a $v\oplus
s$ from the NS and R sectors respectively.  So putting the left- and
right-movers together we finally get for the massless modes \ba\nl
(v\oplus Adj)\otimes (v\oplus s). \ea Thus the bosonic states are
$(v\oplus Adj)\otimes v$ which gives \ba\nl (g_{\mu \nu},B_{\mu
\nu},\phi ; A_{\mu}), \ea where the $A_\mu$ is in the adjoint of
$E_8\times E_8$ or $SO(32)$. Note that in the $SO(32)$ case this is
an {\it identical} spectrum to that of type I strings.

\subsubsection{Summary}
\hspace{1cm} To summarize, we have found 5 consistent strings in
10 dimensions: Type IIA with $N=2$ non-chiral supersymmetry, type
IIB with $N=2$ chiral supersymmetry, type I with N=1 supersymmetry
and gauge symmetry $SO(32)$ and heterotic strings with N=1
supersymmetry with $SO(32)$ or $E_8\times E_8$ gauge symmetry.
Note that as far as the massless modes are concerned we only have
four inequivalent theories, because heterotic $SO(32)$ theory and
Type I theory have the same light degrees of freedom. In
discussing compactifications it is sometimes natural to divide the
discussion between two cases depending on how many supersymmetries
we start with.  In this context we will refer to the type IIA and
B as $N=2$ {\it theories} and Type I and heterotic strings as
$N=1$ {\it theories}.

\subsection{\bf String Compactifications}
\hspace{1cm} So far we have only talked about superstrings
propagating in 10 dimensional Minkowski spacetime.  If we wish to
connect string theory to the observed four dimensional spacetime,
somehow we have to get rid of the extra 6 directions. One way to do
this is by assuming that the extra 6 dimensions are tiny and thus
unobservable in the present day experiments. In such scenarios we
have to understand strings propagating not on ten dimensional
Minkowski spacetime but on four dimensional Minkowski spacetime times
a compact 6 dimensional manifold $K$.  In order to gain more insight
it is convenient to consider compactifications not just to 4
dimensions but to arbitrary dimensional spacetimes, in which case the
dimension of $K$ is variable.

The choice of $K$ and the string theory we choose to start in 10
dimensions will lead to a large number of
 theories in diverse dimensions,
which have different number of supersymmetries and different low
energy effective degrees of freedom. In order to get a handle on such
compactifications it is useful to first classify them according to
how much supersymmetry they preserve.  This is useful because the
higher the number of supersymmetry the less the quantum corrections
there are.

If we consider a general manifold $K$ we find that the
supersymmetry is completely broken.  This is the case we would
really like to understand, but it turns out that string
perturbation theory always breaks down in such a situation;  this
is intimately connected with the fact that typically cosmological
constant is generated by perturbation theory and this destabilizes
the Minkowski solution. For this reason we do not even have a
single example of such a class whose dynamics we understand.
Instead if we choose $K$ to be of a special type we can preserve a
number of supersymmetries.

For this to be the case, we need $K$ to admit some number of
covariantly constant spinors.  This is the case because the number
of supercharges which are `unbroken' by compactification is
related to how many covariantly constant spinors we have. To see
this note that if we wish to define a {\it constant} supersymmetry
transformation, since a space-time spinor, is also a spinor of
internal space, we need in addition a constant spinor in the
internal compact directions. The basic choices are manifolds with
trivial holonomy (flat tori are the only example), $SU(n)$
holonomy (Calabi-Yau n-folds), $Sp(n)$ holonomy (4n dimensional
manifolds),
 7-manifolds of $G_2$ holonomy
and 8-manifolds of $Spin(7)$ holonomy.  The case mostly studied in
physics involves toroidal compactification, $SU(2)=Sp(1)$ holonomy
manifold (the 4-dimensional $K3$), $SU(3)$ holonomy (Calabi-Yau
3-folds). Calabi-Yau 4-folds have also recently appeared in
connection with F-theory compactification to 4 dimensions
\cite{V96,MV962,MV963,S965,S98}. The cases of $Sp(2)$ holonomy
manifolds (8 dimensional) and $G_2$ and $Spin(7)$ end up giving us
compactifications below 4 dimensions.

\subsubsection{Toroidal Compactifications}
\hspace{1cm} The space with maximal number of covariantly constant
spinors is the flat torus $T^d$.  This is also the easiest to
describe the string propagation in. The main modification to the
construction of the Hilbert space from flat non-compact space in
this case involves relaxing the condition $P_L=P_R$ because the
string can wrap around the internal space and so $X$ does not need
to come back to itself as we go around $\sigma$. In particular if
we consider compactification on a circle of radius $R$ we can have
\ba\nl (P_L,P_R)=({n\over 2R}+mR,{n\over 2R}-mR). \ea Here $n$
labels the center of mass momentum of the string along the circle
and $m$ labels how many times the string is winding around the
circle.  Note that the spectrum of allowed $(P_L,P_R)$ is
invariant under $R\rightarrow 1/2R$.  All that we have to do is to
exchange the momentum and winding modes ($n\leftrightarrow m$).
This symmetry is a consequence of what is known as $T$-duality
\cite{GPR94}.

If we compactify on a d-dimensional torus $T^d$ it can be shown that
$(P_L,P_R)$ belong to a 2d dimensional lattice with signature
$(d,d)$.  Moreover this lattice is integral, self-dual and even.
Evenness means, $P_L^2-P_R^2$ is even for each lattice vector.
Self-duality means that any vector which has integral product with
all the vectors in the lattice sits in the lattice as well.  It is an
easy exercise to check these condition in the one dimensional circle
example given above.  Note that we can change the radii of the torus
and this will clearly affect the $(P_L,P_R)$.  Given any choice of a
d-dimensional torus compactifications, all the other ones can be
obtained by doing an $SO(d,d)$ Lorentz boost on $(P_L,P_R)$ vectors.
Of course rotating $(P_L,P_R)$ by an $O(d)\times O(d)$ transformation
does not change the spectrum of the string states, so the totality of
such vectors is given by \ba\nl SO(d,d)\over SO(d)\times SO(d). \ea
Some Lorentz boosts will not change the lattice and amount to
relabeling the states.  These are the boosts that sit in $O(d,d;Z)$
(i.e. boosts with integer coefficients), because they can be undone
by choosing a new basis for the lattice by taking an integral linear
combination of lattice vectors.  So the space of inequivalent choices
are actually given by \ba\nl SO(d,d)\over SO(d)\times SO(d)\times
O(d,d;Z). \ea The $O(d,d;Z)$ generalizes the T-duality considered in
the 1-dimensional case.

\subsubsection{Compactifications on $K3$}
\hspace{1cm} The four dimensional manifold $K3$ is the only
compact four dimensional manifold, besides $T^4$, which admits
covariantly constant spinors.  In fact it has exactly half the
number of covariantly constant spinors as on $T^4$ and thus
preserves half of the supersymmetry that would have been preserved
upon toroidal compactification. More precisely the holonomy of a
generic four manifold is $SO(4)$.  If the holonomy resides in an
$SU(2)$ subgroup of $SO(4)$ which leaves an $SU(2)$ part of
$SO(4)$ untouched, we end up with one chirality of $SO(4)$ spinor
being unaffected by the curvature of $K3$, which allows us to
define supersymmetry transformations as if $K3$ were flat (note a
spinor of $SO(4)$ decomposes as $({\bf 2}, {\bf 1})\oplus ({\bf
1},{\bf 2})$ of $SU(2)\times SU(2)$).

There are a number of realizations of $K3$, which are useful
depending on which question one is interested in. Perhaps the
simplest description of it is in terms of {\it orbifolds}. This
description of $K3$ is very close to toroidal compactification and
differs from it by certain discrete isometries of the $T^4$ which are
used to (generically) identify points which are in the same {\it
orbit} of the discrete group. Another description is as a 19 complex
parameter family of $K3$ defined by an algebraic equation.

Consider a $T^4$ which for simplicity we take to be parametrized by
four real coordinates $x_i$ with $i=1,...,4$, subject to the
identifications $x_i\sim x_i+1$. It is sometimes convenient to think
of this as two complex coordinates $z_1=x_1+ix_2$ and $z_2=x_3+ix_4$
with the obvious identifications.  Now we identify the points on the
torus which are mapped to each other under the $Z_2$ action
(involution) given by reflection in the coordinates $x_i\rightarrow
-x_i$, which is equivalent to \ba\nl z_i\rightarrow -z_i . \ea Note
that this action has $2^4=16$ fixed points given by the choice of
midpoints or the origin in any of the four $x_i$. The resulting space
is \ul{singular} at any of these 16 fixed points because the angular
degree of freedom around each of these points is cut by half. Put
differently, if we consider any primitive loop going `around' any of
these 16 fixed point, it corresponds to an open curve on $T^4$ which
connects pairs of points related by the $Z_2$ involution. Moreover
the parallel transport of vectors along this path, after using the
$Z_2$ identification, results in a flip of the sign of the vector.
This is true no matter how small the curve is.  This shows that we
cannot have a smooth manifold at the fixed points.

When we move away from the orbifold points of $K3$ the description of
the geometry of $K3$ in terms of the properties of the $T^4$ and the
$Z_2$ twist become less relevant, and it is natural to ask about
other ways to think about $K3$.  In general a simple way to define
complex manifolds is by imposing complex equations in a compact space
known as the projective $n$-space ${\bf CP}^n$.  This is the space of
complex variables $(z_1,...,z_{n+1})$ excluding the origin and
subject to the identification \ba\nl (z_1,...,z_{n+1})\sim \lambda
(z_1,...,z_{n+1}) \qquad \lambda \not=0 . \ea One then considers the
vanishing locus of a homogeneous polynomial of degree $d$,
$W_d(z_i)=0$ to obtain an $n-1$ dimensional subspace of ${\bf CP}^n$.
An interesting special case is when the degree is $d=n+1$. In this
case one obtains an $n-1$ complex dimensional manifold which admits a
Ricci-flat metric.  This is the case known as Calabi-Yau. For
example, if we take the case $n=2$, by considering cubics in it
\ba\nl z_1^3+z_2^3+z_3^3+az_1z_2z_3=0 \ea we obtain an elliptic
curve, i.e. a torus of complex dimension 1 or real dimension 2.  The
next case would be $n=3$ in which case, if we consider a quartic
polynomial in ${\bf CP}^3$ we obtain the 2 complex dimensional $K3$
manifold: \ba\nl W=z_1^4+z_2^4+z_3^4+z_4^4 +{\rm deformations}=0 .
\ea There are 19 inequivalent quartic terms we can add.  This gives
us a 19 dimensional complex subspace of 20 dimensional complex moduli
of the $K3$ manifold.  Clearly this way of representing $K3$ makes
the complex structure description of it very manifest, and makes the
Kahler structure description implicit.

Note that for a generic quartic polynomial the $K3$ we obtain is
non-singular.  This is in sharp contrast with the orbifold
construction which led us to 16 singular points.  It is possible to
choose parameters of deformation which lead to singular points for
$K3$. For example if we consider \ba\nl z_1^4+z_2^4+z_3^4+z_4^4+4
z_1z_2z_3z_4=0 \ea it is easy to see that the resulting $K3$ will
have a singularity (one simply looks for non-trivial solutions to
$dW=0$).

There are other ways to construct Calabi-Yau manifolds and in
particular $K3$'s.  One natural generalization to the above
construction is to consider weighted projective spaces where the
$z_i$ are identified under different rescalings. In this case one
considers quasi-homogeneous polynomials to construct submanifolds.

\subsubsection{Calabi-Yau Threefolds}
\hspace{1cm} Calabi-Yau threefolds are manifolds with $SU(3)$
holonomy. The compactification on manifolds of $SU(3)$ holonomy
preserves 1/4 of the supersymmetry.  In particular if we
compactify $N=2$  theories on Calabi-Yau threefolds we obtain
$N=2$ theories in $d=4$, whereas if we consider $N=1$ theories we
obtain $N=1$ theories in $d=4$.

If we wish to construct the Calabi-Yau threefolds as toroidal
orbifolds we need to consider six dimensional tori, three complex
dimensional, which have discrete isometries residing in $SU(3)$
subgroup of the $O(6)=SU(4)$ holonomy group.  A simple example is if
we consider the product of three copies of $T^2$ corresponding to the
Hexagonal lattice and mod out by a simultaneous ${\bf Z}_3$ rotation
on each torus (this is known as the `Z-orbifold'). This $Z_3$
transformation has $27$ fixed points which can be blown up to give
rise to a smooth Calabi-Yau.

We can also consider description of Calabi-Yau threefolds in
algebraic geometry terms for which the complex deformations of the
manifold can be typically realized as changes of coefficients of
defining equations, as in the $K3$ case. For instance we can consider
the projective 4-space ${\bf CP}^4$ defined by 5 complex not all
vanishing coordinates $z_i$ up to overall rescaling, and consider the
vanishing locus of a homogeneous degree 5 polynomial \ba\nl
P_5(z_1,...,z_5)=0. \ea This defines a Calabi-Yau threefold, known as
the quintic three-fold. This can be generalized to the case of
product of several projective spaces with more equations. Or it can
be generalized by taking the coordinates to have different
homogeneity weights. This will give a huge number of Calabi-Yau
manifolds.

\subsection{\bf Solitons in String Theory}
\hspace{1cm} Solitons arise in field theories when the vacuum
configuration of the field has a non-trivial topology which allows
non-trivial wrapping of the field configuration at spatial
infinity around the vacuum manifold.  These will carry certain
topological charge related to the `winding' of the field
configuration around the vacuum configuration. Examples of
solitons include magnetic monopoles in four dimensional
non-abelian gauge theories with unbroken $U(1)$, cosmic strings
and domain walls. The solitons naively play a less fundamental
role than the fundamental fields which describe the quantum field
theory. In some sense we can think of the solitons as `composites'
of more fundamental elementary excitations. However as is well
known, at least in certain cases, this is just an illusion. In
certain cases it turns out that we can reverse the role of what is
fundamental and what is composite by considering a different
regime of parameter.  In such regimes the soliton may be viewed as
the elementary excitation and the previously viewed elementary
excitation can be viewed as a soliton. A well known example of
this phenomenon happens in 2 dimensional field theories. Most
notably the boson/fermion equivalence in the two dimensional
sine-Gordon model, where the fermions may be viewed as solitons of
the sine-Gordon model and the boson can be viewed as a composite
of fermion-anti-fermion excitation.  Another example is the
T-duality we have already discussed in the context of 2
dimensional world sheet of strings which exchanges the radius of
the target space with its inverse.  In this case the winding modes
may be viewed as the solitons of the more elementary excitations
corresponding to the momentum modes.  As discussed before
$R\rightarrow 1/R$ exchanges momentum and winding modes. In
anticipation of generalization of such dualities to string theory,
it is thus important to study various types of solitons that may
appear in string theory.

As already mentioned solitons typically carry some conserved
topological charge.  However in string theory every conserved charge
is a gauge symmetry.  In fact this is to be expected from a theory
which includes quantum gravity.  This is because the global charges
of a black hole will have no influence on the outside and by the time
the black hole disappears  due to Hawking radiation, so does the
global charges it may carry. So the process of formation and
evaporation of black hole leads to a non-conservation of global
charges.  Thus for any soliton, its conserved topological charge must
be a gauge charge.  This may appear to be somewhat puzzling in view
of the fact that solitons may be point-like as well as string-like,
sheet-like etc.  We can understand how to put a charge on a
point-like object and gauge it.  But how about the higher dimensional
extended solitonic states? Note that if we view the higher
dimensional solitons as made of point-like structures the soliton has
no stability criterion as the charge can disintegrate into little
bits.

Let us review how it works for point particles (or point solitons):
We have a 1-form gauge potential $A_{\mu}$ and the coupling of the
particle to the gauge potential involves weighing the world-line
propagating in the space-time with background $A_{\mu}$ by \ba\nl
Z\rightarrow Z {\rm exp}(i \int_\gamma A), \ea where $\gamma$ is the
world line of the particle.  The gauge principle follows from
defining an action in terms of $F=dA$: \ba\label{acti} S=\int F\wedge
*F ,\ea where $*F$ is the dual of the $F$, where we note that
shifting $A\rightarrow d\epsilon$ for arbitrary function $\epsilon$
will not modify the action.

Suppose we now consider instead of a point particle a
$p$-dimensional extended object.  In this convention $p=0$
corresponds to the case of point particles and $p=1$ corresponds
to strings and $p=2$ corresponds to membranes, etc.  We shall
refer to $p$-dimensional extended objects as $p$-branes
(generalizing `membrane').  Note that the world-volume of a
$p$-brane is a $p+1$ dimensional subspace $\gamma_{p+1}$ of
space-time.  To generalize what we did for the case of point
particles we introduce a gauge potential which is a $p+1$ form
$A_{p+1}$ and couple it to the charged $p+1$ dimensional state by
\ba\nl Z\rightarrow Z {\rm exp}(i \int_{\gamma _{p+1}} A_{p+1}).
\ea Just as for the case of the point particles we introduce the
field strength $F=dA$ which is now a totally antisymmetric $p+2$
tensor. Moreover we define the action as in (\ref{acti}), which
possesses the gauge symmetry $A\rightarrow d \epsilon$ where
$\epsilon$ is a totally antisymmetric tensor of rank $p$.

\subsubsection{Magnetically Charged States}
\hspace{1cm} The above charge defines the generalization of
electrical charges for extended objects. Can we generalize the
notion of magnetic charge? Suppose we have an electrically charged
particle in a theory with space-time dimension $D$. Then we
measure the electrical charge by surrounding the point by an
$S^{D-2}$ sphere and integrating $*F$ (which is a $D-2$ form) on
it, i.e. \ba\nl Q_E=\int_{S^{D-2}} *F. \ea Similarly it is natural
to define the magnetic charge.  In the case of $D=4$, i.e. four
dimensional space-time, the magnetically charged point particle
can be surrounded also by a sphere and the magnetic charge is
simply given by \ba\nl Q_M=\int_{S^2} F. \ea Now let us generalize
the notion of magnetic charged states for arbitrary dimensions $D$
of space-time and arbitrary electrically charged $p$-branes.  From
the above description it is clear that the role that $*F$ plays in
measuring the electric charge is played by $F$ in measuring the
magnetic charge.  Note that for a $p$-brane $F$ is $p+2$
dimensional, and $*F$ is $D-p-2$ dimensional.  Moreover, note that
a sphere surrounding a $p$-brane is a sphere of dimension $D-p-2$.
Note also that for $p=0$ this is the usual situation.  For higher
$p$, a $p$-dimensional subspace of the space-time is occupied by
the extended object and so the position of the object is denoted
by a point in the transverse $(D-1)-p$ dimensional space which is
surrounded by an $S^{D-p-2}$ dimensional sphere.

Now for the magnetic states the role of $F$ and $*F$ are
exchanged: \ba\nl F\leftrightarrow *F. \ea To be perfectly
democratic we can also define a magnetic gauge potential $\tilde
A$ with the property that \ba\nl d\tilde A=*F=*dA. \ea In
particular noting that $F$ is a $p+2$ form, we learn that $*F$ is
an $D-p-2$ form and thus $\tilde A$ is an $D-p-3$ form. We thus
deduce that the magnetic state will be an $D-p-4$-brane (i.e. one
dimension lower than the degree of the magnetic gauge potential
${\tilde A}$). Note that this means that if we have an
electrically charged $p$-brane, with a magnetically charged dual
$q$-brane then we have \ba\label{mage} p+q=D-4. \ea This is an
easy sum rule to remember. Note in particular that for a
4-dimensional space-time an electric point charge ($p=0$) will
have a dual magnetic point charge ($q=0$). Moreover this is the
only space-time dimension where both the electric and magnetic
dual can be point-like.

Note that a p-brane wrapped around an r-dimensional compact object
will appear as a $p-r$-brane for the non-compact space-time.  This is
in accord with the fact that if we decompose the $p+1$ gauge
potential into an $(p+1-r)+r$ form consisting of an $r$-form in the
compact direction we will end up with an $p+1-r$ form in the
non-compact directions.  Thus the resulting state is charged under
the left-over part of the gauge potential.  A particular case of this
is when $r=p$ in which case we are wrapping a p-dimensional extended
object about a p-dimensional closed cycle in the compact directions.
This will leave us with point particles in the non-compact directions
carrying ordinary electric charge under the reduced gauge potential
which now is a 1-form.

\subsubsection{String Solitons}
\hspace{1cm} From the above discussion it follows that the charged
states will in principle exist if there are suitable gauge
potentials given by $p+1$-forms.  Let us first consider type II
strings. Recall that from the NS-NS sector we obtained an
anti-symmetric 2-form $B_{\mu\nu}$.  This suggests that there is a
1-dimensional extended object which couples to it by \ba\nl {\rm
exp}(i\int B). \ea But that is precisely how $B$ couples to the
world-sheet of the fundamental string.  We thus conclude that {\it
the fundamental string carries electric charge under the
antisymmetric field} $B$. What about the magnetic dual to the
fundamental string? According to (\ref{mage}) and setting $d=10$
and $p=1$ we learn that the dual magnetic state will be a 5-brane.
Note that as in the field theories, we expect that in the
perturbative regime for the fundamental fields, the solitons be
very massive. This is indeed the case and the 5-brane magnetic
dual can be constructed as a solitonic state of type II strings
with a mass per unit 5-volume going as $1/g_{s}^2$ where $g_{s}$
is the string coupling. Conversely, in the strong coupling regime
these 5-branes are light and at infinite coupling they become
massless, i.e. \ul{tensionless} 5-branes \cite{DKL95}.

Let us also recall that type II strings also have anti-symmetric
fields coming from the R-R sector. In particular for type IIA strings
we have 1-form $A_{\mu}$ and 3-form $C_{\mu\nu \rho}$ gauge
potentials. Note that the corresponding magnetic dual gauge fields
will be 7-forms and 5-forms respectively (which are not independent
degrees of freedom).  We can also include a 9-form potential which
will have trivial dynamics in 10  dimensions.  Thus it is natural to
define a generalized gauge field ${\cal A}$ by taking the sum over
all odd forms and consider the equation ${\cal F}=*{\cal F}$ where
${\cal F}=d{\cal A}$.  A similar statement applies to the type IIB
strings where from the R-R sector we obtain all the even-degree gauge
potentials (the case with degree zero can couple to a ``$-1$-brane''
which can be identified with an instanton, i.e. a point in
space-time). We are thus led to look for p-branes with even $p$ for
type IIA and odd $p$ for type IIB which carry charge under the
corresponding RR gauge field.  It turns out that surprisingly enough
the states in the elementary excitations of string all are neutral
under the RR fields.  We are thus led to look for solitonic states
which carry RR charge. Indeed there are such p-branes and they are
known as $D$-branes \cite{P96,B98}, as we will now review.

\subsubsection{D-Branes}
\hspace{1cm} In the context of field theories constructing
solitons is equivalent to solving classical field equations with
appropriate boundary conditions. For string theory the condition
that we have a classical solution is equivalent to the statement
that propagation of strings in the corresponding background would
still lead to a conformal theory on the worldsheet of strings, as
is the case for free theories.

In search of such stringy p-branes, we are thus led to consider how
could a p-brane modify the string propagation.  Consider an $p+1$
dimensional plane, to be identified with the world-volume of the
$p$-brane. Consider string propagating in this background.  How could
we modify the rules of closed string propagation given this $p+1$
dimensional sheet?  The simplest way turns out to allow closed
strings to open up and end on the $p+1$ dimensional world-volume.  In
other words we allow to have a new sector in the theory corresponding
to open string with ends lying on this $p+1$ dimensional subspace.
This will put Dirichlet boundary conditions on $10-p-1$ coordinates
of string endpoints.  Such $p$-branes are called $D$-branes, with D
reminding us of Dirichlet boundary conditions. In the context of type
IIA,B we also have to specify what boundary conditions are satisfied
by fermions. This turns out to lead to consistent boundary conditions
only for $p$ even for type IIA string and $p$ odd for type IIB. This
is a consequence of the fact that for type IIA(B), left-right
exchange is a symmetry only when accompanied by a $Z_2$ spatial
reflection with determinant -1(+1). Moreover, it turns out that they
do carry the corresponding RR charge \cite{P95}.

Quantizing the new sector of type II strings in the presence of
D-branes is rather straightforward. We simply consider the set of
oscillators as before, but now remember that due to the Dirichlet
boundary conditions on some of the components of string coordinates,
the momentum of the open string lies on the $p+1$ dimensional
world-volume of the D-brane. It is thus straightforward to deduce
that the massless excitations propagating on the D-brane will lead to
the dimensional reduction of $N=1$, $U(1)$ Yang-Mills from $d=10$ to
$p+1$ dimensions. In particular the $10-(p+1)$ scalar fields living
on the D-brane, signify the D-brane excitations in the $10-(p+1)$
transverse dimensions. This tells us that the significance of the new
open string subsector is to quantize the D-brane excitations.

An important property of D-branes is that when $N$ of them coincide
we get a $U(N)$ gauge theory on their world-volume.  This follows
because we have $N^2$ open string subsectors going from one D-brane
to another and in the limit they are on top of each other all will
have massless modes and we thus obtain the reduction of $N=1$ $U(N)$
Yang-Mills from $d=10$ to $d=p+1$.

Another important property of D-branes is that they are BPS states.
A BPS state is a state which preserves a certain number of
supersymmetries and as a consequence of which one can show that their
mass (per unit volume) and charge are equal. This in particular
guarantees their absolute stability against decay.

If we consider the tension of D-branes, it is proportional to
$1/g_{s}$, where $g_{s}$ is the string coupling constant.  Note that
as expected at weak coupling they have a huge tension. At strong
coupling their tension goes to zero and they become \ul{tensionless}.

We have already discussed that in $K3$ compactification of string
theory we end up with singular limits of manifolds when some cycles
shrink to zero size. What is the physical interpretation of this
singularity?

Suppose we consider for concreteness an $n$-dimensional sphere $S^n$
with volume $\epsilon\rightarrow 0$. Then the string perturbation
theory breaks down when $\epsilon << g_{s}$, where $g_{s}$ is the
string coupling constant.  If we have  $n$-brane solitonic states
such as D-branes then we can consider a particular solitonic state
corresponding to wrapping the $n$-brane on the vanishing $S^n$.  The
mass of this state is proportional to $\epsilon$, which implies that
in the limit $\epsilon \rightarrow 0$ we obtain a massless soliton.
An example of this is when we consider type IIA compactification on
$K3$ where we develop a singularity. Then by wrapping D2-branes
around vanishing $S^2$'s of the singularity we obtain massless
states, which are vectors. This in fact implies that in this limit we
obtain enhanced gauge symmetry. Had we been considering type IIB on
$K3$ near the singularity, the lightest mode would be obtained by
wrapping a D3-brane around vanishing $S^2$'s, which leaves us with a
string state with tension of the order of $\epsilon$ \cite{W95},
\cite{DH96}. This kind of regime which exists in other examples of
compactifications as well is called the phase with \ul{tensionless}
strings \cite{S95,S96,GH96,SW96,DLP96,W96,G965,HK96,KS96,D96,BJ96,
M96,CKRRSW97,HOV97,KMV97,S97,DLLP97,MNVW98,KOY98,AO99}.

We could consider higher dimensional D-branes wrapping around the
vanishing cycles and obtain \ul{tensionless} $p$-branes with $p>1$,
but in such cases by dimensional analysis one can see that the
relevant mass scale would be smallest for the smallest dimension
D-brane.

\subsection{\bf From M-Theory to Tensionless Strings }
\hspace{1cm} M-theory is the hypothetical unification of several
types of 10-dimensional strings \cite{HT94,EW95}. Its low-energy
effective description is the 11-dimensional supergravity, and some
valuable information can be extracted from its classical
solutions. The basic dynamical objects of the M-theory are the
2-brane, which is electrically charged under the 3-form gauge
field, and its magnetic dual, the 5-brane. The dynamics governing
the M-brane interactions is by no means well-understood. Some of
its features may be inferred, however, from compactification to
string theory, where the RR charged $p$-branes have a remarkably
simple description in terms of the D-branes \cite{DLP89,P95,B98}.

The D-branes are objects on which the fundamental strings are
allowed to end. There is evidence that a similar phenomenon takes
place in M-theory: the fundamental 2-branes are allowed to have
boundaries on the solitonic 5-branes \cite{T95,S95}. Thus, the
5-brane is the D-object of M-theory. The boundary of a 2-brane is
a string, and the resulting boundary dynamics appears to reduce to
a kind of string theory defined on the $5+1$ dimensional
world-volume. This picture has a number of interesting
implications. Consider, for instance, two parallel 5-branes with a
2-brane stretched between them \cite{S95}. The two boundaries of
the 2-brane give rise to two strings, lying within the first and
second 5-branes respectively. The tension of these strings may be
made arbitrarily small as the 5-branes are brought close together.
In particular, it can be made much smaller than the Planck scale,
which implies that the effective $5+1$ dimensional string theory
is decoupled from gravity \cite{W95}. While it is not clear how to
describe such a string theory in world sheet terms, it has been
suggested that its spectrum is given by the Green-Schwarz approach
\cite{JS96} ($5+1$ is one of the dimensions where the
Green-Schwarz string is classically consistent). In the limit of
coincident 5-branes, we seem to find a theory of \ul{tensionless}
strings. These strings carry $(0,4)$ supersymmetry in $5+1$
dimensions. Strings with $(0,2)$ supersymmetry were explored from
several different points of view in refs. \cite{GH96,SW96,DLP96}
for example.

\newpage
\section{\bf APPEARANCE OF THE NULL BRANES}
\hspace{1cm} In the previous section we saw on examples how
tensionless strings and branes may appear in modern string theory.
Here we are going to consider these and other cases in some details.

\subsection{\bf Compactifications}
\hspace{1cm} In \cite{S95} it is shown that many of the $p$-branes of
type II string theory and $D=11$ supergravity can have boundaries on
other $p$-branes. The rules for when this can and cannot occur are
derived from charge conservation. For example it is found that
membranes in $D=11$ supergravity and IIA string theory can have
boundaries on fivebranes. The boundary dynamics are governed by the
self-dual $D=6$ string. A collection of $N$ parallel fivebranes
contains $N(N-1)/2$ self-dual strings which become tensionless as the
fivebranes approach one another.

In \cite{S96} the author analyzes $M$-theory compactified on
$(K3\times S^1)/Z_2$ where the $Z_2$ changes the sign of the three
form gauge field, acts on $S^1$ as a parity transformation and on
$K3$ as an involution with eight fixed points preserving SU(2)
holonomy. At a generic point in the moduli space the resulting theory
has as its low energy limit $N=1$ supergravity theory in six
dimensions with eight vector, nine tensor and twenty hypermultiplets.
The gauge symmetry can be enhanced ({\it e.g.} to $E_8$) at special
points in the moduli space. At other special points in the moduli
space tensionless strings appear in the theory.

In \cite{GH96} T-duality is used to extract information on an
instanton of zero size in the $E_8\times E_8$ heterotic string.
The authors discuss the possibility of the appearance of a
tensionless anti-self-dual non-critical string through an
implementation of the mechanism suggested by Strominger of two
coincident 5-branes \cite{S95}. It is argued that when an
instanton shrinks to zero size a tensionless non-critical string
appears at the core of the instanton. It is further conjectured
that appearance of tensionless strings in the spectrum leads to
new phase transitions in six dimensions in much the same way as
massless particles do in four dimensions.

The paper \cite{SW96} discusses the singularities in the moduli space
of string compactifications to six dimensions with $N=1$
supersymmetry. Such singularities arise  from either massless
particles or  non-critical tensionless strings.  The points with
tensionless strings are sometimes phase transition points between
different phases of the theory.  These results appear to connect all
known $N=1$ supersymmetric six-dimensional vacua.

Heterotic strings on $R^6 \times K3$ generically appear to undergo
some interesting new phase transition at that value of the string
coupling for which the one of the six-dimensional gauge field kinetic
energies changes sign.  An exception is the $E_8 \times E_8$ string
with equal instanton numbers in the two $E_8$'s, which admits a
heterotic/heterotic self-duality. In \cite{DLP96} the dyonic string
solution of the six-dimensional heterotic string is generalized to
include non-trivial gauge field configurations corresponding to
self-dual Yang-Mills instantons in the four transverse dimensions. It
is found that vacua which undergo a phase transition always admit a
string solution exhibiting a naked singularity, whereas for vacua
admitting a self-duality the solution is always regular.  When there
is a phase transition, there exists a choice of instanton numbers for
which the dyonic string is tensionless and quasi-anti-self-dual at
that critical value of the coupling.  For an infinite subset of the
other choices of instanton number, the string will also be
tensionless, but all at larger values of the coupling.

Phase transitions in $M$-theory and $F$-theory are studied in
\cite{W96}. In $M$-theory compactification to five dimensions on a
Calabi-Yau, there are topology-changing transitions similar to those
seen in conformal field theory, but the non-geometrical phases known
in conformal field theory are absent. At boundaries of moduli space
where such phases might have been expected, the moduli space ends, by
a conventional or unconventional physical mechanism. The
unconventional mechanisms, which roughly involve the appearance of
tensionless strings, can sometimes be better understood in
$F$-theory.

When $N$ five-branes of M-theory coincide, the world-volume theory
contains tensionless strings, according to Strominger's
construction. This suggests a large $N$ limit of tensionless
string theories. For the small $E_8$ instanton theories, the
definition would be a large instanton number. An adiabatic
argument suggests \cite{G965} that in the large $N$ limit an
effective extra uncompactified dimension might be observed. In
\cite{G965} a kind of ``surface-equations'' are also proposed and
might describe correlators in the tensionless string theories. In
these equations, the anti-self-dual two forms of 6D and the
tensionless strings enter on an equal footing.

In \cite{HK96} the authors argue for the existence of phase
transitions in $3+1$ dimensions associated with the appearance of
tensionless strings. The massless spectrum of this theory does not
contain a graviton: it consists of one $N=2$ vector multiplet and one
linear multiplet, in agreement with the light-cone analysis of the
Green-Schwarz string in $3+1$ dimensions. In M-theory the string
decoupled from gravity arises when two 5-branes intersect over a
three-dimensional hyperplane. The two 5-branes may be connected by a
2-brane, whose boundary becomes a tensionless string with $N=2$
supersymmetry in $3+1$ dimensions.

A class of four dimensional N=1 compactifications of the $SO(32)$
heterotic/type I string theory which are destabilized by
nonperturbatively generated superpotentials are described in
\cite{KS96}. In the type I description, the destabilizing
superpotential is generated by a one instanton effect or gaugino
condensation in a nonperturbative $SU(2)$ gauge group.  The dual,
heterotic description involves destabilization due to world-sheet
instanton or $\it half$ world-sheet instanton effects in the two
cases. The analysis performed, also suggests that the tensionless
strings which arise in the $E_8 \times E_8$ theory in six dimensions
when an instanton shrinks to zero size should, in some cases, have
supersymmetry breaking dynamics upon further compactification to four
dimensions.

In the article \cite{D96} the appearance of tensionless strings in
M-theory is examined. These tensionless strings are subsequently
interpreted in a string theory context.  In particular,
tensionless strings appearing in M-theory on $S^{1}$, M-theory on
$S^{1} / {\bf Z}_{2}$, and M-theory on $T^{2}$ are examined. An
interpretation is given for the appearance of such strings in a
string theory context. Then the appearance of some tensionless
strings in string theory is examined. Subsequently the author
interprets these tensionless strings in a M-theory context.

In \cite{BJ96} F-theory \cite{V96} on elliptic threefold
Calabi-Yau near colliding singularities is studied. It is
demonstrated that resolutions of those singularities generically
correspond to transitions  to phases characterized by new tensor
multiplets and enhanced gauge symmetry. These are governed by the
dynamics of tensionless strings.

The work \cite{M96} is devoted to the examination of different
aspects of Calabi--Yau four-folds as compactification manifolds of
F-theory, using  mirror symmetry of toric hypersurfaces. A discussion
is given on the physical properties of the space-time theories, for a
number of examples which are dual to $E_8\times E_8$ heterotic $N=1$
theories. Non-critical strings of various kinds, with low tension for
special values of the moduli, lead to interesting physical effects. A
complete classification is given of those divisors in toric manifolds
that contribute to the non-perturbative four-dimensional
superpotential; the physical singularities associated to it are
related to the appearance of tensionless strings. In some cases
non-perturbative effects generate an everywhere non-zero quantum
tension leading to a combination of a conventional field theory with
light strings hiding at a low energy scale related to supersymmetry
breaking.

Type IIB strings compactified on K3 have a rich structure of
solitonic strings, transforming under $SO(21,5,Z)$. In \cite{DH96}
the BPS tension formula for these strings is derived, and their
properties, in particular, the points in the moduli space where
certain strings become tensionless are discussed. By examining
these tensionless string limits, the authors shed some further
light on the conjectured dual M-theory description of this
compactification.

In \cite{CKRRSW97} the authors study critical points of the  BPS mass
$Z$, the BPS string tension $Z_m$, the black hole potential $V$ and
the gauged central charge potential $P$ for M-theory compactified on
Calabi-Yau three-folds. They first show that the stabilization
equations for $Z$ (determining the black hole entropy) take an
extremely simple form in five dimensions as opposed to  four
dimensions. The stabilization equations for $Z_m$ are also very
simple and determine the size of the infinite $AdS_3$-throat of the
string. The black hole potential in general exhibits two classes of
critical points: supersymmetric critical points which coincide with
those of the central charge and non-supersymmetric critical points.
Then the discussion is generalized to the entire extended K\"ahler
cone encompassing topologically different but birationally equivalent
Calabi-Yau three-folds that are connected via flop transitions. The
behavior of the four potentials is examined to probe the nature of
these phase transitions. It is found that $V$ and $P$ are continuous
but not smooth across the flop transition, while $Z$ and its first
two derivatives, as well as $Z_m$ and its first derivative, are
continuous. This in turn implies that supersymmetric stabilization of
$Z$ and $Z_m$ for a given configuration takes place in at most one
point throughout the entire extended K\"ahler cone. The corresponding
black holes (or string states) interpolate between different
Calabi-Yau  three-folds. At the boundaries of the extended K\"ahler
cone electric states become massless and/or magnetic strings become
tensionless.

In \cite{HOV97} it is shown how Higgs mechanism for non-abelian
$N=2$ gauge theories in four dimensions is geometrically realized
in the context of type II strings as transitions among
compactifications of Calabi-Yau threefolds. This result and
T-duality are used for a further compacitification on a circle to
derive $N=4$, $D=3$ dual field theories.  This reduces dualities
for $N=4$ gauge systems in three dimensions to perturbative
symmetries of string theory. Moreover, the dual of a gauge system
always exists but may or may not correspond to a lagrangian
system. In particular the conjecture of Intriligator and Seiberg
is verified that an ordinary gauge system is dual to
compacitification of exceptional tensionless string theory down to
three dimensions.

In \cite{KMV97} using geometric engineering in the context of type
II strings, the authors obtain exact solutions for the moduli
space of the Coulomb branch of all $N=2$ gauge theories in four
dimensions involving products of $SU$ gauge groups with arbitrary
number of bi-fundamental matter for chosen pairs, as well as an
arbitrary number of fundamental matter for each factor. Asymptotic
freedom restricts the possibilities to $SU$ groups with
bi-fundamental matter chosen according to ADE or affine ADE Dynkin
diagrams. It is found that in certain cases the solution of the
Coulomb branch for $N=2$ gauge theories is given in terms of a
three dimensional complex manifold rather than a Riemann surface.
A new stringy strong coupling fixed points are studied, arising
from the compactification of higher dimensional theories with
tensionless strings. Applications are considered to three
dimensional $N=4$ theories.

In \cite{S97} an analysis is made of the world-volume theory of
multiple Kaluza-Klein monopoles in string and $M$- theory by
identifying the appropriate zero modes of various fields. The results
are consistent with supersymmetry, and all conjectured duality
symmetries. In particular for $M$-theory and type IIA string theory,
the low energy dynamics of $N$ Kaluza-Klein monopoles is described by
supersymmetric $U(N)$ gauge theory, and for type IIB string theory,
the low energy dynamics is described by a (2,0) supersymmetric field
theory in (5+1) dimensions with $N$ tensor multiplets and tensionless
self-dual strings.

In \cite{DLLP97} it is shown that six-dimensional supergravity
coupled to tensor and Yang-Mills multiplets admits not one but two
different theories as global limits, one of which was previously
thought not to arise as a global limit and the other of which is new.
The new theory has the virtue that it admits a global anti-self-dual
string solution obtained as the limit of the curved-space gauge
dyonic string, and can, in particular, describe tensionless strings.
It is speculated that this global model can also represent the
world-volume theory of coincident branes.

Certain properties of six-dimensional tensionless E-strings (arising
from zero size $E_8$ instantons) are studied in \cite{MNVW98}. In
particular, it is shown that $n$ E-strings form a bound string which
carries an $E_8$ level $n$ current algebra as well as a left-over
conformal system with $c=12n-4-248n/(n+30)$, whose characters can be
computed. Moreover, it is shown that the characters of the $n$-string
bound state are captured by $N=4$ $U(n)$ topological Yang-Mills
theory on $K3/2$.

Novel 3+1 dimensional $N=2$ superconformal field theories with
tensionless BPS string solitons are believed to arise when two
sets of M5 branes intersect over a 3+1 dimensional hyperplane. A
DLCQ (discrete light-cone quantization) description \cite{S98} of
these theories is derived in \cite{KOY98} as supersymmetric
quantum mechanics on the Higgs branch of suitable four dimensional
$N=1$ supersymmetric gauge theories. This formulation allows one
to determine the scaling dimensions of certain chiral primary
operators in the conformal field theories.

In \cite{AO99} $N=2$ superconformal theories defined on a $3+1$
dimensional hyperplane intersection of two sets of M5 branes are
considered. These theories have tensionless BPS string solitons. A
dual supergravity formulation is used to deduce some of their
properties via the AdS/CFT correspondence \cite{Mal97,AGMOO99}.

In \cite{BH97} $N=1$ four dimensional gauge theories are studied
as the world-volume theory of D4-branes between NS 5-branes. A
mechanism is proposed for enhanced chiral symmetry in the brane
construction which is associated with tensionless {\it
three-branes} in six dimensions. This mechanism can be explained
as follows. As is, by now well known, quantization of open strings
lead to massless hypermultiplets whenever two D-branes (which
break to 1/4 of the supersymmetry) meet in space. It is not known
however what are the states which get massless when a D brane
meets a NS brane. The analysis performed in \cite{BH97} predicts
that the states are vector multiplets. The only virtual states
which end on both a NS brane and a D6 brane are D4 branes. There
is no other brane which has this property. Thus we can have
virtual open D4-branes which have three-brane boundaries which
propagate on the world-volume of the D6 and NS branes. When these
two branes touch, the tension of the three-branes vanishes. The
world-volume of the D4-branes consists of 0123 and a real line in
the 456 space which connects the NS and D6 branes. A
supersymmetric configuration which is consistent with the
supersymmetries in this problem implies that the D6 and NS branes
will have identical 45 positions and different $x^6$ positions.
This gives a special case to the point where in the field theory
the masses are zero and chiral symmetry is expected. Thus one may
predict that quantization of tensionless three-branes in six
dimensions gives rise to massless vector multiplet in six
dimensions.

A simple proof is given in \cite{K99} of the known S-duality of
heterotic string theory compactified on a $T^{6}$. Using this
S-duality the tensions for a class of BPS 5-branes in heterotic
string theory on a $S^{1}$ is calculated. One of these, the
Kaluza-Klein monopole, becomes tensionless when the radius of the
$S^{1}$ is equal to the string length. Then the question of
stability of the heterotic NS 5-brane with a transverse circle is
studied. It turns out that for large radii the NS 5-brane is
absolutely stable.

\subsection{\bf High Energy Limit}
\hspace*{1cm} The characteristic scale of string theory is given
by the string tension $T_1 =(2\pi\alpha ')^{-1}$. At energies of
the order of $\sqrt{T_1}$ or higher, string physics truly
distinguishes itself from point particle physics and various high
energy limits have been studied to gain insight into the elusive
physical basis of the theory. One may view the high energy limit
of string theory as a {\it zero tension limit}, since only the
energy measured in string units, $E/\sqrt{T_1}$, is relevant.
There are different ways of taking the limit $T_1\mapsto 0$ in the
full theory. A natural choice is to take $T_1\mapsto 0$ in the
string action. One may than ask if the resulting null string can
be quantized and if interactions can be introduced. All this
applies also to the higher dimensional string generalizations -
the null $p$-branes of different kind.

Another reason for studying the zero tension limit of $p$-branes is
the duality between strings and 5-branes \cite{DKL95}. As we have
already discussed in the Introduction, the string theory admits
5-branes as solitonic solution, and vice versa, the 5-brane theory
admits strings as solitonic solutions. Moreover, the 5-brane tension
$T_5$ and the string tension $T_1$ are related through a Dirac type
quantization rule \ba\nl \kappa^2 T_1 T_5 = n\pi , \ea where $\kappa$
is the gravitational constant and $n$ is an integer. We see from this
relation that the small tension region of 5-branes is related to the
large tension region of strings, so studying the zero tension limit
of $p$-branes could actually teach us more about strings.

A further reason why it could be interesting to study such
tensionless extended objects is the appearance of null $Dp$-branes
in the high energy limit. As is known, static and moving
$Dp$-branes can either be perceived as space-like hyper-surfaces
on which open strings can end, or they can equivalently be
described by $Dp$-brane boundary states into which closed strings
can disappear. $Dp$-branes are characterized by a tension $T_{Dp}$
$(\neq 0)$ which can be computed by considering the exchange of
closed strings between two $Dp$-branes \cite{P95}. Because of this
non-vanishing tension these branes are either static, or they move
with velocities which are less than the speed of light. A natural
problem to pose is whether it is possible to extend the notion of
$Dp$-branes above to also include $Dp$-branes which move at the
speed of light, or equivalently $Dp$-branes which have a vanishing
tension $T_{Dp}=0$, i.e. tensionless $Dp$-branes.

The expression for $T_{Dp}$ in terms of the inverse of the
fundamental string tension $2\pi\alpha '$, and the string coupling
$g_{s}$, is given by
\begin{equation}
T_{Dp}=(2\pi)^{-(p-1)/2} g_{s}^{-1}(2\pi\alpha ')^{-(p+1)/2}\equiv
(2\pi)^{-(p-1)/2} g_{s}^{-1}T_p ,
\end{equation}
where $p\geq 0$ is the number of space-like directions in the
brane. When this expression is extrapolated to arbitrarily large
values of $g_{s}$, $T_{Dp}$ can attain values which are
arbitrarily close to zero. Hence, in the strongly coupled regime
$Dp$-branes become light states (since their masses $M_{Dp}$
behaves as $M_{Dp}\sim g_{s}^{-1}$), and when $g_{s}=\infty$ the
$Dp$-branes will become tensionless. This behaviour lies at the
very foundation of M(atrix)-theory \cite{BFSS96}.

\subsection{\bf Gravity and Cosmology}
\hspace{1cm} The investigations in the domain of classical and
quantum string propagation in curved space-times are relevant for the
physics of quantum gravitation as well as for the understanding of
the cosmic string models in cosmology \cite{LWC99}.

As we have already mention, strings are characterized by an energy
scale ${\sqrt T_1}$. The frequencies of the string modes are
proportional to $T_1$ and the length of the string scales with $1/
\sqrt T_1$. The gravitational field provides another length scale,
the curvature radius of the space-time $R_c$. For a string moving
in a gravitational field a useful parameter is the dimensionless
constant $C=R_c\sqrt T_1$. Large values of $C$ imply weak
gravitational field ( the metric does not change appreciably over
distances of the order of the string length). We may reach large
values of $C$ by letting $T_1 \to {\infty}$. In this limit the
string shrinks to a point. In the opposite limit, small values of
$C$, we encounter strong gravitational fields and it is
appropriate to consider $T_1 \to 0$, i.e. null or tensionless
strings.

It turns out that the concept of null strings also appear in the
strong coupling limit of {\it two dimensional} gravity \cite{G94}.
The gravitational constant $\kappa$ is the analog of the Regge
slope ($\alpha^{'}$) and when $\kappa\rightarrow \infty$, 2
dimensional quantum gravity can be understood as a tensionless
string theory embedded in a two-dimensional target space. The
temporal coordinate of the target space play the role of the time
and the wave function can be interpreted as in standard quantum
mechanics.

\subsubsection{The Null String Approach}
\hspace{1cm} The string equations of motion and constraints in
curved space-time are highly nonlinear and, in general, not
exactly solvable. There are different methods available to solve
the string equations of motion and constraints in curved
space-times (for a review see for example \cite{dVS95}). These are
the string perturbation approach, the null string approach, the
$\tau$-expansion, and the construction of global solutions (for
instance by solitonic and inverse scattering methods). A general
approximation method is the null string approach \cite{dVN92}. In
such approach the string equations of motion and constraints are
systematically expanded in powers of $c$ (the speed of light in
the world-sheet). This corresponds to a small string tension
expansion. At zeroth order, the string is effectively equivalent
to a continuous beam of massless particles labeled by the
world-sheet spatial parameter $\sigma$. The points on the string
do not interact between them but they interact with the
gravitational background.

In \cite{dVGN94} quantum bosonic strings in strong gravitational
fields are studied. Within the systematic expansion introduced in
\cite{dVN92} one obtains to zeroth order the null string , while
the first order correction incorporates the string dynamics. This
formalism is applied to quantum null strings in de Sitter
space-time. After a reparametrization of the world-sheet
coordinates, the equations of motion are simplified. The quantum
algebra generated by the constraints is considered, ordering the
momentum operators to the right of the coordinate operators. No
critical dimension appears. It is anticipated however that the
conformal anomaly will appear when the first order corrections
proportional to string tension $T_1$ are introduced.

The classical dynamics of a bosonic string in the $D$--dimensional
Friedmann--Robertson--Walker (with $k=0$) and Schwarzschild
backgrounds is investigated in \cite{LS96}. This is done by making
a perturbative development in the string coordinates around a null
string configuration. The background geometry is taken into
account exactly.

The dynamics of a string near a Kaluza-Klein black hole is studied
in \cite{JMS99}. Solutions to the classical string equations of
motion are obtained using the world sheet velocity of light as an
expansion parameter, i.e. the null string expansion. The
electrically and magnetically charged cases are considered
separately. Solutions for string coordinates are obtained in terms
of the world-sheet coordinate $\tau$. It is shown that the
Kaluza-Klein radius increases/decreases with $\tau$ for
electrically/magnetically charged black hole.

The tension as a perturbative parameter in the nonlinear string
equations of motion in curved space-times is also considered in
\cite{G90,Z96,RZ98,DZ98}.

\subsubsection{Null Branes in Cosmology}
\hspace{1cm} In general, in any cosmological model based on string
theory, one has to face a regime of strong coupling and large
curvatures when approaching the big-bang singularity. In the
strong coupling regime, D-branes are the fundamental players.
D-branes may give new insight into the understanding of the
cosmological evolution of the Universe at early epochs. The
authors of \cite{MR98} analyze the dynamics of D-branes in curved
backgrounds and discuss the parameter space of M-theory as a
function of the coupling constant and of the curvature of the
Universe. They show that D-branes may be efficiently produced by
gravitational effects. Furthermore, in curved spacetimes the
transverse fluctuations of the D-branes develop a tachyonic mode
and when the fluctuations grow larger than the horizon, the branes
become tensionless and break up. This signals a transition to a
new regime. They also comment on possible implications for the
so-called  {\it brane world} scenario, where the Standard Model
gauge and matter fields live inside some branes while gravitons
live in the bulk.

The paper \cite{RZ96} considers the dynamics of null bosonic
$p$-branes in curved spaces. It is shown that their motion
equations can be linearized and exactly solved (in the case of
Robertson-Walker space-time with $k=0$, i.e. with flat space-like
section) in contrast to the case of tensile $p$-branes. It is
found that the perfect fluid of null $p$-branes is an alternative
dominant source of gravity in the Hilbert-Einstein equations for
$D$-dimensional Friedmann universe with flat space-like section.
This work is a generalization for $p>1$ of the results obtained
earlier in \cite{RZ95}.

\newpage
\section{\bf CLASSICAL AND QUANTUM PROPERTIES \\ OF THE NULL BRANES}
\hspace{1cm} Until now we learn how null branes of different
dimension may appear in the context of contemporary string theory
and in the connected with it theory of gravity and cosmology. At
the same time, this explains why we are interested in more careful
investigation of the classical and quantum properties of such
extended relativistic objects.

In this section we begin the description of different models for
tensionless $p$-branes. The particular case $p=1$, i.e. the
tensionless {\it strings}, which is much more studied by now, will
not be considered here. We will concentrate our attention on the
known results about the general case.

A Lagrangian which could describe under certain conditions null
bosonic branes in $D$-dimensional Minkowski space-time was first
proposed in \cite{BLS862}. An action for a tensionless $p$-brane
with space-time supersymmetry was first given in \cite{Z882,Z881}.
Since then, other types of actions and Hamiltonians (with and
without supersymmetry) have been introduced and studied in the
literature \cite{G92,BZ93,ILST93,HLU94,PS97,B986,B991,B992}. Owing
to their zero tension, the world-volume of the null $p$-branes is
a light-like, $(p+1)$-dimensional hypersurface, imbedded in the
Minkowski space-time. Correspondingly, the determinant of the
induced metric is zero. As in the tensile case, the null brane
actions can be written in reparametrization and space-time
conformally invariant form. However, their distinguishing feature
is that at the classical level they may have any number of global
space-time supersymmetries and be $\kappa$-invariant in all
dimensions, which support Majorana (or Weyl) spinors. At the
quantum level, they are anomaly free and do not exhibit any
critical dimension, when appropriately chosen operator ordering is
applied \cite{BZ89,G92,BZ93,PS97,B9711}. The only exception are
the tensionless branes with manifest conformal invariance, with
critical dimension $D=2$ for the bosonic case and $D=2-2N$ for the
spinning case, $N$ being the number of world-volume
supersymmetries \cite{PS97}.

Let us mention also the paper \cite{GRRA892}, which is devoted to the
construction of field theory propagators of null strings and
$p$-branes, as well as the corresponding spinning versions.

Almost all of the above investigations deal with {\it free} null
branes moving in {\it flat} background (a qualitative
consideration of null $p$-brane interacting with a scalar field
has been done in \cite{BZ93}). The interaction of tensionless
membranes ($p=2$) with antisymmetric background tensor field in
four dimensional Minkowski space, described by means of
Wess-Zumino-like action, is studied in \cite{Z8990}. The resulting
equations of motion are integrated exactly.

To our knowledge, the only papers till now devoted to the classical
dynamics of {\it null} $p$-branes ($p\geq 2$) moving in {\it curved}
space-times are \cite{RZ96,B993,B994,B995}.

In \cite{RZ96}, the null $p$-branes living in $D$-dimensional
Friedmann-Robertson-Walker space-time with flat space-like section
(${\it k}=0$) have been investigated. The corresponding equations of
motion have been solved exactly. It was argued that an ideal fluid of
null $p$-branes may be considered as a source of gravity for
Friedmann-Robertson-Walker universes.

In \cite{B993}, the classical mechanics of the null branes in a
gravity background was formulated. The Batalin-Fradkin-Vilkovisky
(BFV) approach in its Hamiltonian version was applied to the
considered dynamical system. Some exact solutions of the equations of
motion and of the constraints for the null membrane in general
stationary axially symmetric four dimensional gravity background were
found. The examples of Minkowski, de Sitter, Schwarzschild, Taub-NUT
and Kerr space-times were considered. Another exact solution, for the
Demianski-Newman background, can be found in \cite{B994}.

The article \cite{B995} considers null bosonic $p$-branes moving in
curved space-times and a method for solving their equations of motion
and constraints, which is suitable for string theory backgrounds is
developed. As an application, an explicit exact solution for the ten
dimensional solitonic five-brane gravity background is given.

Now, we are going to describe in more detail the results of works on
null $p$-branes previous to ours.

To begin with, let us write down the Lagrangian density for the
bosonic $p$-brane given in \cite{BLS862}: \ba\label{pbc} L =
S_{\mu_0\mu_1 ... \mu_p}(\xi)\Sigma^{\mu_0\mu_1 ... \mu_p}(\xi),\h
\xi=(\xi^0,\xi^1,...,\xi^p).\ea Here \ba\nl \Sigma^{\mu_0\mu_1 ...
\mu_p}(\xi) = \epsilon^{J_{0} J_{1} ... J_{p}}\p_{J_{0}}
x^{\mu_0}\p_{J_{1}} x^{\mu_1} ... \p_{J_{p}} x^{\mu_p},\ea
$\epsilon$ being the $p$-dimensional Levi-Civita symbol. In
(\ref{pbc}), the totally antisymmetric tensor $S_{\mu_0\mu_1 ...
\mu_p}(\xi)$ is restricted to lie on a single orbit of $SO(1,d)_0$
in the space of $p+1$ dimensional antisymmetric tensors. For
suitable choices of the orbits for $S(\xi)$, one expects
(\ref{pbc}) to describe physically sensible $p$-dimensional
objects including null ones.

In \cite{Z882,Z881}, an action (\ref{8}) for a space-time
supersymmetric null $p$-brane is proposed in the following form
\ba\label{10} S_p = C\int d^{p+1}\xi
det\left(w^{\mu}_{J}w^{\nu}_{K}\eta_{\mu\nu}\right)/2E ,
\\ \nl w^{\mu}_{J} = \pJ x^\mu - i\bar \theta\gamma^{\mu}\pJ \theta
,\h (J,K = 0,1,...,p),\ea where $E(\xi)$ is a Lagrange multiplier.
It is argued there that this action should possess Siegel
$\kappa$-invariance \cite{WS83} in {\it any} space-time dimension
and for {\it any} number of supersymmetries, contrary to the
tensionful superstrings and super $p$-branes.

In the paper \cite{GRRA892}, field theory propagators for null
strings and $p$-branes, as well as for their generalizations to the
spinning case are constructed by analogy with the massless
relativistic particles. The propagator for the bosonic null $p$-brane
is given by \ba\nl G_b\left(x_2(\us),x_1(\us)\right) = \int {\cal
D}\nu(\us)\nu(\us)^{-D/2}\exp\left(-\int d^p \sigma\frac{(\Delta
x)^2}{2\nu(\us)}\right),\\ \nl x^{\mu}(\tau,\us) = x^{\mu}(\tau_1) +
\frac{\Delta x^{\mu}}{\Delta\tau}(\tau-\tau_1) + Y^{\mu}(\tau,\us),\h
\us = \left(\sigma^1,...,\sigma^p\right) \ea where $x_1$ and $x_2$
are the initial and final brane configurations. The corresponding
propagator for the spinning null $p$-brane is obtained to be \ba\nl
G_s\left(x_2(\us),x_1(\us)\right) = \int {\cal
D}\nu(\us)\nu(\us)^{-1-D/2}\gamma^{\mu}\Delta x_{\mu}(\us)
\exp\left(-\int d^p \sigma\frac{(\Delta x)^2}{2\nu(\us)}\right),\ea
where $\gamma^{\mu}$ are the $D$-dimensional Dirac matrices,
realizing the zero modes of the fermionic constraints. $G_b$ and
$G_s$ satisfy the equation \ba\nl \p^2 G =
\delta\left(x_2-x_1\right),\ea $\p^2$ being the $(p+1)$-dimensional
d'alembertian, as can be easily checked explicitly.

Let us now describe the main results obtained in \cite{G92}. These
are:

1. It is proven that the null spinning $p$-brane is a system of rank
1.

2. An explicit expression for the propagator in the momentum space is
found.

3. The question is discussed about how the critical dimensions are
modified if the boundary conditions are changed.

4. It is argued that the functional diffusion equation is not
modified when fermionic corrections are considered.

Now we are going to briefly explain how these results are
obtained. The hamiltonian constraints for the spinning $p$-brane
can be written as \ba\nl T_0=\eta^{\mu\nu}p_\mu p_\nu = 0 ,\h
T_j=p_\nu\pj x^\nu + \frac{i}{2}\sum_{J=0}^{p}\Gamma^{\mu}_{J}\p_j
\Gamma_{J\mu} = 0, \h S_J = \Gamma^{\mu}_{J}p_{\mu} = 0,\\ \nl
J=(0,j). \ea Here $\Gamma^{\mu}_{J}(\xi)$ are real variables and
transform as spinors in the world-volume and as vectors in
space-time. It is also supposed that they satisfy the Clifford
algebra \ba\nl \{\Gamma^{\mu}_{J}\left(\us_1\right),
\Gamma^{\nu}_{K}\left(\us_2\right)\} =
i\delta_{JK}\eta^{\mu\nu}\delta^p\left(\us_1-\us_2\right).\ea As a
consequence, the corresponding canonical Hamiltonian turns out to
be \ba\nl H_c=\int d^p\sigma\bigl(N^J T_J + i\bar\lambda^J S_J
\bigr) = 0, \ea where $N^J$ and $\bar\lambda^J$ are Lagrange
multipliers.

The constraints $T_J$, $S_J$ are first class and satisfy the algebra
\ba \nl [T_0(\ul \sigma_1),T_{j}(\ul \sigma_2)]&=&- [T_0(\ul
\sigma_1) + T_0(\ul \sigma_2)] \p_j \delta^p (\ul \sigma_1 - \ul
\sigma_2) ,
\\ \nl
[T_{j}(\ul \sigma_1),T_{k}(\ul \sigma_2)]&=& [\delta_{j}^{l}T_{k}(\ul
\sigma_1) + \delta_{k}^{l}T_{j}(\ul \sigma_2)]\p_l\delta^p(\ul
\sigma_1-\ul \sigma_2),\\ \nl  [T_0(\ul \sigma_1),T_0(\ul
\sigma_2)]&=&0, \\ \nl  [S_J(\ul \sigma_1),T_0(\ul \sigma_2)]&=&0,
\\ \nl  [S_J(\ul \sigma_1),T_k(\ul \sigma_2)]&=&
\frac{1}{2}\left[S_J(\us_1) + 2S_J(\us_2)\right]\p_k \delta^p (\ul
\sigma_1 - \ul \sigma_2),
\\ \nl \{S_J(\us_1),S_K(\us_2)\}&=& 2i\delta_{JK}T_0(\us_1)
\delta^p (\ul \sigma_1 - \ul \sigma_2).\ea

Then the proper time gauge \cite{T8082} is chosen \ba\nl \dot N_0 =
0,\h N_j = 0,\h \dot \lambda_0 =0,\h \lambda_j = 0\ea  and the author
proceed with applying the BFV formalism \cite{FV75,BV77,FF78,H85}.
Correspondingly, the following extended phase space is introduced
\ba\nl \left(p_\mu,x^\mu,\Gamma^{\mu}_{J}\right)\oplus
\left(\Pi_J,N^J,\rho_J,\lambda^J\right)\oplus \left(\eta_J,\bar{\cal
P}^J,\bar\eta_J,{\cal P}^J,C_J,\bar b^J,\bar C_J,b^J \right),\ea
where $(\Pi_J,\rho_J)$ are the canonical momenta associated with the
Lagrange multipliers $(N^J,\lambda^J)$ while the dynamics remains
unchanged provided we impose $\Pi_J$ and $\rho_J$ as new constraints,
i.e. \ba\nl \Pi_J = 0,\h\rho_J = 0. \ea The variables
$(\eta_J,\bar\eta_J)$ and $(C_J,\bar C_J)$ represent anticommuting
and commuting ghosts respectively, with $(\bar{\cal P}^J,{\cal
P}^J,\bar b^J,b^J)$ as their conjugated momenta.

Using the constraint algebra, the Becchi-Rouet-Stora-Tyutin (BRST)
charge is obtained to be \ba\nl \Omega &=& \int d^p\sigma[\eta^J
T_J + C^J S_J + {\cal P}^J \Pi_J + b^J \rho_J + \bar{\cal
P}_0\left( \p_j\eta^0 \eta^j + \frac{1}{2}\eta^0\p_j\eta^j\right)
\\ \nl &+& \bar{\cal P}_j\left(\p_k\eta^j\eta^k + \p_k\eta^k\eta^j
+ \eta^k\p_k\eta^j\right) - \bar{\cal P}_0\left( \p_j\eta^j C^0 +
\frac{1}{2}\eta^j\p_jC^0- iC^0 C^0 - iC_j C^j\right) \\ \nl &-&
\bar b_0 \left(\p_j\eta^j C^0 + \frac{1}{2}\eta^j\p_j C^0\right) -
\bar b_k\left(\p_j\eta^j C^k - \frac{1}{2}\p_k C^j\eta^k\right) +
\bar b_0\left(\p_j C^0\eta^j + \frac{1}{2}C^0 \p_j\eta^j\right) \\
\nl &-& \frac{1}{2}\bar b_k \eta^j\p_jC^k + \bar b_j\p_k
\eta^k\eta^j + \bar b_j C^k\p_k\eta^j].\ea This result means that
the null spinning $p$-brane is a system of rank 1.

In the BFV formalism, the gauge fixing appears associated with the
choice of the gauge fermion $\Psi$, which in \cite{G92} is taken
to be \ba\label{gf} \Psi = \int d^p\s\left[{\cal P}_J N^J + \bar
b_J \lambda^J + \frac{1}{\varepsilon}N^j \bar\eta_j +
\frac{1}{\varepsilon}\lambda^j \bar C_j \right],\ea where
$\varepsilon$ is an arbitrary parameter that is set to zero at the
end of the calculations.

In order to integrate the expression (\ref{gf}) one can impose the
following boundary conditions \ba\nl x_1^\mu(\us,\tau_1) =
x^\mu(\us), \h  x_2^\mu(\us,\tau_2) = x_2^\mu(\us),\\ \label{bc}
\Gamma_{J}^{\mu}(\us,\tau_1) + \Gamma_{J}^{\mu}(\us,\tau_2) =
2\gamma_{J}^{\mu}(\us),\\ \nl \eta_J(\us,\tau_1) = \eta_J(\us,\tau_2)
= 0, \h \bar\eta_J(\us,\tau_1) =\bar\eta_J(\us,\tau_2) = 0, \\ \nl
C_J(\us,\tau_1) = C_J(\us,\tau_2) = 0,\h \bar C_J(\us,\tau_1) = \bar
C_J(\us,\tau_2) = 0. \ea

Using the Fradkin-Vilkovisky theorem, after some calculations, one
obtains \ba\nl Z = \int {\cal D}\mu\exp\left(iS_{eff}\right)
\\ \label{prop} = \int\prod_{\us}dp^\mu(\us)
\frac{\int d^p\s\gamma_{\mu 0}(\us)p^\mu(\us) \exp\left(i\int d^p\s
p^\mu \Delta x_\mu(\us)\right)}{\int d^p\s p^2(\us)},\ea which is the
propagator for a null spinning $p$-brane in the momentum space. In
receiving (\ref{prop}), it is supposed that \ba\nl
\Gamma_{J}^{\mu}(\us,\tau) = \gamma_{J}^{\mu}(\us) +
\psi^{\mu}_{J}(\us,\tau),\ea where $\psi^{\mu}_{J}$ is a quantum
fluctuation that fulfils antiperiodic boundary conditions \ba\nl
\psi^{\mu}_{J}(\us,\tau_1) + \psi^{\mu}_{J}(\us,\tau_2) = 0,\ea in
accordance with (\ref{bc}).

The expression (\ref{prop}) shows that the spectrum of the null
spinning $p$-brane is continuous, so that this model has no
critical dimensions. However, a note of caution is necessary: the
change of boundary conditions, or equivalently {\it the change of
the operator ordering}, may introduce critical dimensions. More
precisely, if we modify the boundary conditions within the
ordering prescription for the operators proposed in
\cite{GRRA891,GRRA90}, the critical dimensions that appear for the
bosonic null $p$-branes for example are \ba\nl D = \frac{5p +
8}{p}.\ea We note that this is in disagreement with the result in
\cite{MS88} for tensile $p$-branes, which is \ba\nl D = \frac{5p +
8}{p}\left(p+1\right),\ea contrary to the author's claim.

Finally, it is pointed out in \cite{G92} that it is possible to
show that the functional diffusion equation for the null spinning
$p$-brane reads \ba\nl \frac{\p G\left[M,M_0\right]}{\p V} =
\frac{\delta^2 G\left[M,M_0\right]}{\delta x^2(\us)},\ea where $V$
is the world-volume, $G\left[M,M_0\right]$ is the propagator and
$M$, $M_0$ represent the final and initial configurations of the
null spinning $p$-brane respectively.

Now we turn to the considerations devoted to tensionless branes made
in the review article \cite{BZ93}. The results of
\cite{BZ89,BZ90,BZ911,BZ912,BZ913} are contained there, so we will
not pay separate attention to them. As a matter of fact, \cite{BZ912}
and \cite{BZ913} are extended versions of \cite{BZ90,BZ911}.

We start with the {\it bosonic} null $p$-brane case. The
corresponding action may be represented in the following simple,
reparametrization invariant form \cite{Z881} \ba\label{1} S_{0,p} =
\frac{1}{2}\int d^{p+1}\xi\frac{\det(\p_J x^{\mu}\p_K x_{\mu})}{E},
\ea where $E(\tau,\us)$ is a hyper-sheet density which plays the role
of Lagrange multiplier. The action (\ref{1}) is invariant under the
following (infinitesimal) conformal transformations \ba\nl \delta_D
x^\mu &=& - \lambda x^\mu, \h \delta_D E = -2(p+1)\lambda E,\\ \nl
\delta_c x^\mu &=& -c^\mu x^2 - 2(cx)x^\mu, \h \delta_c E =
-4(p+1)(cx)E.\ea

For comparison, we write down also the action for the tensile brane
in a form in which the limit $T = (\alpha')^{-(p+1)/2} \to 0$ can be
taken \ba\label{2}  S_{p} = \frac{1}{2}\int
d^{p+1}\xi\left[\frac{|\det(\p_J x^{\mu}\p_K x_{\mu})|}{E} +
\frac{1}{(\alpha')^{p+1}}E\right].\ea After substitution of the
general solution \ba\nl E = \left((\alpha')^{p+1}|\det(\p_J
x^{\mu}\p_K x_{\mu})|\right)^{1/2}\ea of the equation of motion for
$E$ into the action (\ref{2}), the latter takes the Dirac-Nambu form
\ba\label{3} S'_{p} = \frac{-1}{(\alpha')^{(p+1)/2}}\int
d^{p+1}\xi\sqrt{|\det(\p_J x^{\mu}\p_K x_{\mu})|}.\ea

The action (\ref{1}) is characterized by the following constraints
\ba\nl T_0\equiv\eta^{\mu\nu}p_\mu p_\nu \approx 0 ,\h T_j\equiv
p_\nu\pj x^\nu \approx 0,\ea which originate from the arbitrariness
in the choice of the world-volume parametrization. Their (equal
$\tau$) Poisson bracket algebra reads  \ba \nl \{T_0(\ul
\sigma_1),T_0(\ul \sigma_2)\}&=&0,
\\ \nl
\{T_0(\ul \sigma_1),T_{j}(\ul \sigma_2)\}&=& -[T_0(\ul \sigma_1) +
T_0(\ul \sigma_2)] \p_j \delta^p (\ul \sigma_1 - \ul \sigma_2) ,
\\ \nl
\{T_{j}(\ul \sigma_1),T_{k}(\ul \sigma_2)\}&=& T_{j}(\ul
\sigma_2)\p_k^2\delta^p(\ul \sigma_1-\ul \sigma_2) -
\left(1\leftrightarrow 2,j\leftrightarrow k\right). \ea

Let us suppose that there exists a scalar field $\varphi(x)$, with
vacuum expectation value (VEV) \ba\nl\langle\varphi\rangle\propto
M_{Planck} = \left(\frac{hc}{G}\right)^{1/2}\approx 10^{19}GeV/c^2\ea
in four dimensional space-time. Then free $p$-branes and null
$p$-branes may be considered as the theories corresponding to
different vacuum states of $p$-brane interacting with a background
scalar field $\varphi(x)$. The corresponding action in 4-dimensions
may be chosen in the form \ba\label{8}  S_{p} =
\frac{\gamma_p}{2}\int d^{p+1}\xi\left[\frac{\det(\p_J x^{\mu}\p_K
x_{\mu})}{E} + \lambda E \varphi^{2p+2} -
\frac{1}{\alpha'}E\varphi^{2p} + \ldots\right],\ea where $\lambda$ is
dimensionless coupling constant. Here the scalar field potential has
been chosen in such a way that it coincide with the Higgs potential
$\propto \lambda\varphi^4 - \mu^2\varphi^2$ when $p=1$.

The VEV's $\langle\varphi\rangle$ corresponding to different extreme
of the potential energy of the field $\varphi$ are given by the
following expressions \ba\label{9a} \langle\varphi\rangle_0 = 0, \\
\label{9b} \langle\varphi\rangle_{\pm} =
\pm\left(\frac{p}{(p+1)\lambda\alpha'}\right)^{1/2}.\ea The equality
(\ref{9a}) describes the symmetric phase and the equality (\ref{9b})
describes the phase with broken conformal and discrete $(\varphi\to -
\varphi)$ symmetries. In the neighborhood of the solution (\ref{9a}),
the action (\ref{8}) coincides with the null $p$-brane action
(\ref{1}). Correspondingly, in the neighborhood of the solution
(\ref{9b}), the action (\ref{8}) coincides with the $p$-brane action
(\ref{3}) after the exclusion of the auxiliary field $E$. This
qualitative consideration may be viewed as an illustration of a
possible mechanism producing nonzero tension for the null $p$-brane.
This mechanism is similar to the Higgs mechanism and rebuilds the
tensionless branes into the Dirac-Nambu $p$-branes with tension
$\propto \langle\varphi\rangle^{p+1}$.

Now let us consider the BRST quantization of the null $p$-brane
theory and reproduce the result obtained in \cite{BZ93} that there
is no critical dimensions.

At first, one studies the equations of motion generated by the
action (\ref{1}). The determinant
$\tilde{G}\equiv\det\tilde{G}_{JK}$ of the induced world-volume
metric $\tilde{G}_{JK}$ \ba\nl \tilde{G}_{JK} &=& \p_J x^\mu \p_K
x_\mu = \left(\begin{array}{cc}\dot x^\mu\dot x_\mu & \dot x^\mu
\p_k x_\mu \\ \p_j x^\mu \dot x_\mu & G_{jk}\end{array} \right),
\\ \label{31} G_{jk}&=& \p_j x^\mu \p_k x_\mu \ea in the integrand
of $S_{0,p}$ may be presented in the form \ba\label{32}
\tilde{G}=G\dot x^\mu \Pi^{\nu}_{\mu}\dot x_\nu ,\h G = \det
G_{jk}, \ea where the matrix $\Pi^{\nu}_{\mu}$ is defined by the
relations \ba\label{33} \Pi^{\nu}_{\mu} = \delta_{\mu}^{\nu} -
\p_j x_\mu \left(G^{-1}\right)^{jk}\p_k x^\nu , \h
\Pi^{\lambda}_{\mu}\Pi^{\nu}_{\lambda}=\Pi^{\nu}_{\mu}.\ea The
representation (\ref{32}) follows from the well known relation for
the determinant of a block matrix \ba\nl
\det\left(\begin{array}{cc}
  A & C \\
  B & D
\end{array}\right) = \det\left(A-BD^{-1}C\right)\det D.\ea

From the variation of the action (\ref{1}) with respect to $x^\mu$
and $E$, one receives the equations of motion \ba\label{34a}
2\left[\frac{\p}{\p\tau}p^{\tau}_{\lambda} +
\frac{\p}{\p\s^l}p^{l}_{\lambda}\right] = 0,\\ \label{34b} G\left[
\dot{x}^2 - \left( \dot{x}_\mu \p_j x^\mu\right)
\left(G^{-1}\right)^{jk}\left(\p_k x^\nu \dot{x}_\nu \right)
\right] = 0,\ea where \ba\nl p^{\tau}_{\lambda}&=&
\frac{G}{E}\left[ \dot{x}_{\lambda} - \p_j x_{\lambda}
\left(G^{-1}\right)^{jk}\left(\p_k x^\nu \dot{x}_\nu \right)
\right],
\\ \nl p^{l}_{\lambda} &=& \frac{\dot{x}^2}{2E}
\frac{\delta G}{\delta\p_l x^{\lambda}}\\ \nl &-&
\frac{\left(\dot{x}^\mu \p_k x_\mu\right)}{2E} \left[
\left(G^{-1}\right)^{kj}\left(\p_j x^\nu \dot{x}_\nu
\right)\frac{\delta G}{\delta\p_l x^{\lambda}} +
2\left(G^{-1}\right)^{kl}G \dot{x}_{\lambda} + G \left( \dot{x}_\mu
\p_j x^\mu\right)\frac{\delta \left(G^{-1}\right)^{jk}}{\delta\p_l
x^{\lambda}}\right].\ea If we fix $p$ gauge degrees of freedom from
$(p+1)$ ones contained in the reparametrization symmetry group by the
following gauge conditions \ba\label{35a}  \dot{x}_\mu \p_j x^\mu =
0,\ea then the equations (\ref{34a}), (\ref{34b}) take a more simple
form \ba\label{34'} \frac{\p}{\p\tau}\left(
\frac{G}{E}\dot{x}_{\lambda}\right)=0,\h G\dot{x}^2 = 0.\ea The
momentum density of null $p$-brane in the gauge (\ref{35a}) equals to
\ba\label{36} p_{\lambda}=\frac{G}{E}\dot{x}_{\lambda}\ea and
$p_{\lambda}=0$ if $G=0$, so the solution $G=0$ of equations
(\ref{34'}) corresponds to trivial dynamics. That is why one chooses
$G\ne 0$ and equations (\ref{34'}) become equivalent to the equations
\ba\label{34''} \frac{\p}{\p\tau}\left(
\frac{G}{E}\dot{x}_{\lambda}\right)=0,\h \dot{x}^2 = 0.\ea

Taking into account that additional gauge condition may be added to
(\ref{35a}) the authors choose this condition in the form
\ba\label{35b}\frac{\p}{\p\tau}\left( \frac{G}{E}\right)=0.\ea Then
the equations of motion (\ref{34''}) take a linear form \ba\label{37}
\ddot{x}^{\mu}=0,\h \dot{x}^2 = 0.\ea

Of course, the variation of the action (\ref{1}) have to be
supplemented with appropriate boundary conditions for open or closed
null $p$-branes. In what follows, the discussion is restricted to the
closed case only.

The following step is the quantization of null bosonic $p$-brane. In
order to simplify the technical aspects of the BRST-quantization
procedure, the gauge condition \ba\label{35c} \frac{G}{E} = \gamma_p
= const \ea is used instead of (\ref{35b}). This gauge gives the
possibility to consider the minimal extension of the original phase
space. It may be constructed by extending the initial phase space
with the canonically conjugated Grassmannian ghost pairs $\bigl
(\eta^J,{\cal P}_J\bigr )$ for the constraints $T_J$.

So, the classical BRST charge \cite{FV75,BV77,SH83,H85} for the
closed null $p$-brane in the minimal sector has the following form
\cite{BZ89} \ba\label{311} \Omega^{min}=\int
d^p\s\{T_0\eta^0+T_j\eta^j+ {\cal P}_0 [(\p_j\eta^0)\eta^j -
\eta^0(\p_j\eta^j)] - {\cal P}_k \eta^j (\p_j\eta^k) \} .\ea It is
easy to see that the BRST charge (\ref{311}) satisfy the equation
\ba\label{312} \{\Omega^{min},\Omega^{min}\} = 0.\ea In the quantum
case the charge $\Omega^{min}$ transforms into the quantum BRST
operator $\hat{\Omega}^{min}$. Then the nilpotency condition
\ba\label{313} \left(\hat{\Omega}^{min}\right)^2 = 0\ea ought to be
fulfilled for the self consistency of the quantum theory. However,
anomalies can appear in the right hand side of equation (\ref{313})
as a result of quantum ordering process. Taking into account the
point-like character of the first equation in (\ref{37}) for the null
$p$-brane it is natural to choose the initial data for ordinary phase
space variables $\hat{q}_0$ and $\hat{p}_0$ as the "physical
variables" in terms of which quantum ordering must be done. Remind
that in the string case quantum ordering in terms of the creation and
annihilation oscillator operators $a$ and $a^\dag$ is more preferable
as a consequence of the oscillator character of the string equation
of motion \ba\nl \ddot{x}^\mu (\tau,\s) - x''^\mu (\tau,\s) = 0.\ea
Thus one chooses the $\hat{q}_0\hat{p}_0$-ordering in the BRST
generator $\hat{\Omega}^{min}$. This ordering does not give rise to
anomalous terms in the calculation of
$\left(\hat{\Omega}^{min}\right)^2$. As a consequence, the quantum
nilpotency condition is satisfied.

Let us discuss the case of $\hat{q}\hat{p}$-ordering, which is
formulated in terms of the hole operators of phase variables.
Actually, $\hat{\Omega}^{min}$ contains only such combinations of the
canonically conjugated operators $\hat{q}$ and $\hat{p}$, for which
the ordering is preserved in the process of the anticommutator
$\{\hat{\Omega}^{min},\hat{\Omega}^{min}\}$ calculation, and it is
not difficult to prove this \cite{BZ93}.

After transition from $\hat{q}$, $\hat{p}$ to their initial data
$\hat{q}_0$, $\hat{p}_0$ the considered quantum
$\hat{q}_0,\hat{p}_0$-ordering also will be conserved. To clear up
this statement it is enough to analyze the equations of motion for
the ghost sector only, because the equations for $\hat{x}^\mu$ and
$\hat{p}_\mu$ are linear ones. The BRST-invariant Hamiltonian which
generates these equations of motion depends on the choice of the
gauge fermion $\Psi$, which in the minimal sector has the general
form \ba\nl \Psi = \int d^p\s\left( \frac{\lambda^0}{2}{\cal P}_0 -
\zeta^k {\cal P}_k \right).\ea Using the arbitrariness in the choice
of the Lagrange multipliers $\lambda^0$ and $\zeta^k$, they are
chosen in \cite{BZ93} as \ba\nl \lambda^0 = \gamma^{-1}_p,\h \zeta^k
= 0,\h \Rightarrow \Psi_0 = \frac{1}{2\gamma_p}\int d^p\s{\cal
P}_0.\ea The corresponding Hamiltonian is obtained to be \ba\nl
H_{\Psi_0} = -\frac{1}{2\gamma_p}\int d^p\s\left(p^\mu p_\mu + {\cal
P}_0\p_k\eta^k \right).\ea

The equations of motion $\dot{f}=\{f,H_{\Psi_0}\}$ generated by
$H_{\Psi_0}$ have the form \ba\nl \dot{x}^\mu &=&
\frac{1}{\gamma_p}p^\mu,\h \dot{p}^\mu = 0,\h \dot{\eta}^0 =
\frac{1}{2\gamma_p}\p_k \eta^k,\\ \nl \dot{\eta}^k &=& 0,\h
\dot{{\cal P}_0} = 0,\h \dot{{\cal P}_k} = \frac{1}{2\gamma_p}\p_k
{\cal P}_0 .\ea The general solution of this system of equations is
\ba\nl x^\mu (\tau,\us)&=&x_0^\mu(\us) + \frac{\tau}{\gamma_p}p_0^\mu
(\us),\h p_\mu (\tau,\us)=p_{0\mu}(\us),\\ \nl \eta^0 (\tau,\us)&=&
\eta^0_0 (\us) + \frac{\tau}{2\gamma_p}\p_k \eta_0^k (\us),\h {\cal
P}_0 (\tau,\us) = {\cal P}_{00}(\us),\\ \nl \eta^k
(\tau,\us)&=&\eta^k_0(\us),\h {\cal P}_k (\tau,\us) = {\cal
P}_{0k}(\us) + \frac{\tau}{2\gamma_p}\p_k {\cal P}_0.\ea

We see that the transition from the $\hat{q},\hat{p}$ variables to
their initial data $\hat{q}_0,\hat{p}_0$ is a linear transformation.
Owing to this fact the proof of nilpotency condition for
$\hat{\Omega}^{min}$ carried out in the terms of
$\hat{q}\hat{p}$-ordering does not change after transition to
$\hat{q}_0\hat{p}_0$-ordering.

Here in \cite{BZ93} the conclusion is drown that the critical
dimensions are absent in the null bosonic $p$-brane theory and this
theory is quantum mechanically self consistent in flat space-time of
arbitrary dimension. However, it is noted that the question about
presence or absence of a critical dimension in the quantum theory
essentially depends on the choice of initial operator set in terms of
which the procedure of operator ordering is defined.

Finally, let us describe the dynamics of null $p$-brane in the light
cone gauge. At first we note that in the gauge (\ref{35a}) the
induced metric $\tilde{G}_{JK} $ takes the form \ba\label{324}
\tilde{G}_{JK}\left(\xi\right) = \left( \begin{array}{cc}
  0 & 0 \\
  0 & G_{JK}\left(\xi\right)
\end{array}\right).\ea The representation is invariant under
diffeomorphisms \ba\label{327}\s^j\left(\tau ',\us '\right) =
\s^j\left(\us '\right).\ea The gauge condition (\ref{35c}), which
linearizes the null $p$-brane equations of motion, is conserved under
the diffeomorphism transformations \ba\label{329} \tau\left(\tau
',\us '\right) = a\left(\us '\right) +
\tau'\det\left(\frac{\p\s^j(\us')}{\p\s^{'k}}\right).\ea

The equations of motion $\ddot{x}^{\mu}=0$ are invariant under the
transformations (\ref{327}) and (\ref{329}). Therefore, it is
possible to use the gauge symmetry (\ref{327}), (\ref{329}) for
fixing light-cone gauge condition defined as \ba\label{330} x^+
(\tau,\us)=c\tau,\h c=\frac{p_{0}^{+}}{\gamma} =
\frac{P_{0}^{+}}{N_p\gamma},\h P^{+}_{0}= \int d^p\s
p^{+}_{0}(\us).\ea This gauge condition is conserved under the
diffeomorphisms (\ref{327}) and (\ref{329}) restricted by the
condition $\tau'=\tau$. We see that the gauge conditions (\ref{35a}),
(\ref{35c}) and (\ref{330}) are characterized by a residual gauge
symmetry comprising the so called area-preserving transformations,
which are defined by the infinitesimal relations \ba\nl \delta\tau=0,
\h \delta\sigma^{j_1} = \varepsilon^{j_1 j_2 ...
j_p}\p_{j_2}\Lambda_{j_3 ... j_p}.\ea

In the light-cone gauge, the general solution of (\ref{37}) is \ba\nl
x^m (\tau,\us) &=& x_0^m (\us) + p_0^m (\us)\gamma^{-1}_p \tau, \h (m
= 1,...,D-1)\\ \nl x^{-} (\tau,\us) &=& x_0^{-} (\us) +
\frac{N_p}{2\gamma P_0^{+}}p_{0m}^2(\us)\\ \nl p_0^{-}(\us) &=&
\frac{N_p}{2P_0^{+}}p_{0m}^2(\us).\ea Then for the mass operator of
null $p$-brane defined by the relation \ba\nl M^2\equiv P_0^\mu
P_{0\mu}\equiv 2P_0^+ P_0^- - P_{0m}^2,\h P_{0\mu} = \int d^p\s
p_{0\mu}(\us)\ea we get the representation \ba\nl M^2 = N_p \int
d^p\s p_{0m}^2(\us) - \left(\int d^p\s p_{0m}(\us)\right)^2.\ea It
can be shown that $M^2\ge 0$ and it is independent on the center of
mass variables $q_{0m}$, $P_{0m}$.

The canonical Hamiltonian in the light-cone gauge has the form
\ba\label{350} H_{lc} = \frac{P_0^+}{\gamma N_p}\int d^p\s p_0^-(\us)
= \frac{1}{2\gamma}\int d^p\s p_{0m}^2(\us) = \frac{1}{2\gamma N_p}
\left[p_{0m}^2(\us) + M^2\right].\ea The eigenfunctions
$\Psi_{|k_{0m};\{k_{ma}\}>}$ associated with the Hamiltonian
(\ref{350}) are generalized "plane waves" \ba\label{351}
\Psi_{|k_{0m};\{k_{ma}\}>} = \exp\left(i\sum_a k_{ma}x_m^a\right) =
\exp\left(ik_{0m}q_{0m}\right) \exp\left(i\sum_{a\ne 0}
k_{ma}x_m^a\right),\ea which describe the coherent motion of infinite
number of quasi particles. Here the index $a$ marks a complete
orthonormal basis of functions $Y_a(\us)$ on the (null) $p$-brane
\cite{BZ93,WHN88,WLN89}. The physical subspace can be extracted from
the set of vectors (\ref{351}) by the annihilation conditions \ba\nl
\sum_{a,b\ne 0}f^{de}_{abc}x_m^a P_m^b \Psi^{phys} = 0.\ea Then,
supposing that there exist nonzero solutions of the above equalities,
it is not difficult to show that {\it the spectrum of null $p$-brane
is continuous} \cite{BZ93}.

Now we turn to the null super $p$-brane case in four dimensions
\cite{BZ93}. The dynamics of null super $p$-brane is described by
the action (\ref{10}). However, the covariant quantization in this
formulation is a very complicated problem because of an infinite
reducibility of the fermionic constraints analogous to those of
the superstring. In order to covariantly quantize the null super
$p$-brane theory in $D=4$, a new twistor-like Lorentz harmonic
formulation has been proved to be effective \cite{BZ90,BZ911}. It
uses the Lorentz harmonic superspace \ba\nl z^M = (x^\mu,
\theta^\alpha_A,\bar{\theta}^{\dot{\alpha}
A};v^\mp_\alpha,\bar{v}^\pm_{\dot{\alpha}}),\h \alpha =1,2;\h
A=1,...,N\ea as a target space. This space is an extension of the
usual ($N$-extended) superspace
$(x^\mu,\theta^\alpha_A,\bar{\theta}^{\dot{\alpha}A})$ obtained by
adding the commuting spinor variables $(v^\mp_\alpha,
\bar{v}^\pm_{\dot{\alpha}})$, which coincide with the
Newman-Penrose diades \cite{PR86} in four dimensions. In this
formulation, the action for null super $p$-brane theory is
\ba\label{42} S_{0,p} = -\frac{1}{2}\int d^p\xi \varrho^{(-|+)J}
v^-_\alpha \bar{v}^+_{\dot{\alpha}}
\omega_J^{\dot{\alpha}\alpha},\ea where $\varrho^{(-|+)J}$ is a
world-volume density, \ba\nl \omega_J^{\dot{\alpha}\alpha} =
\tilde{\s}_\mu^{\dot{\alpha}\alpha} \left(\p_J x^\mu
-i\p_j\theta^\alpha_A
\s^\mu_{\alpha\dot{\alpha}}\bar{\theta}^{\dot{\alpha}A} +
i\theta^\alpha_A \s^\mu_{\alpha\dot{\alpha}}
\p_J\bar{\theta}^{\dot{\alpha}A}\right) \equiv \omega_J^\mu
\tilde{\s}_\mu^{\dot{\alpha}\alpha}\ea is the invariant Cartan
form for ordinary superspace and $v^\mp_\alpha$,
$\bar{v}^\pm_{\dot{\alpha}}$ are restricted by the conditions
\ba\nl \Xi(\tau,\us)\equiv v^{\alpha -}(\tau,\us)v^+_\alpha
(\tau,\us)-1=0,
\\ \nl \bar{\Xi}(\tau,\us)\equiv \bar{v}^{\dot{\alpha}+}(\tau,\us)
\bar{v}^-_{\dot{\alpha}}(\tau,\us) - 1 = 0.\ea

The action (\ref{42}) is invariant under the following local
$\kappa$-transformations \ba\nl \delta\theta_{\alpha A} &=&
ip_{\alpha\dot{\alpha}}\bar{\kappa}^{\dot{\alpha}}_{A},\h
\delta\bar{\theta}^A_{\dot{\alpha}} = -i\kappa^{\alpha
A}p_{\alpha\dot{\alpha}},\\ \nl \delta x_{\alpha\dot{\alpha}} &=&
2\left(\kappa^{\beta A}p_{\beta\dot{\alpha}}\theta_{\alpha A} +
\bar{\theta}^A_{\dot{\alpha}}p_{\alpha\dot{\gamma}}
\bar{\kappa}^{\dot{\gamma}}_A \right),\\ \nl \delta\varrho^J &=& 0,
\h \delta v^\mp_\alpha = 0,\h \delta\bar{v}^\pm_{\dot{\alpha}}=0, \ea
where \ba\nl p_{\alpha\dot{\alpha}} = \varrho^0(\tau,\us)v^-_\alpha
\bar{v}^+_{\dot{\alpha}}\ea is the momentum density of null super
$p$-brane.

The variation of the action (\ref{42}) with respect to $x^\mu$,
$\theta^{\alpha A}$, $\bar{\theta}^{\dot{\alpha}A}$, $v^-_\alpha$
and $\bar{v}^+_{\dot{\alpha}}$ leads to the equations \ba\nl \p_J
\left(\varrho^J v^-_\alpha \bar{v}^+_{\dot{\alpha}}\right) = 0\h
\mbox{or}\h \dot{p}_{\alpha\dot{\alpha}} = \p_j \left(\varrho^j
v^-_\alpha \bar{v}^+_{\dot{\alpha}}\right),\\ \nl \varrho^J\p_J
\theta_{\alpha A}= \zeta_A(\tau,\us)v^-_\alpha,\h
\varrho^J\p_J\bar{\theta}^A_{\dot{\alpha}} =
\bar{\zeta}^A(\tau,\us)\bar{v}^+_{\dot{\alpha}},\\ \nl
\varrho^J\omega_{J\alpha\dot{\alpha}} = e(\tau,\us)v^-_\alpha
\bar{v}^+_{\dot{\alpha}}\h\mbox{or}\h \omega_{0\alpha\dot{\alpha}}
= \frac{e}{\left(\varrho^0\right)^2}p_{\alpha\dot{\alpha}} -
\frac{\varrho^j}{\varrho^0}\omega_{j\alpha\dot{\alpha}}.\ea
Finally, the $\varrho^J$-variation generates the constraints
\ba\nl v^-_\alpha
\bar{v}^+_{\dot{\alpha}}\omega_J^{\dot{\alpha}\alpha}\approx  0\h
\mbox{or}\h p_{\alpha\dot{\alpha}}p^{\dot{\alpha}\alpha}\approx
0,\h p_{\alpha\dot{\alpha}}\omega_j^{\dot{\alpha}\alpha}\approx
0.\ea The arbitrary functions $\varrho^J(\tau,\us)$,
$\zeta_A(\tau,\us)$, $\bar{\zeta}^A(\tau,\us)$ and $e(\tau,\us)$
in the above equations show the invariance of the action
(\ref{42}) under reparametrizations and $\kappa$-transformations.

It can be proved \cite{BZ90,BZ911,BZ93} that the presence of the
auxiliary variables $v^\mp_{\alpha}$, $\bar{v}^{\pm}_{\dot{\alpha}}$
and $\varrho^{(-|+)J}$ in (\ref{42}) does not increase the number of
physical degrees of freedom and the formulations (\ref{10}),
(\ref{42}) for null super $p$-brane theory are equivalent at the
classical level.

However, the harmonics $v^\mp_{\alpha}$ and their complex conjugated
$\bar{v}^{\pm}_{\dot{\alpha}}$ play an exceptional role in the
Lorentz covariant division of null super $p$-brane constraints into
irreducible once of first and second class. This is carried out by
projecting the Grassmann constraints \ba\nl D^A_\alpha =
-\pi^A_\alpha + ip_{\alpha\dot{\alpha}}
\bar{\theta}^{\dot{\alpha}A}\ea into first class $D^{-A}=v^{\alpha
-}D^A_\alpha$ and second class $D^{+A} = v^{\alpha +}D^A_\alpha$
constraints.

In terms of the coordinates $(z^M,\varrho^{(-|+)J})$ and their
conjugated momenta $(p_M,p_J^\varrho) =
(p_\mu,\pi^A_\alpha,\bar{\pi}_{\dot{\alpha}A};p^\pm_\alpha,
\bar{p}^\mp_{\dot{\alpha}},p_J^\varrho)$, the first class constraints
\ba\nl Y_\Lambda = \left(Y_{\Lambda'},T_j\right) =
\left(D^{-A},\bar{D}^+_A,p^{+-},p_j^\varrho,\nabla^0,\bar{\nabla}^0,
\nabla^{-2},\bar{\nabla}^{+2},T_j\right)\ea of null super $p$-brane
theory have the form \ba\nl D^{-A}&\equiv & v^{\alpha
-}D^A_\alpha\equiv v^{\alpha -}\left(-\pi^A_\alpha +
ip_{\alpha\dot{\alpha}}\bar{\theta}^{\dot{\alpha}A}\right)\approx
0,\\ \nl p^{+-}&\equiv & v^{\alpha
-}\sigma^\mu_{\alpha\dot{\alpha}}p_\mu \approx 0,\h p_j^\varrho
\approx 0,\\ \nl \nabla^0 &\equiv & v^{\alpha +}p^-_\alpha -
v^{\alpha -}p^+_\alpha + \varrho^{0(-|+)}p_J^\varrho \approx 0,\h
\nabla^{-2}\equiv v^{\alpha -}p^-_\alpha\approx 0,\\ \nl T_j&\equiv &
\frac{1}{2}\omega^{\alpha\dot{\alpha}}_{j}p_{\alpha\dot{\alpha}} +
\left(\p_j \theta^{\alpha}_{A}D^A_{\alpha} + \p_j
v^{\alpha\mp}p^\pm_\alpha + c.c.\right) - \p_j \varrho^{0(-|+)}
p_J^\varrho\approx 0,\ea and complex conjugated to
$D^{-A},\nabla^0,\nabla^{-2}$ constraints $\bar{D}^+_A,-
\bar{\nabla}^0,\bar{\nabla}^{+2}$ ({\bf Remark}: there are obvious
mistakes in the expressions for $\nabla^0$ and $T_j$). They form the
following Poisson bracket algebra (we list the nonzero brackets only)
\ba\nl \{T_{j}(\ul \sigma_1),T_{k}(\ul \sigma_2)\}&=& T_{j}(\ul
\sigma_2)\p_k^2\delta^p(\ul \sigma_1-\ul \sigma_2) -
\left(1\leftrightarrow 2,j\leftrightarrow k\right),\\ \nl \{T_{j}(\ul
\sigma_1),Y_{\Lambda'}(\ul \sigma_2)\}&=& -Y_{\Lambda'}(\ul
\sigma_1)\p_j^2\delta^p(\ul \sigma_1-\ul \sigma_2),\\ \label{431}
\{\nabla^0(\ul \sigma_1),Y_{\Lambda'}(\ul \sigma_2)\}&=&
q_{R}\left(Y_{\Lambda'}\right)\delta^p(\ul \sigma_1-\ul \sigma_2), \\
\nl \{\bar{\nabla^0}(\ul \sigma_1),Y_{\Lambda'}(\ul \sigma_2)\}&=&
q_{L}\left(Y_{\Lambda'}\right)\delta^p(\ul \sigma_1-\ul \sigma_2), \\
\nl \{D^{-A}(\ul \sigma_1),\bar{D}^+_B(\ul \sigma_2)\}&=&
-2ip^{+-}\delta^A_B \delta^p(\ul \sigma_1-\ul \sigma_2),\ea where
$q_{L,R}\left(Y_{\Lambda'}\right)$ are the charges of $Y_{\Lambda'}$
under $U_{L,R}(1)$ symmetry groups.

The constraints $p^{+-}$ and $T_j$ correspond to reparametrization
symmetry of the action (\ref{42}), $D^{-A}$ and $\bar{D}^+_A$
describe the fermionic $\kappa$-symmetry, whereas $\nabla^0$ and
$\bar{\nabla}^0$ are related to the local $U_L(1)\times U_R(1) \cong
SO(1,1)\times SO(2)$ symmetry and $\nabla^{-2}$, $\bar{\nabla}^{+2}$
are connected with the local shifts of $v^{\alpha +}$ and
$\bar{v}^{\dot{\alpha}-}$.

The second class constraints $S_f$ in the twistor-like formulation
(\ref{42}) appear to be the pairs \ba\nl \Xi &\approx & 0, \\ \nl
\chi &\equiv & v^{\alpha -}p_\alpha^+ + v^{\alpha +}p^-_{\alpha}+
\varrho^0 p^\varrho_0 \approx 0,\ea \ba\nl \bar{\Xi}&\approx &0,\\
\nl \bar{\chi}&\approx &0 ,\ea \ba\nl D^{+A}&\equiv &v^{\alpha
+}D^A_\alpha\equiv v^{\alpha +}\left(-\pi^A_\alpha +
ip_{\alpha\dot{\alpha}}\bar{\theta}^{\dot{\alpha}A}\right)\approx 0,
\\ \nl \bar{D}^{-A}&\equiv &\bar{v}^{\dot{\alpha}-}
\bar{D}_{\dot{\alpha}A}\equiv
\bar{v}^{\dot{\alpha}-}\left(\bar{\pi}_{\dot{\alpha}A} -
i\theta^\alpha_A p_{\alpha\dot{\alpha}}\right)\approx 0,\ea \ba\nl
\nabla^{+2} &\equiv& v^{\alpha +}p^+_\alpha\approx 0,\\ \nl
p^{--}&\equiv& -v^{\alpha
-}\sigma^\mu_{\alpha\dot{\alpha}}\bar{v}^{\dot{\alpha}-}p_\mu \approx
0,\ea \ba\nl \bar{\nabla}^{-2}&\equiv&
\bar{v}^{\dot{\alpha}-}\bar{p}^-_{\dot{\alpha}}\approx 0,\\ \nl
p^{++}&\equiv& -v^{\alpha
+}\sigma^\mu_{\alpha\dot{\alpha}}\bar{v}^{\dot{\alpha}+}p_\mu \approx
0,\ea \ba\nl p_0^\varrho &\approx 0&,\\ \nl \varrho^{0(-|+)} -
p^{(-|+)}&\equiv&  \varrho^{0(-|+)} - v^{\alpha
+}\sigma^\mu_{\alpha\dot{\alpha}}\bar{v}^{\dot{\alpha}-}p_\mu \approx
0.\ea

The Poisson bracket matrix of the constraints $S_f$ may be
transformed to block-diagonal form if certain redefinitions are
performed. The following step is to use the method for conversion
of second class constraints into effective first class abelian
$\mathcal{A}_f$ ones \cite{FS86,BFF89,EM93}. The conversion
procedure suggests the introduction of new additional set of
canonically conjugated variables $(q^R,p_R)$, so that any definite
pair of the original second class constraints corresponds to some
pair of the newly introduced phase space variables. Now the
transition from the initial first class constraints $Y_{\Lambda}$
to the effective ones $\tilde{Y}_{\Lambda}$, depending also on
$(q^R,p_R)$, must be done. However, it turns out that the
resulting algebra essentially complicates the evaluation procedure
for the BRST generator $\Omega$. For this reason, a new set of
effective first class constraints $\tilde{\tilde{Y}}_{\Lambda}$
have been introduced in \cite{BZ93}. Their algebra coincides with
the original $Y_{\Lambda}$ algebra (\ref{431}) with the exception
of the Poisson brackets \ba\nl
\{\tilde{\tilde{\nabla}}^{-2}(\us_1),
\tilde{\tilde{\bar{\nabla}}}^{+2}(\us_2)\} = \mathcal{E}(\us_1)
\delta^p(\ul \sigma_1-\ul \sigma_2),\ea where \ba\nl \mathcal{E} =
\frac{i}{2J}\left[ \tilde{\tilde{D}}^{-A}
\tilde{\tilde{\bar{D}}}^+_A + \frac{2}{i} \tilde{\tilde{p}}^{+-}
\left( \tilde{\tilde{\nabla}}^0 + \tilde{\tilde{\bar{\nabla}}}^0
\right) \right]\ea is a first class constraint ({\bf Remark}: the
quantity $J$ is not defined in \cite{BZ93}). As a result, the
total algebra of the {\it effective first class constraints} for
null super $p$-brane is found to be \ba\label{452}
\{\mathcal{A}_f,\mathcal{A}_g\}=0,\h
\{\tilde{\tilde{Y}}_{\Lambda},\mathcal{A}_f\}=0,\h
\{\tilde{\tilde{Y}}_{\Lambda},\tilde{\tilde{Y}}_{\Sigma}\} =
C_{\Lambda\Sigma}^{\Pi}\tilde{\tilde{Y}}_{\Pi} +
\mathcal{E}_{\Lambda\Sigma},\ea with the same structure constants
$C_{\Lambda\Sigma}^{\Pi}$ as in the original algebra (\ref{431})
and \ba\nl \mathcal{E}_{\Lambda\Sigma} =
\left[\delta_{\tilde{\tilde{Y}}_{\Lambda},
\tilde{\tilde{\nabla}}^{-2}}
\delta_{\tilde{\tilde{Y}}_{\Sigma},\tilde{\tilde{\bar{\nabla}}}^{+2}}
- \left(\Lambda\leftrightarrow\Sigma\right)\right]\mathcal{E}.\ea

The classical BRST charge for the algebra (\ref{452}) in the minimal
sector equals to $\Omega_{min}=\Omega' + \mathcal{A}_f c''^{f}$ with
$\Omega'$ defined as \ba\label{456} \Omega'&=&\int d^p\s
[Y^{Mod}_{\Lambda}c^{\Lambda} - 2i\pi^{+-}c^+_A \bar{c}^{-A} -
\frac{i}{J}\left(\pi^0 +
\bar{\pi}^0\right)\tilde{\tilde{p}}^{+-}\zeta^{+2}\bar{\zeta}^{-2}\\
\nl &+& \left( \frac{i}{4J}p^{-A}\tilde{\tilde{\bar{D}}}^+_A
\zeta^{+2}\bar{\zeta}^{-2} + c.c.\right) + \left(
\frac{1}{2J}\pi^{+-}\bar{p}^+_A
\bar{c}^{-A}\zeta^{+2}\bar{\zeta}^{-2} + c.c.\right)].\ea The
modified constraints $Y^{Mod}_{\Lambda}$ entering into (\ref{456})
coincide with $\tilde{\tilde{Y}}_{\Lambda}$ except the constraints
$T^{Mod}_j,\nabla^{0 Mod},\bar{\nabla}^{0 Mod}$.

In order to construct the quantum BRST operator \ba\nl
\hat{\Omega}_{min}= \hat{\Omega}'+\hat{\mathcal{A}}_f \hat{c}''^f,\ea
a generalized $\hat{q}\hat{p}$-ordering is chosen in \cite{BZ93}. The
term "generalized" means that one may include some momentum variables
into a $\hat{q}$-set and some coordinate variables into a
$\hat{p}$-set. Then, in analogy with the bosonic null $p$-brane case,
it is proven that this ordering is preserved in the process of
computation of the anticommutator $\{\hat{\Omega}',\hat{\Omega}'\}$
and this leads to the fulfillment of the condition $(\hat{\Omega}')^2
=0$. This consideration permits to the authors of \cite{BZ93} to
conclude that the null super $p$-brane theories are self-consistent
in $D=4$.

A number of classically equivalent actions for $p$-branes are
derived and their tensionless limits are discussed in
\cite{ILST93,HLU94}. The starting point is the Nambu-Goto-Dirac
world-volume action \ba\label{pact} S=T\int{ d^{p+1}\xi
\sqrt{-det\gamma _{JK}} } \ea where $X^\mu=X^\mu (\xi)$ and
\ba\label{indm} \gamma _{JK}\equiv
\partial _J X^\mu \partial _K X^\nu \eta_{\mu\nu} \ea is the
metric induced on the world-volume from the Minkowski space-time
metric $\eta_{\mu\nu}$. The generalized momenta derived from the
Lagrangian in (\ref{pact}) are \ba P_\mu =T\sqrt{-\gamma}\gamma^{J
0}\partial _J X_\mu . \ea where $\gamma ^{J K}$ is the inverse of
$\gamma _{J K}$. They satisfy the constraints \ba\label{cons}
P^2+T^2\gamma \gamma ^{00}=0\cr P_\mu
\partial_aX^\mu=0,\quad a=1,...,p. \ea Here $\gamma \equiv
det\gamma_{JK}$.  As usual for a diffeomorphism invariant theory,
the naive Hamiltonian vanishes and the total Hamiltonian consists
of the sum of the constraints (\ref{cons}) multiplied by Lagrange
multipliers, which we shall call $\lambda$ and $\rho ^a$:
\ba\label{hami} {\cal H}=\lambda (P^2+T^2\gamma
\gamma^{00})+\rho^aP\cdot\partial_aX \ea The phase space action
thus becomes \ba\label{phact} S^{PS}=\int{d^{p+1}\xi
\left\{{P\cdot \dot{X}-\lambda (P^2+T^2\gamma
\gamma^{00})-\rho^aP\cdot\partial_aX}\right\}}. \ea One integrates
out the momenta to find the configuration space action
\ba\label{cact} S^{CS}=\frac{1}{2}\int{d^{p+1}\xi {1 \over
{2\lambda}}\left\{{\dot{X}^2-2\rho ^a \dot X^\mu\partial _aX_\mu
+\rho ^a\rho^b\partial _bX^\mu \partial _aX_\mu -4\lambda
^2T^2\gamma\gamma^{00}}\right\}}. \ea For $p=1$, the following
identification may be done \ba\label{poneg} g^{J K}= \left(
\matrix{-1\quad\hfill \rho \cr
 \rho \quad\hfill-\rho ^2+4\lambda^2T^2\cr}
\right),\ea which leads to the usual Weyl invariant tensile string
action \ba\label{ggact} S=-\frac{1}{2}
T\int{d^{2}\xi\sqrt{-g}{g^{JK}\partial _J X^\mu
\partial _K X^\nu \eta_{\mu\nu}}}. \ea For $p>1$ it is not
possible to directly identify the geometric fields in
(\ref{cact}). One first has to rewrite it as  \ba\label{Gact}
S^{CS}= \frac{1}{2}\int{d^{p+1}\xi \left\{{
{{h^{JK}\gamma_{JK}}\over {2\lambda}} -2\lambda T^2G(p-1)+2\lambda
T^2GG^{ab}\gamma_{ab} }\right\}},\ea where \ba h^{JK}=\left(
\matrix{1\quad\hfill -\rho ^a\cr -\rho ^a\quad\hfill\rho ^a\rho
^b\cr} \right) \ea is a rank 1 auxiliary matrix and $G_{ij}$ is a
$p$-dimensional auxiliary metric with determinant $G$.
(Integrating out $G_{ij}$ one recovers (\ref{cact}).) Now the
identification \ba\label{gdef} g^{JK}= {1 \over
4}T^{-2}\lambda^{-2}G^{-1}\left( \matrix{-1\quad\hfill \rho ^a\cr
\rho ^a\quad\hfill-\rho ^a\rho ^b+4\lambda^2T^2GG^{ab}\cr} \right)
\ea produces the usual $p$-brane action involving the world-volume
metric $g_{JK}$: \ba\label{gact} S=-\frac{1}{2}
T\int{d^{p+1}\xi\sqrt{-g}\left\{{g^{JK}\partial _J X^\mu
\partial _K X^\nu \eta_{\mu\nu}-(p-1)}\right\}}. \ea The identification (\ref{gdef}) tells us
the transformation properties of the Lagrange multipliers. Note that
for $p=0,1$ the auxiliary metric $G_{ij}$ never appears in
(\ref{Gact}), and the configuration space action is the usual
manifestly reparametrization invariant massive point-particle action
and reparametrization invariant tensile string action respectively.

It is clear from the above procedure that we may take the limit $T\to
0$ anywhere between (\ref{hami}) and (\ref{gdef}). The identification
(\ref{gdef}) will differ in that limit, however. The metric density
$T\sqrt{-g}g^{J K}$ becomes degenerate and gets replaced by a rank 1
matrix which can be written as $V^J V^K$ in terms of the vector
density $V^J$ \ba V^J \leftrightarrow {1 \over
{\sqrt{2}\lambda}}(1,\rho ^a). \ea In fact, using this prescription
the $T\to 0$ limit of the $p$-brane action is \ba S=\int{d^{p+1}\xi
V^J V^K\partial _J X^\mu \partial _K X^\nu \eta_{\mu\nu}}. \ea

Another type of null bosonic $p$-brane action is given in
\cite{HLU94} \ba\nl S = \int d^{p+1}\xi \left(\Delta\cdot\mathbf{P} -
\frac{1}{2}V(\xi) \mathbf{P}\cdot\mathbf{P}\right),\ea where
$\Delta^{\mu_0 ... \mu_p}$ is the following antisymmetric space-time
tensor \ba\nl \Delta^{\mu_0 ... \mu_p} =
\frac{1}{\left[(p+1)!\right]^{1/2}} \epsilon^{J_0 ...
J_p}\p_{J_0}x^{\mu_0}\ldots\p_{J_p}x^{\mu_p}\ea and
$\mathbf{P}_{\mu_0 ... \mu_p}$ are totally antisymmetric "generalized
momenta" satisfying the constraint $\mathbf{P}^2 = - T^2$.

The quantization of different types of tensionless p-branes is
also discussed in \cite{PS97}. Now we are going to describe the
results of this paper. The constraints are written there in the
form \be \phi^{-1}(\sigma_1,\ldots,\sigma_p)
&=&P^{\mu}P_{\mu}(\sigma_1,\ldots,\sigma_p)=0\nonumber\\
L^{\alpha}(\sigma_1,\ldots,\sigma_p)&=&P^{\mu}\partial
_{\alpha}X_{\mu} (\sigma_1,\ldots,\sigma_p)=0,\nonumber \ee  Note
that  the greek index $\alpha$ runs from 1 to $p$. In Fourier
modes the constraints read (for simplicity closed $p$-branes are
considered)
\be
\phi^{-1}_{m_1,\ldots ,m_p}&=&\half \sum_{k_1,\ldots, k_p
=-\infty}^{+\infty}p_{m_1 - k_1,\ldots, m_p - k_p} \cdot
p_{k_1,\ldots ,k_p}= 0,\label{ccon1}\\ L^{\alpha}_{m_1,\ldots
,m_p}&=&-i\sum_{k_1,\ldots, k_p =-\infty}^{+\infty}
 k_{\alpha}p_{m_1 -
k_1,\ldots, m_p - k_p}\cdot x_{k_1,\ldots ,k_p}= 0\label{ccon2} \ee
and they satisfy the following algebra
\be
\left[ {\phi}^{-1}_{m_1,\ldots ,m_p},{L}^{\alpha}_{n_1,\ldots
,n_p}\right]   &=& (m_{\alpha}-n_{\alpha}){\phi}^{-1}_{m_1
+n_1,\ldots,m_p+n_p},\label{excont}\\
  \left[ {L}^{\alpha}_{m_1,\ldots
,m_p},{L}^{\beta}_{n_1,\ldots ,n_p}\right]
&=&m_{\beta}L^{\alpha}_{m_1 +n_1,\ldots,m_p+n_p}-
n_{\alpha}L^{\beta}_{m_1 +n_1,\ldots,m_p+n_p}\nonumber\\
&&+A^{\alpha\beta}(m_1,\ldots
,m_p)\delta_{m_1+n_1}\ldots\delta_{m_p+n_p}. \label{d12}\ee The right
hand side of equation (\ref{d12}), when $m_1 +n_1 = \ldots = m_p +
n_p =0$, is expressed in terms of $L^{\alpha}_{0,\ldots,0}$. But this
operator is not well defined since it depends on the different
orderings of $x^{\mu}_{m_1,\ldots ,m_p}$ and $p^{\mu}_{m_1,\ldots
,m_p}$. Taking into account this ambiguity the possible central
extensions in the right hand side of the commutators (\ref{d12}) are
included. The values of these central extensions are constrained by
the Jacobi identities and the antisymmetry of the commutators. It is
found that for $p>1$
\be
A^{\alpha\beta}(m_1,\ldots
,m_p)=A^{\alpha\beta}(m_{\alpha},m_{\beta})=\half\left
(m_{\beta}d^{\alpha}+m_{\alpha}d^{\beta}\right ),\nonumber \ee where
$d^{\alpha}$ are constants.

In order to clarify the implications of the last relation  one takes
$\alpha=\beta$ in (\ref{d12}). Then
\be
\left[ {L}^{\alpha}_{m_1,\ldots ,m_p},{L}^{\alpha}_{n_1,\ldots
,n_p}\right] &=&(m_{\alpha}-n_{\alpha})L^{\alpha}_{m_1
+n_1,\ldots,m_p+n_p}+m_{\alpha}d^{\alpha}
\delta_{m_1+n_1}\ldots\delta_{m_p+n_p}.\nonumber \ee The
corresponding relation for the string ($p=1$) is
\be
\left[ {L}_{m},{L}_{n}\right] &=&(m-n)
    {\phi}^{L}_{m+n}+({d}_{3}m^{3}+{d}_{1}m)\delta_{m+n}.
\nonumber \ee Thus one finds that in the case of the tensionless
$p$-branes with $p>1$, the central extensions in the algebra of
the constraints become "smoother" since their cubic terms have to
vanish due to the Jacobi identities. It is interesting to note
that the same cancellation will also occur in the case of the
usual $p$-branes. The constraints $L^{\alpha}$ are not modified by
the nonzero tension and so the results obtained here for the
subalgebra (\ref{d12}) are also valid for the tensile $p$-brane.

The investigation of the quantum theory of this model within the
framework of a BRST quantization requires the introduction   of new
operators. To every constraint, one introduces a ghost pair
$c^{A}_{m_1,\ldots,m_p}$, $b^{A}_{m_1,\ldots,m_p}$,
$A\in\{-1,L_{\alpha}\}$, that is fermionic. The generator of BRST
transformations, the BRST charge is found to be
\be
Q& = &\sum_{k_1,\ldots,k_p}
\phi_{-k_1,\ldots,-k_p}^{-1}c_{k_1,\ldots,k_p}^{-1}+ \sum_{\alpha
=1}^{p} \sum_{k_1,\ldots,k_p}{L^{\alpha}}_{-k_1,\ldots,-k_p}
c_{k_1,\ldots,k_p}^{L_{\alpha}}\nonumber\\ &&- \sum_{\alpha
=1}^{p}\sum_{k_1,\ldots,k_p}\sum_{l_1,\ldots,l_p}
(k_{\alpha}-l_{\alpha})c_{-k_1,\ldots,-k_p}^{-1}
c_{-l_1,\ldots,-l_p}^{L_{\alpha}} b_{k_1+ l_1,\ldots,k_p +
l_p}^{-1}\nonumber\\ &&-\half \sum_{\alpha,\beta
=1}^{p}\sum_{k_1,\ldots,k_p}\sum_{l_1,\ldots,l_p}
k_{\beta}c_{-k_1,\ldots,-k_p}^{L_{\alpha}}
c_{-l_1,\ldots,-l_p}^{L_{\beta}} b^{L_{\alpha}}_{k_1+ l_1,\ldots,k_p
+ l_p}\nonumber\\ &&+\half \sum_{\alpha,\beta
=1}^{p}\sum_{k_1,\ldots,k_p}\sum_{l_1,\ldots,l_p}
l_{\alpha}c^{L_{\alpha}}_{-k_1,\ldots,-k_p}
c_{-l_1,\ldots,-l_p}^{L_{\beta}} b_{k_1+ l_1,\ldots,k_p +
l_p}^{L_{\beta}} .\label{Q} \ee The classical nilpotency, ${Q}^2=0$,
is guaranteed by construction. To check the nilpotency of the quantum
${\cal Q}= \frac{1}{2}(Q + Q^{\dagger})$, which is constructed to be
hermitian, one uses the extended constraints
$\tilde{\phi}^{I}_{m_1,\ldots,m_p}$. These are BRST invariant
extensions of  the original constraints and they satisfy the same
algebra as the original ones for first rank systems. They are defined
by  the equation
\be
\tilde{\phi}^{I}_{m_1,\ldots,m_p}\equiv \{
b^{I}_{m_1,\ldots,m_p},{\cal Q}\},\nonumber \ee We can now calculate
the BRST anomaly using the method described in \cite{RM87}. There it
is shown that
\be
{\cal Q}^{2}=\half \sum_{I,J}\sum_{m_1,\ldots,m_p}{\tilde
{d}}^{IJ}_{m_1,\ldots,m_p}c^{I}_{m_1,\ldots,m_p}
c^{J}_{-m_1,\ldots,-m_p},\nonumber \ee where ${\tilde {d}}^{IJ}$ are
the central extensions of the extended constraints  algebra. This
means that
 \be {\cal Q}^{2} =
\half\sum_{\alpha}\tilde{d}^{\alpha}m_{\beta}
c^{L_{\alpha}}_{m_1,\ldots,m_p}
c^{L_{\beta}}_{-m_1,\ldots,-m_p}.\label{as} \ee The exact values of
$\tilde{d}^{\alpha}$ depend on the  vacuum and ordering we use. The
simplest and safest method to determine these constants is to
calculate the vacuum expectation value of the commutators (\ref{d12})
for the extended constraints.

According to arguments presented in \cite{ILST93}, the vacuum
suitable for tensionless strings is not one annihilated by the
positive modes of the operators but  one annihilated  by the momenta
\be
  p^{\mu}_{m}|0\rangle_{p} =0 \quad \forall m.\label{vac}
\ee In the case of the tensionless $p$-brane, the vacuum is
defined also by (\ref{vac}) in \cite{PS97}.

Following the prescription of \cite{HM89}, one takes  the {\em ket }
states to be built from our vacuum of choice, $|0\rangle_{p}$, and
the {\em bra } states to be built from $\mbox{}_{x}\langle 0|$
satisfying $\mbox{}_{x}\langle 0|0\rangle_{p}=1$.

For the vacuum (\ref{vac}) and from the requirement that the BRST
charge (\ref{Q}) should annihilate the vacuum, one obtains further
requirements on the ghost part of the vacuum. Doing this one finds
that the vacuum has to satisfy the following conditions
\be
 p^{\mu}_{m_1,\ldots,m_p}|0\rangle
=b^{-1}_{m_1,\ldots,m_p}|0\rangle = 0,\nonumber\\
  \langle 0|x^{\mu}_{m_1,\ldots,m_p}  =\langle
0|c^{-1}_{m_1,\ldots,m_p}=0.\nonumber \ee The expectation value of
the commutator (\ref{d12}) is \be\label{fcon1} \left\langle 0\left
|\left[\tilde{L}^{\alpha}_{m_1,\ldots
,m_p},\tilde{L}^{\alpha}_{-m_1,\ldots ,-m_p}\right]\right
|0\right\rangle =2m_{\alpha}\left\langle 0\left
|\tilde{L}^{\alpha}_{0,\ldots,0}\right|0\right\rangle
+m_{\alpha}\tilde{d}^{\alpha}\nonumber\\ \Rightarrow
0=2m_{\alpha}a^{\alpha}_L+m_{\alpha}\tilde{d}^{\alpha} \Rightarrow
\tilde{d}^{\alpha}=-2a^{\alpha}_L\label{alpha} \ee where
$a^{\alpha}_L\equiv \left\langle 0\left
|\tilde{L}^{\alpha}_{0,\ldots,0}\right|0\right\rangle$. But for a
hermitian BRST charge ${\cal{Q}}$ one will has
\be
a^{\alpha}_L= -\half(D+1+p)\sum_{k_{\alpha}=-\infty}^{+\infty}
k_{\alpha}\sum_{k_1, \ldots,k_{\alpha -1},k_{\alpha
+1},\ldots,k_p}1=0.\nonumber \ee Thus from the relations
(\ref{as}) and (\ref{alpha}) one deduces that the BRST charge is
nilpotent for any space-time dimension $D$. So in the theory of
tensionless $p$-branes, just as in the theory of tensionless
strings the critical dimension is absent and the theory is quantum
mechanically consistent for any dimension $D$.

If we choose the vacuum to be annihilated by the positive modes, as
is the case for the usual string, the commutators (\ref{d12}) will
give
\be
&&\left\langle 0\left |\left[\tilde{L}^{\alpha}_{m_1,\ldots
,m_p},\tilde{L}^{\alpha}_{-m_1,\ldots ,-m_p}\right]\right
|0\right\rangle =2m_{\alpha}\left\langle 0\left
|\tilde{L}^{\alpha}_{0,\ldots,0}\right|0\right\rangle
+m_{\alpha}\tilde{d}^{\alpha}\nonumber\\ &\Rightarrow &
[(D-25-p)m_{\alpha}^3- (D-1-p)m_{\alpha}]\sum_{k_1,
\ldots,k_{\alpha -1},k_{\alpha +1},\ldots,k_p} \frac{1}{6}=
2m_{\alpha}a^{\alpha}_L+ m_{\alpha}\tilde{d}^{\alpha}\nonumber\\
&\Rightarrow &D=25+p.\label{crst} \ee This results agrees with the
critical dimension of $D=27$ for the membrane ($p=2$) which was
given in \cite{MS88}, and can be generalized to the case of null
spinning $p$-brane  \cite{PS97} \be
D=\frac{100+4p-22N}{4+N}\label{crd} ,\ee $N$ being the number of
world-volume supersymmetries.

The isomorphism $C_{D-1,1}\simeq O(D,2)$ for $D\geq 3$ makes it
possible to construct a theory in two extra dimensions such that
the previous model corresponds to a particular gauge fixing of the
latter and the conformal symmetry is manifest and linearly
realized. This {\it conformal} tensionless $p$-brane action can be
given by \be S&=&\int d^{p+1}\xi \{V^{a}(\partial_{a}+W_{a}) X^{A}
  V^{b}(\partial_{b}+W_{b})
X_{A} +\Phi X^{A}X_{A}\},\label{action2} \ee where $A=0,\ldots,d+1$
and the new metric has the form
\be
\eta _{AB} = \left( \matrix{\eta_{\mu\nu}\quad  \hfill 0 \quad 0  \cr
 0...0 \quad \hfill 1 \quad 0 \cr 0...0 \quad \hfill 0 -1
\cr} \right).\nonumber \ee $W_{a}$ is the gauge field for scale
transformations and $\Phi $ is a Lagrange multiplier field.

We can check that by imposing two gauge fixing conditions $P^{+}=0$,
$X^{+}=1$ the generators of the Lorentz transformations in the
extended space become the generators of the conformal group in the
original space. Thus rotations in the extended space correspond to
conformal transformations in the original space.

Going to the Hamiltonian formulation we  find in exactly the same
manner that the Hamiltonian is again a linear combination of the
constraints. In addition to the original constraints
(\ref{ccon1})-(\ref{ccon2}) we will have two new ones which in
Fourier modes can be written as follows
\be
\phi^{1}_{m_1,\ldots ,m_p}&=&\half \sum_{k_1,\ldots, k_p
=-\infty}^{+\infty}x_{m_1 - k_1,\ldots, m_p - k_p} \cdot
x_{k_1,\ldots ,k_p}= 0,\label{ccon3}\\ \phi^{0}_{m_1,\ldots
,m_p}&=&\half\sum_{k_1,\ldots, k_p =-\infty}^{+\infty}
 p_{m_1 -
k_1,\ldots, m_p - k_p}\cdot x_{k_1,\ldots ,k_p}= 0\label{ccon4}. \ee
The constraint algebra with the central extensions included, for
$p>1$, will be given by
\be
\left[ {\phi}^{-1}_{m_1,\ldots ,m_p},\phi^{1}_{n_1,\ldots
,n_p}\right]   &=& -2i{\phi}^{0}_{m_1 +n_1,\ldots,m_p+n_p}-2ic
\delta_{m_1+n_1}\ldots\delta_{m_p+n_p},\label{excont1}\\
 \left[
{\phi}^{-1}_{m_1,\ldots ,m_p},\phi^{0}_{n_1,\ldots ,n_p}\right]   &=&
-i{\phi}^{-1}_{m_1 +n_1,\ldots,m_p+n_p},\label{excont2}\\
 \left[
{\phi}^{-1}_{m_1,\ldots ,m_p},{L}^{\alpha}_{n_1,\ldots ,n_p}\right]
&=& (m_{\alpha}-n_{\alpha}){\phi}^{-1}_{m_1
+n_1,\ldots,m_p+n_p},\label{excont3}\\
 \left[
{\phi}^{1}_{m_1,\ldots ,m_p},\phi^{0}_{n_1,\ldots ,n_p}\right]   &=&
i{\phi}^{1}_{m_1 +n_1,\ldots,m_p+n_p},\label{excont4}\\
 \left[
{\phi}^{1}_{m_1,\ldots ,m_p},{L}^{\alpha}_{n_1,\ldots ,n_p}\right]
&=& (m_{\alpha}+n_{\alpha}){\phi}^{1}_{m_1
+n_1,\ldots,m_p+n_p},\label{excont5}\\
 \left[
{\phi}^{0}_{m_1,\ldots ,m_p},{L}^{\alpha}_{n_1,\ldots ,n_p}\right]
&=& m_{\alpha}{\phi}^{0}_{m_1 +n_1,\ldots,m_p+n_p}+cm_{\alpha}
\delta_{m_1+n_1}\ldots\delta_{m_p+n_p},\label{excont6}\\ \left[
{L}^{\alpha}_{m_1,\ldots ,m_p},{L}^{\beta}_{n_1,\ldots ,n_p}\right]
&=&m_{\beta}L^{\alpha}_{m_1 +n_1,\ldots,m_p+n_p}-
n_{\alpha}L^{\beta}_{m_1 +n_1,\ldots,m_p+n_p}\nonumber\\
&&+\half\left (m_{\beta}d^{\alpha}-n_{\alpha}d^{\beta}\right )
\delta_{m_1+n_1}\ldots\delta_{m_p+n_p}. \ee With the constraint
algebra at hand one finds the BRST charge to be
\be
 Q& = &\sum_{k_1,\ldots,k_p}\left [
\phi_{-k_1,\ldots,-k_p}^{-1}c_{k_1,\ldots,k_p}^{-1}+ \right
.\sum_{\alpha =1}^{p} {L^{\alpha}}_{-k_1,\ldots,-k_p}
c_{k_1,\ldots,k_p}^{L_{\alpha}}\nonumber\\ &&\left .+
\phi_{-k_1,\ldots,-k_p}^{0}c_{k_1,\ldots,k_p}^{0} +
\phi_{-k_1,\ldots,-k_p}^{1} c_{k_1,\ldots,k_p}^{1}\right ]\nonumber\\
&&+\sum_{k_1,\ldots,k_p}\sum_{l_1,\ldots,l_p}\left [ 2i
c_{-k_1,\ldots,-k_p}^{-1} c_{-l_1,\ldots,-l_p}^{1} b^{0}_{k_1+
l_1,\ldots,k_p + l_p}\right .\nonumber\\ &&+
ic_{-k_1,\ldots,-k_p}^{-1} c_{-l_1,\ldots,-l_p}^{0} b^{-1}_{k_1+
l_1,\ldots,k_p + l_p}-i c_{-k_1,\ldots,-k_p}^{1}
c_{-l_1,\ldots,-l_p}^{0} b^{1}_{k_1+ l_1,\ldots,k_p + l_p}\nonumber\\
&&-\half \sum_{\alpha,\beta =1}^{p}
k_{\beta}c_{-k_1,\ldots,-k_p}^{L_{\alpha}}
c_{-l_1,\ldots,-l_p}^{L_{\beta}} b^{L_{\alpha}}_{k_1+ l_1,\ldots,k_p
+ l_p}\nonumber\\ &&+\half\sum_{\alpha,\beta =1}^{p}
l_{\alpha}c^{L_{\alpha}}_{-k_1,\ldots,-k_p}
c_{-l_1,\ldots,-l_p}^{L_{\beta}} b_{k_1+ l_1,\ldots,k_p +
l_p}^{L_{\beta}}\nonumber\\ &&- \sum_{\alpha =1}^{p}
(k_{\alpha}-l_{\alpha})c_{-k_1,\ldots,-k_p}^{1}
c_{-l_1,\ldots,-l_p}^{L_{\alpha}} b^{1}_{k_1+ l_1,\ldots,k_p +
l_p}\nonumber\\ &&- \sum_{\alpha =1}^{p}
(k_{\alpha}-l_{\alpha})c_{-k_1,\ldots,-k_p}^{-1}
c_{-l_1,\ldots,-l_p}^{L_{\alpha}} b_{k_1+ l_1,\ldots,k_p +
l_p}^{-1}\nonumber\\ &&-\sum_{\alpha =1}^{p}\left .
k_{\alpha}c_{-k_1,\ldots,-k_p}^{0} c_{-l_1,\ldots,-l_p}^{L_{\alpha}}
b^{0}_{k_1+ l_1,\ldots,k_p + l_p}\right ].\label{Q1} \ee Again one
can check the nilpotency of $\cal{Q}$ with the use of the extended
constraints. They satisfy the same algebra as the original
constraints. The only thing that remains now is to calculate the
values of the constants $\tilde{d}^{\alpha}$, $\alpha = 1,\ldots,p$
and $\tilde{c}$ for the vacuum and ordering introduced previously. In
this case, the condition $ p^{A}_{m_1,\ldots,m_p}|0\rangle =0$
together with the requirement that the BRST charge (\ref{Q1}) should
annihilate the vacuum gives the following consistency conditions
$\forall m_1,\ldots,m_p$
\be
  p^{A}_{m_1,\ldots,m_p}|0\rangle
=c^{1}_{m_1,\ldots,m_p}|0\rangle= b^{-1}_{m_1,\ldots,m_p}|0\rangle
=0,\nonumber\\
  \langle 0|x^{A}_{m_1,\ldots,m_p}
=\langle 0|c^{-1}_{m_1,\ldots,m_p}=
 \langle 0|b^{1}_{m_1,\ldots,m_p} =0.\nonumber
\ee In the same way as was done previously, one finds that
$\tilde{d}^{\alpha}=0$. The value of $\tilde{c}$ is found by
calculating the expectation value of the commutator $\left\langle
0\right|[\tilde{\phi}^{0}_{m_1,\ldots
,m_p},\tilde{\phi}^{L}_{-m_1,\ldots ,-m_p}]\left|
  0\right\rangle $. This gives
\be
\label{fcon2}
 &\left\langle
0\right|[\tilde{\phi}^{0}_{m_1,\ldots
,m_p},\tilde{\phi}^{L}_{-m_1,\ldots ,-m_p}]\left|
  0\right\rangle =m_{\alpha}\left\langle
0\left |\tilde{\phi}^{0}_{0,\ldots,0}\right |
  0\right\rangle
+\tilde{c}m_{\alpha}&\nonumber\\
  &\Rightarrow
0=m_{\alpha}a_{0}+\tilde{c}m_{\alpha}=0&\nonumber\\
  &\Rightarrow
  \tilde{c}=-a_{0}.&\nonumber
\ee But for a Hermitian BRST charge $\cal{Q}$ we will have
\be
\alpha_{0}\equiv \left\langle 0\right|\tilde{\phi}^{0}_{0}\left|
  0\right\rangle =-\frac{i}{4}[D-2]\left(
 \sum_{k_1,\ldots,k_p}1
\right).\nonumber \ee

The BRST charge is nilpotent when all the constants $\tilde{d}_{a}$
and $\tilde{c}$ are equal to 0. In particular we must have
\be
\tilde{c}=0 \Rightarrow D=2.\nonumber \ee Thus one finds a
critical dimension of $D=2$ for all the tensionless conformal
$p$-branes. It should be stated here that the result is valid for
$p\neq 0$ since the quantum theory of the  conformal particle is
consistent in any dimension.

The above results can be easily generalized to the case of the
conformal tensionless spinning $p$-brane. Using the techniques
presented here, one finds for the $p$-brane  a negative critical
dimension
\be
D=2-2N,\quad \forall p\geq 1,\nonumber \ee $N$, being the number
of supersymmetries. Again this result is valid for $p\neq 0$ since
for the conformal spinning particle the same analysis reveals
consistency in any dimension.

The first work on the classical dynamics of tensionless branes
$(p>1)$ in curved background is \cite{RZ96} (unpublished). Because of
this, we will consider the results obtained in it in some detail.

The action for null $p$-branes in a cosmological background
$~G_{MN}(x)~$ may be written as \ba\label{rz1} S=\int d^{p+1}\xi
\; {det(\partial_{\mu} x^M \; G_{MN}(x) \;
\partial_{\nu} x^N)\over E(\tau,\sigma^n )},
\\ \nl (M,N=0,...,D-1), \h (m,n=1,...,p),\ea where
$~E(\tau,\sigma^n)~$ is a $~(p+1)~$--dimensional world-volume
density. The determinant $~g~$ of the induced null $p$-brane
metric $~g_{\mu\nu}~$ \ba\nl g_{\mu\nu}= \partial_{\mu}x^M\;
G_{MN}(x)\;\partial_{\nu} x^N =
    \left(\matrix{\dot{x}^A \; G_{AB}(x) \;\dot{x}^B
    & \dot{x}^A \; G_{AB}(x)\;
    \partial_{n}x^B \cr \partial_{m}x^A \; G_{AB}(x)
    \;\dot{x}^B &\hat{g}_{mn}(x)\cr }\right) ,\ea
\ba\nl\hat{g}_{mn}(x) = \partial_{m}x^A \; G_{AB}(x) \;
\partial_{n}x^B, \ea may be presented in a factorized
form \ba\nl g=\dot{x}^M \;\tilde{\Pi}_{MN}(x)
\;\dot{x}^N\;\hat{g},\h \hat{g}=\det\hat{g}_{mn},\ea where point
denotes differentiation with respect to $~\tau~$. The matrix
\ba\nl \tilde{\Pi}_{MN}= G_{MN}- G_{MB} \;\partial_{m}x^B
\;\hat{g}^{-1mn} \;\partial_{n}x^L \;G_{LN} \ea has the properties
of a projection operator. Therefore, the action (\ref{rz1}) can be
written in the following form: \ba\nl S= \int d^{p+1}\xi \;
{\dot{x}^M \;\tilde{\Pi}_{MN}(x) \;\dot{x}^N \;\hat{g} \over
E(\tau,\sigma^n )}.\ea

The variation of this action with respect to $~E~$ generates the
degeneracy condition for the induced metric $~g_{\mu\nu}$ \ba\nl
g=\det g_{\mu\nu}=0 , \ea which separates the class of
$~(p+1)~$--dimensional isotropic geodesic hypersurfaces
characterized by the null volume. In the gauge \ba\nl \dot{x}^M
G_{MN} \partial _m x^N=0 ; ~~~ \left({\hat{g} \over E}
\right)^\bullet =0 \ea one finds the equations of motion and
constraints in the following form $$ \ddot{x}^M + \Gamma_{PQ}^M \;
\dot{x}^P \dot{x}^Q = 0 $$ \ba\label{rz8} \dot{x}^M
G_{MN}\dot{x}^N =0,~~~~~~~~ \dot{x}^M G_{MN}\partial_{m}x^N=0 \ea

Now consider the case of $D$-dimensional Friedmann universe with
$~k=0~$ (flat spatial section) described by the metric form
\ba\label{rz9} ds^2=G_{MN}dx^Mdx^N=(dx^0)^2-R^2(x^0)
\;dx^i\delta_{ik}dx^k . \ea It is convenient to transform
equations (\ref{rz8}) to the conformal time $~\tilde{x}^0(\tau
,\sigma^n)~$, defined by \ba\label{rz10}
dx^0=C(\tilde{x}^0)d\tilde{x}^0,~~~~~C(\tilde{x}^0) = R(x^0),~~~~~
\tilde{x}^i = x^i \ea In the gauge of conformal time, the metric
(\ref{rz9}) is presented in the conformal-flat form
\ba\label{rz11} ds^2=C(\tilde{x}^0) \eta_{MN}d\tilde{x}^M
d\tilde{x}^N,~~~~~~~~~ \eta_{MN}= diag(1, -\delta_{ij}) \ea with
Christoffel symbols $~\tilde{\Gamma}_{PQ}^M(\tilde{x})~~$
\ba\label{rz12} \tilde{\Gamma}_{PQ}^M(\tilde{x})=
C^{-1}(\tilde{x})[\delta_P^M \tilde{\partial}_QC+\delta_Q^M
\tilde{\partial}_PC - \eta_{PQ}\tilde{\partial}^MC].\ea Taking
into account the relations (\ref{rz10}), (\ref{rz12}), one can
transform equations (\ref{rz8}) to the form
 \ba\nl
 \ddot{\tilde{x}}^M +
 2C^{-1}\dot{C}\dot{\tilde{x}}^M=0
 \ea
 \ba\nl
\eta_{MN}\dot{\tilde{x}}^M\dot{\tilde{x}}^N=0,~~~~~~~~\eta_{MN}
 \dot{\tilde{x}}^M\partial_{m}{\tilde{x}}^N=0.
 \ea
The first integration of these  equations leads to the following
first order equations
 \ba\label{rz15}
 H^\ast C^2\dot{\tilde{x}}^0 = \psi^0(\sigma^1,\sigma^2,..,\sigma^p),
 ~~~~~~~ H^\ast C^2 \dot{\tilde{x}}^i= \psi^i(\sigma^1,\sigma^2,..,
\sigma^p ), \ea the solutions of which have the form
\ba\label{rz16} \tau = H^\ast \psi_0^{-1} \int_{t_o}^t dt \; R(t),
\ea
 $$ x^i(\tau ,\sigma^1,\sigma^2,..,\sigma^p)) = H^{\ast
 -1} \psi^i \int_0^\tau d\tau \; R^{-2}(t),$$
where $~H^{\ast}~$ is a metric constant with dimension $~L^{-1}~$,
$~t_0 \equiv x^0(0,\sigma^1,\sigma^2,..,\sigma^p),~$ and
$~x^i(0,\sigma^1,\sigma^2,..,\sigma^p) )~$, $\psi ^M(\sigma^n)$
are the initial data. The solution (\ref{rz16}) for the space
world
 coordinates $~x^i(t) (i=1,..., D-1)~$ as a function of the cosmic time $
 ~t=x^0~$, may be written in the equivalent form as
 \ba\nl
 x^i(t,\sigma^1,\sigma^2,..,\sigma^p)=x^i(t_0,\sigma^1,\sigma^2,..,\sigma^p)
+\nu^i(\sigma^1,\sigma^2,..,\sigma^p )\int_{t_o}^t~ dt \;
R^{-1}(t), \ea where $~\nu^i(\sigma^n ) \equiv \psi_0^{-1}\psi^i$.
 The explicit form of the solutions (\ref{rz16}) allows to transform the
 constraints (\ref{rz11}) into those for the Cauchy initial data:
\ba\label{rz18} \nu^i(\sigma^m)\nu^k(\sigma^n)\delta_{ik} =1,~~
\ea \ba\label{rz19}
\partial_{m}x^0(0,\sigma^m)=
R(x^0(0,\sigma^m) \;
\nu^i(\sigma^n)\delta_{ik}\partial_{m}x^k(0,\sigma^m), \ea where
$\nu^k(\sigma^1,\sigma^2,..,\sigma^p)=\psi^{i} \psi_0^{-1}$. Note
that the constraints (\ref{rz18}) and (\ref{rz19}) produce
additional constraints , which are their integrability conditions
\ba\nl
\partial_{m}(\nu_i(\sigma^n)\partial_{n}x^{i}(0,\sigma^l)) -
\partial_{n}(\nu_{i}(\sigma^n)\partial_{m}{x}^{i}(0,\sigma^l))=0.
\ea

Now  one can show that the null p-branes may be considered as
dominant gravity sources of the Friedmann universes. To this end,
one assumes that the perfect fluid of these null p-branes is
homogenious and isotropic. The energy density $~\rho (t)~$ and the
pressure $~p(t)~$ of this fluid and its energy -momentum$~\langle
T_{MN}\rangle~$ are connected by the standard relations
\ba\label{rz21} \langle{T_ 0 }^0\rangle =\rho(t),~~~~~~~ \langle
{T_i }^j \rangle = -p(t){\delta_i} ^j = - { {\delta_i}^j \over
D-1} ~{A\over R^D(t)}, \ea The tensor $~\langle T_{MN}\rangle~$ is
derived from the momentum--energy tensor $~T_{MN}~$ of a null
$p$-brane by means of its space averaging when a set of null
$p$-branes is introduced instead of a separate null $p$-brane. The
energy -momentum tensor $~ T^{MN}(x)~$ of null $p$-brane is
defined by the variation of the action (\ref{rz1}) with respect to
$~G_{MN}(x)~$ \ba\label{rz22} T^{MN}(X)= {1 \over \pi\gamma^\ast
\sqrt{ _{|G|}}} \int ~d\tau d^{p}\xi \;
\dot{x}^M\dot{x}^N\delta^D(X^M-x^M) \ea After the substitution of
the velocities $\tilde{x}^m$ (\ref{rz15}) and subsequent
integration with respect to $~\tau~$, the non-zero components of
$~T_{MN}~$ (\ref{rz22}) take the following form \ba\nl
T^{00}(X)=\frac{1}{\pi\gamma^* H^*}R^{-D}(t)\int d\tau d^{p}{\xi}
\ \psi_0(\sigma^{m})\delta^{D-1}(X^i-x^i(\tau,\sigma^{m})),\ea
\ba\nl T^{ik}(X)=\frac{1}{\pi\gamma^*H^*}R^{-(D+2)}\int d\tau
{d^{p}\xi}\{\nu^i(\sigma)\nu^k(\sigma)\psi_0(\sigma)
\delta^{D-1}\left(X^i-x^i(\tau,\sigma^{m})\right),\ea where the
time dependence in $~T^{MN}~$ is factorized and accumulated in the
scale factor $~R(t).~$ The constraint
$~\nu^i(\sigma^m)\nu^k(\sigma^n)\delta_{ik} =1~~$ gives rise to
the following relation between the components of the tensor
$~\langle T_{MN}\rangle~$ \ba\nl Tr~T ={T_0}^0+G_{ij}T^{ij}=0 .
\ea As a result of the space averaging, one finds the non-zero
components of $~\langle T_{MN}\rangle~$ to be equal to
\ba\label{rz25} \langle {T_0 }^0 \rangle = \rho(t) =  {A \over
R^D(t)},~~~~~~ \langle {T_i }^j \rangle = -p(t){\delta_i} ^j , \ea
where $~A~$ is a constant with dimension $~L^{-D}~$. The equations
(\ref{rz25}) show that the equation of state of null $p$-branes
fluid is just the equation of state for a gas of massless
particles \ba\label{rz26} \langle Tr~T\rangle = \langle {T_M} ^M
\rangle =0 \Longleftrightarrow \rho = (D-1)p. \ea

Now assume that the fluid of null $p$-branes is a dominant source
of the Friedmann-Robertson-Walker (FRW) gravity (\ref{rz9}). For
the validity of the last conjecture it is necessary that the
Hilbert-Einstein (HE) equations \ba\label{rz28}  {R_M} ^N =8\pi
G_D\langle {T_M}^N \rangle \ea with non-zero Ricci tensor $~{R_M}
^N~$ components defined by $~{G_M} ^N~$ (\ref{rz9}) $$~ {R_0} ^0 =
-{D-1 \over R}~{d^2R\over dt^2}, $$ \ba\nl {R_i} ^k = -{\delta _i}
^k \left[{1\over R}~ {d^2R\over dt^2} + {D-2\over R^2}
\left({dR\over dt}\right)^2\right] \ea should contain the tensor
$\langle{T_M} ^N\rangle$ (\ref{rz21}) as a source of the FRW
gravity.  Moreover, the constraints (\ref{rz25}), i.e. $$ \rho R^D
- A = 0, $$ must be an integral of motion for the HE system
(\ref{rz28}). It is actually realized because $$ {d\over dt}(\rho
R^D) = - {D-2\over 16 \pi G_D } R^{D-1}~{dR\over dt } {R_M} ^M =
0, $$ since the trace $~{R_M}^M \sim \langle{T_M}^M\rangle=0~ $
(see (\ref{rz26})). In view of this fact it is enough to consider
only one equation of the system (\ref{rz28}) \ba\label{rz30}
\left({1\over R}~{dR \over dt}\right)^2 = {16 \pi G_D
\over(D-1)(D-2)} ~{A\over R^D}\ \ \ea which defines the scale
factor R(t) of the FRW metric (\ref{rz9}). Note that when $D=4$,
the equation (\ref{rz30}) transforms into the well-known Friedmann
equation for the energy density in the radiation dominated
universe with $~k = 0~$.  The solutions of equation (\ref{rz30})
are $$ R_I(t) = [q(t_c -t)]^{2/D}, ~~~~~~~ t < t_c, $$ $$
R_{II}(t) =[q(t -t_c)]^{2/D},~~~~~~ t > t_c , $$ where $q=[4\pi
G_D A / (D-1)(D-2)]^{1/2}~~$ and $t_c$ is a constant of
integration which is a singular point of the metric. The solution
$~R_I~$ describes the stage of negatively accelerated contraction
of $D$-dimensional FRW universe. The second solution $~R_{II}~$
describes the stage of negatively accelerated expansion of the FRW
universe from a state with space volume equal to "zero". Thus we
see that the perfect fluid of non-interacting null $p$-branes may
be considered as an alternative source of the gravity in FRW
universes with $~k = 0~$.

Now let us give the Hamiltonian for null $p$-brane theory in
curved space-time in this formulation. The canonical momentum of
null $p$-brane $~{\cal P}_M~$ conjugated to its world coordinate
$x^M$ is \ba\nl {\cal P}_M = 2E^{-1}\; \hat{g}
\;\tilde{\Pi}_{MN}(x) \;\dot{x}^N .\ea Then one finds the
following primary constraints \ba\nl
 G_{MN} \partial _m x^N {\cal P}^M = 0  ~~~
.\ea The Hamiltonian density produced by the action functional
(\ref{rz1}) is \ba\nl {\cal H}_0 = {1\over 4}E \;\hat{g}^{-1}
G^{MN} {\cal P}_M{\cal P}_N .\ea The condition for conservation of
the primary constraint \ba\nl
        {\cal P}_{(E)} = 0,
\ea where $~{\cal P}_{(E)}~$ is the canonical momentum conjugated
to $~E,~$ generates the following condition \ba\nl \dot{\cal
P}_{(E)} = {\int d^{p}\xi  \;\{\cal H}_{0} ,{\cal P}_{(E)}\} = -
{1\over{4\hat g}}G^{MN}{\cal P}_M{\cal P}_N = 0 .\ea The latter
produces a secondary constraint \ba\nl G^{MN}{\cal P}_M{\cal P}_N
= 0 .\ea Additional constraints do not appear, so the total
Hamiltonian of null $p$-brane is given by \ba\nl
 H = \int d^{p}\xi \  [\lambda^{m}(G_{MN} \partial _m x^N {\cal P}^M) +
{E\over{4\hat g}}G^{MN}{\cal P}_M{\cal P}_N  + \omega{\cal
P}_{(E)}] , \ea This hamiltonian and the reparametrization
constraints may be used for the quantization of null $p$-branes in
a curved space-time.

Finally, let us consider the interaction of null bosonic membranes
$(p=2)$ with antisymmetric tensor field $T_{\mu\nu\lambda}$ in four
space-time dimensions \cite{Z8990}.

In the general case of null $p$-brane living in $D$-dimensional
space, the interaction with the antisymmetric field $T_{\mu\nu\ldots
\lambda}$ may be introduced in the action as follows \cite{Z8990}
\ba\label{t4} S = \int d\xi^{p+1}\left[ \frac{\det(\p_J x^\mu \p_K
x_\mu)}{2E} - \tilde{\lambda}\epsilon^{J_1\ldots J_l}
\p_{J_1}x^{\mu_1}\ldots \p_{J_l}x^{\mu_l}T_{\mu_1\ldots \mu_l}
\right].\ea In the gauge \ba\nl \dot{x}^\nu\p_j x_\nu = 0,\h
E=e(\s^j)\det(\p_j x^\mu \p_k x_\mu),\ea one obtains from (\ref{t4})
the equations \ba\nl \ddot{x}_\nu + \lambda\epsilon^{J_1\ldots J_l}
\p_{J_1}x^{\mu_1}\ldots \p_{J_l}x^{\mu_l} \p_{[\nu}T_{\mu_1\ldots
\mu_l]}=0,\h \dot{x}^2 = 0, \ea where \ba\nl  \p_{[\nu}T_{\mu_1\ldots
\mu_l]}\equiv \p_\nu T_{\mu_1\ldots\mu_l} -
\p_{\mu_1}T_{\nu\mu_2\ldots\mu_l} - \ldots -
\p_{\mu_l}T_{\mu_1\ldots\mu_{l-1}\nu}\ne 0.\ea

From now on, we restrict ourselves to the particular case $D=4$,
$p=2$. The world vector $x^\mu$ of the null membrane interacting
with the field $T_{\lambda\mu\nu}$ which is dual to the vector
field $T_\kappa$ $(T_{\lambda\mu\nu} = T^\kappa
\epsilon_{\kappa\lambda\mu\nu})$ is defined as \ba\label{t17a}
\dot{x}^\mu \dot{x}_\mu = 0,\h \dot{x}^\mu\p_j x_\mu = 0,\h
(j=1,2),\\ \label{t17b} \ddot{x}_{\kappa} + \lambda
\epsilon_{\kappa\mu\nu\lambda}\p_J x^\mu \p_K x^\nu \p_L x^\lambda
\epsilon^{JKL}\p_\rho T^\rho(x) = 0,\h \lambda =
e(\s^j)\tilde{\lambda}.\ea At first, we consider equation
(\ref{t17a}) and using the isotropic character of $\dot{x}^\nu$
present this vector, as well as the vectors $\p_j x^\nu$ which are
orthogonal to $\dot{x}^\nu$, in the spinor form \ba\nl
\dot{x}_{A\dot{A}} = u_A \bar{u}_{\dot{A}},\\ \label{t18} \p_1
x_{A\dot{A}} = u_A \bar{v}_{\dot{A}} + v_A \bar{u}_{\dot{A}},\\
\nl \p_2 x_{A\dot{A}} = u_A \bar{w}_{\dot{A}} + w_A
\bar{u}_{\dot{A}}.\ea The spinor basis used in (\ref{t18}) is
defined as \ba\nl \p_J x^\mu = -\frac{1}{2} \left(
\tilde{\s}^\mu\right)^{\dot{A}A}\p_J x_{A\dot{A}},\h \p_J
x_{A\dot{A}} = \s^{\mu}_{A\dot{A}}\p_J x_\mu.\ea

In the spinor basis, it is evident that the representation
(\ref{t18}) is the general solution of (\ref{t17a}), because
\ba\nl u^A u_A = 0 = \bar{u}_{\dot{A}}\bar{u}_{\dot{A}},\h u^A =
\epsilon^{AB}u_B,\h \bar{u}^{\dot{A}} =
\epsilon^{\dot{A}\dot{B}}\bar{u}_{\dot{B}},\\ \nl
\epsilon^{AB}=-\epsilon^{BA},\h \epsilon^{12}=1.\ea

After solving two of three integrability conditions $(\p_j
x^\nu)^{.} = \p_j \dot{x}^\nu$, one finds the following
representations for the spinors $u^A$, $v^A$, $w^A$: \ba\nl u_A =
\exp\left[ r(\tau,\sigma^j)\right]\alpha_A(\s^j),\h
\bar{u}_{\dot{A}} =
\exp\left[\bar{r}(\tau,\sigma^j)\right]\bar{\alpha}_{\dot{A}}(\s^j),\\
\label{t20} v_A = \exp\left(-\bar{r}\right)\left\{\beta_A (\s^j) +
\int_0^\tau d\tilde{\tau}\left[ \p_1 u_A +
i\mu(\tilde{\tau},\s^j)u_A\right]\exp(\bar{r}) \right\},\\ \nl w_A
= \exp\left(-\bar{r}\right)\left\{\gamma_A (\s^j) + \int_0^\tau
d\tilde{\tau}\left[ \p_2 u_A +
i\nu(\tilde{\tau},\s^j)u_A\right]\exp(\bar{r}) \right\}.\ea Using
(\ref{t20}), we conclude that the vectors $\p_J x_{A\dot{A}}$
constrained by the equation (\ref{t17a}) can be presented in the
form \ba\nl \dot{x}_{A\dot{A}} =
R(\tau,\s^j)\alpha_A\bar{\alpha}_{\dot{A}},\h R\equiv \int_0^\tau
d\tilde{\tau}\exp(r+\bar{r}),\\ \label{t21a} \p_1 x_{A\dot{A}} = (
\alpha_A \bar{\beta}_{\dot{A}} + \beta_A \bar{\alpha}_{\dot{A}}) +
\p_1 \left( R\alpha_A\bar{\alpha}_{\dot{A}}\right),\\ \nl \p_2
x_{A\dot{A}} = \left( \alpha_A \bar{\gamma}_{\dot{A}} + \gamma_A
\bar{\alpha}_{\dot{A}}\right) + \p_2 \left(
R\alpha_A\bar{\alpha}_{\dot{A}}\right).\ea The last integrability
condition $\p_1(\p_2 x^\nu) = \p_2(\p_1 x^\nu)$ may be written in
the form \ba\label{t21b} \p_1 \left( \alpha_A
\bar{\gamma}_{\dot{A}} + \gamma_A \bar{\alpha}_{\dot{A}}\right) =
\p_2 ( \alpha_A \bar{\beta}_{\dot{A}} + \beta_A
\bar{\alpha}_{\dot{A}}).\ea In particular, equation (\ref{t21b})
has the solution $\beta_A = \p_1 \alpha_A$, $\gamma_A =
\p_2\alpha_A$, which generates the following world vector
\ba\label{t21c} x^\mu (\tau,\s^j) = \left[1 +
R(\tau,\s^j)\right]\left( \bar{\alpha}\sigma^\mu \alpha\right) +
q^\mu.\ea If we introduce the vectors $a^\mu = -\frac{1}{2}\left(
\bar{\alpha}\sigma^\mu \alpha\right)$ and $b^\mu(\s^j)$ composed
of the spinors $\alpha_A$, $\beta_A$, $\gamma_A$ and satisfying
the following relations \ba\label{t21d} a^\mu a_\mu = 0,\h
a^\nu\p_jb_\nu=0,\ea the general solution of (\ref{t17a}) may be
written in the form \ba\label{t21e} x^\mu(\tau,\s^j) =
R(\tau,\s^j) a^\mu(\s^j) + b^\mu(\s^j).\ea Taking into account the
invariance of (\ref{t17a}) under world-volume reparametrizations
defined by the relations $\tau\to \tilde{\tau}(\tau,\s^j)$,
$\s^j\to \tilde{\s}^j(\s^k)$, one can choose $R$ as a new proper
time: $\tilde{\tau}=R$.

In the new coordinates ($R,\s^j$), the general solution
(\ref{t21e}) takes the form of the general solution of free null
membrane equations of motion. Hence, the null membrane
interactions with the $T_{\lambda\mu\nu}$ field can be compensated
by the reparametrizations of its world-volume. The $R$-function is
defined by the left hand side of (\ref{t17b}), which may be
written in the following form \ba\nl \ddot{x}_\mu -
6\lambda\Omega_\mu \p_\nu T^\nu = 0,\ea where $\Omega_\mu$ is the
four-vector of an element of the light-like volume \ba\nl
\Omega^\lambda = -\frac{1}{8}\dot{x}_{A\dot{A}}\p_1
x_{B\dot{B}}\p_2 x_{C\dot{C}}
\epsilon^{\lambda\mu\nu\rho}\tilde{\s}^{\dot{A}A}_{\mu}
\tilde{\s}^{\dot{B}B}_{\nu}\tilde{\s}^{\dot{C}C}_{\rho},\ea
written in the spinor basis. $\Omega_\mu$ can be presented also in
the form \ba\label{t22} \Omega^\lambda = \frac{i}{4}\left[
\epsilon^{AB}\epsilon^{\dot{A}\dot{C}}\left(
\tilde{\s}^\lambda\right)^{\dot{B}C} -
\epsilon^{AC}\epsilon^{\dot{A}\dot{B}}\left(
\tilde{\s}^\lambda\right)^{\dot{C}B}\right] \dot{x}_{A\dot{A}}\p_1
x_{B\dot{B}}\p_2 x_{C\dot{C}}.\ea When equations (\ref{t21a}) for
$\p_J x_{A\dot{A}}$ are substituted into (\ref{t22}), one finds
that the four-vector $\Omega^\lambda$ is collinear  to the
light-like vector $a^\lambda(\s^j)$ \ba\label{t23} \Omega^\lambda
= -\frac{1}{2} \dot{R}\left( AR^2 + BR
+D\right)a^\lambda(\s^j),\ea where the coefficients $A(\s^j)$,
$B(\s^j)$ and $D(\s^j)$ are defined as \ba\nl A(\s^j) &=& i\left[
\left( \alpha\cdot\beta\right) \left(
\bar{\alpha}\cdot\bar{\gamma}\right) - c.c.\right],\\ \nl B(\s^j)
&=& i\left[ \left( \alpha\cdot\p_1 \alpha\right) \left(
\bar{\alpha}\cdot\bar{\gamma}\right) - \left(
\bar{\alpha}\cdot\p_2 \bar{\alpha}\right) \left(
\alpha\cdot\beta\right) - c.c.\right],\\ \nl D(\s^j) &=& i\left[
\left( \alpha\cdot\p_1 \alpha\right)  \left( \bar{\alpha}\cdot\p_2
\bar{\alpha}\right) - c.c. \right].\ea

Taking into account the equations (\ref{t21a})-(\ref{t21e}) and
(\ref{t23}), one finds that (\ref{t17b}) is reduced to the single
equation for the function $R$ \ba\nl \ddot{R} + 3\lambda\dot{R}\left(
AR^2 + BR +D \right)\p_\nu T^\nu (R,a^\mu,b^\rho)=0,\ea which can be
integrated. Its general solution may be presented as \ba\nl \tau &=&
\int\frac{dR}{f(R)} + \tilde{C}_2(\s^j),\\ \nl f(R) &=& -\lambda\int
dR \left( AR^2 + BR + D\right)\p_\nu T^\nu (R,a^\mu,b^\rho) +
\tilde{C}_1(\s^j).\ea

In the particular case when $T_{\mu\nu\lambda} =
x^\rho\epsilon_{\rho\mu\nu\lambda}$, this solution is expressed by
the table integral \ba\label{t27} \lambda\tau + C_2(\s^j) =
-\frac{1}{4}\int\frac{dR}{\frac{1}{2}AR^3 + \frac{1}{2}BR^2 + DR +
C_1(\s^j)}.\ea The relation (\ref{t27}) establishes a connection
between the old proper time $\tau$ and the new one $R$. The usage
of the new world-volume coordinates $\tilde{\tau}=R$, $\s^j$ leads
to the exclusion of the field $T_{\lambda\mu\nu}$ from equations
(\ref{t17a}), (\ref{t17b}) in the considered particular case.

\newpage
\vspace*{.5cm}
\section{\bf
 TENSIONLESS BRANES AND THE NULL STRING CRITICAL DIMENSION}
\hspace{1cm} In this section we perform BRST quantization of the
null bosonic $p$-branes using different types of operator
ordering. It is shown that one can or can not obtain critical
dimension for the null string $(p=1)$, depending on the choice of
the operator ordering and corresponding vacuum states. When $p>1$,
operator orderings leading to critical dimension in the $p=1$ case
are not allowed. Admissible orderings give no restrictions on the
dimension of the embedding space-time. The results described here
are obtained in the paper \cite{B9711}.

\subsection{\bf Classical theory}
\hspace{1cm} To begin with, we first write down the Hamiltonian
for the null $p$-branes living in $D$-dimensional Minkowski
space-time. It can be cast in the form: \ba \nl
H=\int{d^p\sigma\left(\mu^0\varphi_0+\mu^j\varphi_j\right)},\ea
where $\mu^0 , \mu^j$ are Lagrange multipliers being arbitrary
functions of the time parameter $\tau$ and volume coordinates
$\sigma^1,...,\sigma^p$ . The constraints $\varphi_0,\varphi_j$
are defined by the equalities: \ba \nl \vf_0=p^{\mu}p_{\mu}, \h
\vf_j&=&\eta_{\mu\nu}p^{\mu}\p_j x^{\nu}.\ea Here $x^\mu$ and
$p_\mu$ are canonically conjugated coordinates and momenta and
$\eta_{\mu\nu}$ $=$ diag$(-1,1,...,1)$.

$\vf_0$ and $\vf_j$ obey the Poisson bracket algebra \ba \nl
\{\varphi_0(\underline{\sigma_1}), \varphi_0(\underline{\sigma_2})\}&
= &0 ,
\\
\{\varphi_0(\underline{\sigma_1}), \varphi_j(\underline{\sigma_2})\}&
= &[\vf_0(\ul{\sigma_1})+\vf_0(\ul{\sigma_2})]\partial_j \delta^p
(\ul{\sigma_1}-\ul{\sigma_2}) ,
\\
\nl \{\vf_j(\ul{\sigma_1}),\vf_k(\ul{\sigma_2})\}&=
&[\delta_{j}^{l}\vf_k(\ul{\sigma_1})+
\delta_{k}^{l}\vf_j(\ul{\sigma_2})]
\partial_l\delta^p(\ul{\sigma_1}-\ul{\sigma_2}) ,
\ea which means, that they are first class quantities. The Dirac
consistency conditions \cite{D64} \ba \nl \{\vf_0,H\}\approx{0}
\hspace{1cm}, \hspace{1cm} \{\vf_j,H\}\approx{0} , \ea do not
place any restrictions on the Lagrange multipliers $\mu^0,\mu^j$.

Following the BFV-BRST method for quantization of constrained
systems \cite{FV75,BV77,FF78,H85}, we now introduce for each
constraint $\vf_0$, $\vf_j$ a pair of anticommuting ghost
variables $(\eta^0,P_0)$, $(\eta^j,P_j)$ respectively, which are
canonically conjugated. Then the BRST charge is \ba \nl
Q=\int{d^p\sigma\{\vf_0\eta^0+\vf_j\eta^j+
P_0[(\p_j\eta^j)\eta^0+(\p_j\eta^0)\eta^j]+
P_k(\p_j\eta^k)\eta^j\}} \ea and it has the property \ba \nl
\{Q,Q\}_{pb}=0 \ea where $\{.,.\}_{pb}$ is the Poisson bracket in
the extended phase space $(x^\nu,p_\mu;\eta^0,P_0;\eta^j,P_k)$.

In the new phase space, the constraints are given by the following
brackets: \ba \nl \vf_0^{tot}&=&\{Q,P_0\}_{pb}=\vf_0+2P_0\p_j\eta^j+
(\p_j P_0)\eta^j=\vf_0+\vf_0^{gh} ,
\\
\nl \vf_j^{tot}&=&\{Q,P_j\}_{pb}=\vf_j+2P_0(\p_j\eta^0)+ (\p_j
P_0)\eta^0+P_j\p_k\eta^k+ P_k(\p_j\eta^k)+(\p_k P_j)\eta^k
\\
\nl &=&\vf_j+\vf_j^{gh} \ea and they are first class. The BRST
invariant Hamiltonian is \ba \nl H_\chi=\{Q,\chi\}_{pb} \hspace{1cm},
\hspace{1cm} \{Q,H_\chi\}_{pb}=0 , \ea where $\chi$ is arbitrary,
anticommuting, gauge fixing function. We choose \ba \nl
\chi=\Lambda^0\int{d^p\sigma P_0} + \Lambda^j\int{d^p\sigma P_j} ,
\hspace{1cm} \Lambda^0 , \Lambda^j - const \ea and obtain:
\bq\label{974} H_\chi=\int{d^p\sigma\bigl[\Lambda^0\vf_0^{tot}+
\Lambda^j\vf_j^{tot}\bigr]} . \eq

Let us note that additional set of canonically conjugated ghosts
$(\bar{\eta_0},\bar{P^0}) , (\bar{\eta_j},\bar{P^j})$ must be
added if we wish to write down the corresponding BRST invariant
Lagrangian. If so, $Q$ and $\chi$ have to be modified as follows
\ba \nl \tilde {Q}=Q+\int d^p\sigma(M_0\bar{P^0}+M_j\bar{P^j}) ,
\ea \ba \nl \tilde {\chi}=\chi+\int
d^p\sigma\left[\bar{\eta_0}(\chi^0+ \frac{\rho_0}{2}M^0)+
\bar{\eta_j}(\chi^j+\frac{\rho_{(j)}}{2}M^j)\right] , \ea where
$M_0 , M_j$ are the momenta, canonically conjugated to $\mu^0$ and
$\mu^j$ respectively, $\chi^0$ and $\chi^j$ are gauge fixing
conditions \cite{M81,M82} for $\vf_0$ and $\vf_j$, $\rho_0$ and
$\rho_{(j)}$ are parameters. All the above results in the
Lagrangian density $(\p_\tau=\p/\p\tau)$: \ba \nl
L_{\tilde{\chi}}=L+L_{GF}+L_{GH} , \ea where \ba \nl
L=(1/4\mu^0)(\p_\tau x-\mu^j\p_j x)^2 , \ea the gauge fixing part
is \ba \nl L_{GF}=\frac{1}{2\rho_0}(\p_\tau \mu^0-\chi^0)(\p_\tau
\mu_0-\chi_0)+ \frac{1}{2\rho_{(j)}}(\p_\tau \mu^j-\chi^j)
(\p_\tau \mu_j-\chi_j) \ea and the ghost part is \ba \nl
L_{GH}=-\p_\tau\bar{\eta_0}\p_\tau\eta^0-\p_\tau\bar{\eta_j}
\p_\tau\eta^j+\mu^0[2\pu\bar{\eta_0}\p_j\eta^j+
(\p_j\pu\bar{\eta_0})\eta^j]
\\
\nl +\mu^j[2\pu\bar{\eta_0}\p_j\eta^0+
(\p_j\pu\bar{\eta_0})\eta^0+\pu\bar{\eta_k}\p_j\eta^k+
\pu\bar{\eta_j}\p_k\eta^k+ (\p_k\pu\bar{\eta_j})\eta^k]
\\
\nl +\int
d^p\sigma'\{\bar{\eta_0}(\us')[\{\vf_0,\chi^0(\us')\}_{pb}\eta^0
+\{\vf_j,\chi^0(\us')\}_{pb}\eta^j]
\\
\nl +\bar{\eta_j}(\us')[\{\vf_0,\chi^j(\us')\}_{pb}\eta^0
+\{\vf_k,\chi^j(\us')\}_{pb}\eta^k]\} . \ea

Let us now go back to the Hamiltonian picture. The Hamiltonian
(\ref{974}) leads to equations of motion with the following
general solution for the bosonic variables \ba \nl
x^\nu&=&y^\nu(\ul{z})+2g(\tau)p^\nu(\ul{z}),\h p_\nu=p_\nu(\ul{z})
, \ea and for the ghosts \cite{B96} \ba\nl
\eta^0&=&\zeta^0(\ul{z})+g(\tau)\p_j\eta^j(\ul{z}),\\ \label{975}
P_0&=&P_0(\ul{z}),\h \eta^j=\eta^j(\ul{z}),\\ \nl
P_j&=&\Pi_j(\ul{z})+g(\tau)\p_j P_0(\ul{z}) . \ea Here $y^\nu ,
p_\nu , \zeta^0 , P_0 , \eta^j$ and $\Pi_j$ are arbitrary
functions of the variables $z^j$ , \ba \nl z^j=\Lambda^j\tau +
\sigma^j \hspace{1cm} \mbox{and} \hspace{1cm}
g(\tau)=\Lambda^0\tau . \ea

On the solutions (\ref{975}) the BRST charge $Q$ takes the form
\cite{B96} \ba \nl Q^S=\int{d^pz\{\phi_0\zeta^0+\phi_j\eta^j+
P_0[(\p_j\eta^j)\zeta^0+(\p_j\zeta^0)\eta^j]+
\Pi_k(\p_j\eta^k)\eta^j\}} , \ea where $\phi_0=p^2(\ul{z})$ ,
$\phi_j=p_\nu(\ul{z})\p_j y^\nu(\ul{z})$. Now the constraints are \ba
\nl \phi_0^{tot}(\ul{z})=\{Q^S,P_0(\ul{z})\}_{pb} \h, \h
\phi_j^{tot}(\ul{z})=\{Q^S,\Pi_j(\ul{z})\}_{pb} , \ea and they are
connected with $\vf_0^{tot} , \vf_j^{tot}$ by the equalities \ba \nl
\vf_0^{tot}(\ul{z})=\phi_0^{tot}(\ul{z}) \h, \h
\vf_j^{tot}=\phi_j^{tot}(\ul{z})+ g(\tau)\p_j\phi_0^{tot}(\ul{z}) .
\ea

From now on, we confine ourselves to the case of periodic boundary
conditions when our phase-space variables admit Fourier series
expansions. Let us denote the Fourier components of $y^\nu, p^\nu,
\zeta^0, P_0, \eta^j$ and $\Pi_j$ with $x^{\nu}_{\ul k},
p^{\nu}_{\ul k}, c_{\ul k}, b_{\ul k}, \bar{c}^{j}_{\ul k}$ and
$\bar{b}_{j ,\ul k}$ respectively. For the zero modes of $p^\nu$
and $x^\nu$, we introduce the notations \ba \nl
P^\mu=(2\pi)^pp^{\mu}_{\ul 0} \h , \h
q^\nu=\frac{-i}{(2\pi)^p}x^{\nu}_{\ul 0} . \ea Then we have the
following non-zero Poisson brackets: \ba\nl
\{P^\mu,q^\nu\}_{pb}&=&-\eta^{\mu\nu}, \h \{p^{\mu}_{\ul
k},x^{\nu}_{\ul n}\}_{pb}= -i\eta^{\mu\nu}\delta_{\ul k +\ul n,\ul
0},\\ \label{976} \{c_{\ul k},b_{\ul n}\}_{pb}&=&-i\delta_{\ul k
+\ul n,\ul 0},\h \{\bar{c}^{j}_{\ul k},\bar{b}_{k,\ul n}\}_{pb}=
-i\delta^{j}_{k}\delta_{\ul k +\ul n,\ul 0} . \ea

The Fourier expansions for the constraints $\phi_0^{tot}$ and
$\phi_j^{tot}$ are \ba \nl
\phi_0^{tot}(\ul{z})=\frac{1}{(2\pi)^p}\sum_{\ul{m}\in Z^p}
C_{\ul{m}}^{tot}e^{-i\ul{m}\ul{z}}, \h
\phi_j^{tot}(\ul{z})=\frac{1}{(2\pi)^p}\sum_{\ul{m}\in Z^p}
D_{j,\ul m}^{tot}e^{-i\ul{m}\ul{z}} . \ea Here \bq\label{977}
C_{\ul{n}}^{tot}=i\{Q^S,b_{\ul{n}}\}_{pb}=
C_{\ul{n}}+C_{\ul{n}}^{gh}, \h D_{j,\ul
n}^{tot}=i\{Q^S,\bar{b}_{j,\ul n}\}_{pb}= D_{j,\ul n}+D_{j,\ul
n}^{gh}, \eq where \ba \nl Q^S&=&\sum_{\ul{n}\in
Z^p}\{[C_{\ul{n}}+(1/2)C_{\ul{n}}^{gh}]c_{-\ul{n}}+ [D_{j,\ul
n}+(1/2)D_{j,\ul n}^{gh}]\bar{c}_{-\ul{n}}^j\} ,
\\
\nl
C_{\ul n}&=&(2\pi)^p\sum_{\ul k\in Z^p}p^{\nu}_{\ul k}p_{\nu,\ul n-\ul k},
\\ \label{978}
D_{j,\ul n}&=&-\sum_{\ul k\in Z^p}(n_j-k_j)p^{\nu}_{\ul k} x_{\nu,\ul
n-\ul k} ,
\\
\nl C^{gh}_{\ul n}&=&\sum_{\ul k\in Z^p}(n_j - k_j)b_{\ul n +\ul k}
\bar{c}^{j}_{-\ul k} ,
\\
\nl D^{gh}_{j,\ul n}&=&\sum_{\ul k\in Z^P}[(n_j - k_j) b_{\ul n +\ul
k}c_{-\ul k} + (\delta_{j}^{l}n_k - \delta_{k}^{l}k_j)\bar{b}_{l,\ul
n +\ul k} \bar{c}^{k}_{-\ul k}] \ea

Using expressions (\ref{976})-(\ref{978}), one obtains the algebra
of the total generators (\ref{977}) \ba \nl
\{C_{\ul{n}}^{tot},C_{\ul{m}}^{tot}\}_{pb}&=&0 ,
\\ \label{979} \{C_{\ul{n}}^{tot},D_{j,\ul m}^{tot}\}_{pb}&=
&-i(n_{j}-m_{j})C_{\ul{n}+\ul{m}}^{tot} ,
\\ \nl
\{D_{j,\ul n}^{tot} , D_{k,\ul m}^{tot}\}_{pb}&=
&-i(\delta^{l}_{j}n_{k}- \delta^{l}_{k}m_{j})D_{l,\ul n +\ul m}^{tot}
. \ea

\subsection{\bf Quantization}
\hspace{1cm} Going to the quantum theory according to the rule
$i\{.,.\}_{pb}$ $\rightarrow$ (anti)commutator, we define $Q^S$ by
introducing the renormalized operators ($\alpha,\beta_j$ are
constants) \bq\label{9710}
C_{\ul{n}}^{tot}=C_{\ul{n}}+C_{\ul{n}}^{gh}-\alpha\delta_{\ul{n},\ul{0}}
\h , \h D_{j,\ul{n}}^{tot}=D_{j,\ul{n}}+D_{j,\ul{n}}^{gh}-
\beta_j\delta_{\ul{n},\ul{0}} \eq and postulating \cite{B96} \ba
\nl Q^S=\sum_{\ul{n}\in Z^p}:\{[C_{\ul{n}}+(1/2)C_{\ul{n}}^{gh}
-\alpha\delta_{\ul{n},\ul{0}}]c_{-\ul{n}}
\\
\nl +[D_{j,\ul{n}}+(1/2)D_{j,\ul{n}}^{gh}-
\beta_j\delta_{\ul{n},\ul{0}}]\bar{c}_{-\ul{n}}^j\}: , \ea where
:...: represents operator ordering and operator ordering in
$C_{\ul{n}},...,D_{j,\ul{n}}^{gh}$ is also assumed.

Let us turn to the question about the critical dimensions which might
appear in the model under consideration. As is well known, the critical
dimension arises as a necessary condition for nilpotency of the BRST
charge operator. In turn, this is connected with the vanishing of the
central charges in the quantum constraint algebra. Because of that,
we are going to find out the central terms which appear in our
quantum gauge algebra for different values of $p$ (the most general form
of central extension, which is compatible with the Jacobi identities is
written in the Appendix).

We start with the case $p=1$, which corresponds to a closed
string. In this case $j=k=1$ and one defines the operator ordering
with respect to $p_{-n}^\nu,...,\bar{c}_{-n}$ and
$p_n^\nu,...,\bar{c}_n , (n>0)$, so that \ba \nl p_{-n}^{\nu}\mid
0> = ... =\bar{c}_{-n}\mid 0> = 0 , \h <0\mid p_{n}^{\nu} = ...
=<0\mid \bar{c}_{n} = 0. \ea We call this ordering
"$string-like$". Using the explicit expressions for the
constraints (\ref{978}), one obtains that central terms appear in
the commutators $[D_n,D_m]$, $[D_n^{gh},D_m^{gh}]$ and they are
respectively \ba \nl c=(D/6)(n^2-1)n\delta_{n+m,0} \h , \h
c^{gh}=-(1/3)(13n^2-1)n\delta_{n+m,0} . \ea Therefore, the quantum
constraint algebra has the form \ba \nl [C_n^{tot},C_m^{tot}]&=&0
,
\\
\nl
[C_n^{tot},D_m^{tot}]&=&(n-m)C_{n+m}^{tot}+2\alpha n\delta_{n+m,0} ,
\\
\nl
[D_n^{tot},D_m^{tot}]&=&(n-m)D_{n+m}^{tot}+
(1/6)[(D-26)n^2+(12\beta-D+2)]n\delta_{n+m,0} .
\ea
This means that the conditions for the nilpotency of the BRST charge
operator $Q^S$ are
\ba
\nl
(D-26)n^2+(12\beta-D+2)=0
\h ,
\h
\alpha=0 ,
\ea
which leads to the well known result $D=26,\beta=2$. Obviously,
this reproduces one of the basic features of the quantized tensionful closed
bosonic string - its critical dimension.

Going to the case $p>1$, one natural generalization of the creation
and annihilation operators definition is \ba \nl p_{\ul{n}}^\nu\mid
0>_j=0 , \h _j<0\mid p_{-\ul{n}}^\nu=0 , \h \mbox{for} \h
\sum_{j=1}^{p}n_{a} > 0 \ea and analogously for the operators
$x_{\ul{n}}^\nu,...,\bar{c}_{\ul{n}}^j$. However, it turns out that
such definition does not agree with the Jacobi identities for the
quantum constraint algebra (except for $p=1$). That is why, we
introduce the creation $(+)$ and annihilation $(-)$ operators as
follows \cite{B96} \bq\label{9711}
p_{\ul{n}}^\nu=(1/\sqrt{2})(p_{\ul{n}}^{\nu+}+p_{-\ul{n}}^{\nu-}) ,
...,\bar{c}_{\ul{n}}^j=(1/\sqrt{2})(\bar{c}_{\ul{n}}^{j+}+
\bar{c}_{-\ul{n}}^{j-}) \eq and respectively new vacuum states \ba
\nl p_{\ul n}^{\nu-}\mid vac> = ... = \bar{c}_{\ul n}^{j-}\mid vac> =
0 \hspace{.5cm}, \hspace{.5cm} <vac\mid p_{\ul n}^{\nu+} = ... =
<vac\mid \bar{c}_{\ul n}^{j+} = 0 . \ea This choice of the creation
and annihilation operators corresponds to the representation of all
phase-space variables $p^\nu,...,\bar{c}^j$ as sums of frequency
parts which are conjugated to each other and satisfy the same
equation of motion as the corresponding dynamical variable.

By direct computation one shows that with operator product defined
with respect to the introduced creation and annihilation operators
(\ref{9711}) (we shall refer to as "$normal$ $ordering$"), a
central extension of the algebra of the gauge generators
(\ref{9710}) does not appear, i.e. $\alpha=0 , \beta_j=0$.
Consequently, the BRST charge operator $Q^S$ is automatically
nilpotent in this case and there is no restriction on the
dimension of the background space-time for $p>1$.

The impossibility to introduce a $string-like$ operator ordering
when $p>1$ leads to the problem of finding those operator
orderings which are possible for $p=1$ as well as for $p>1$. First
of all, we check the consistency of the (already used for $p>1$)
$normal$ $ordering$ for $p=1$. It turns out to be consistent, but
now critical dimension for the null string does not appear. The
same result - absence of critical dimension for every value of
$p$, one obtains when uses the so called $particle-like$ operator
ordering. Now the $ket$ vacuum is annihilated by momentum-type
operators and the $bra$ vacuum is annihilated by coordinate-type
ones: \ba \nl p_{\ul n}^{\mu}\mid 0>_M &=& b_{\ul n}\mid 0>_M =
\bar{b}_{\ul n}\mid 0>_M = 0 ,
\\
\nl
_C<0\mid x_{\ul n}^{\mu} &=& _C<0\mid c_{\ul n}
= _C<0\mid \bar{c}_{\ul n} = 0
\hspace{.5cm} ,
\hspace{.5cm}
\forall \ul n \in Z^p .
\ea
Further, we check the case when $Weyl$ $ordering$ is applied. Now it turns out,
that in the null string case ($p=1$) this leads to critical dimension $D=26$,
but for the null brane ($p>1$) this ordering is inconsistent, as was the
$string-like$ one.

As a final result, we checked four types of operator orderings.
Two of them are valid for the string as well as for the brane and
then we do not receive any critical dimension. The other two types
of orderings give critical dimension $D=26$ for the string and are
not applicable for the brane. Our opinion is that the right
operator ordering is the one applicable for all $p=1,2,...$. In
other words, our viewpoint is that neither null strings nor null
branes have critical dimensions. The same point of view is
presented in \cite{PS97}.

Let us discuss in more details the impossibility to introduce at
$p>1$ an operator ordering which at $p=1$ gives critical dimension.
This is connected with the fact that the constraint algebra, as is
shown in the Appendix A, does not possess non-trivial central
extension when $p>1$ (see also \cite{SS90}, \cite{PS97}). As a matter
of fact, the string critical dimension appears in front of $n^3$,
i.e. in the non-trivial part of the constraint algebra central
extension, which can not be taken away by simply redefining the
generators $D_n$, in contrast to the trivial part $\sim n$. Because
of the nonexistence of non-trivial central extension when $p>1$, any
critical dimension arising is impossible in view of the Jacobi
identities. Therefore, if the quantum null brane constraint algebra
is given by (up to trivial central extensions) \ba \nl
[C_{\ul{n}}^{tot},C_{\ul{m}}^{tot}]&=&0 ,
\\
\nl [C_{\ul{n}}^{tot},D_{j,\ul m}^{tot}]&=&
(n_{j}-m_{j})C_{\ul{n}+\ul{m}}^{tot} ,
\\
\nl [D_{j,\ul n}^{tot} , D_{k,\ul m}^{tot}]&=&
(\delta^{l}_{j}n_{k}- \delta^{l}_{k}m_{j})D_{l,\ul n +\ul m}^{tot}
, \ea then the latter has no critical dimension and exists in any
D-dimensional space-time, when embedding of the $p+1$- dimensional
world-volume of the $p$-brane is possible.

Finally, note that in each of the $p$ subalgebras (at fixed $j$) of
the constraint algebra, one can obtain non-trivial central extension
and consequently - critical dimension (see Appendix A). For example,
taking $string-like$ or $Weyl$ $ordering$, one derives $D=25+p$. The
same critical dimensions are obtained also in \cite{PS97}. However,
the considered quantum dynamical system is described by the ${\it
full}$ constraint algebra, where only trivial central extensions are
possible.

\subsection{\bf Comments}
\hspace{1cm}The observation, that there is an operator ordering
which is valid $\forall p\in Z_{+}$ and another one, which is
admissible only for $p=1$ \cite{B96}, leads to the problem of
finding those orderings which are possible for every positive
integer value of $p$. We applied here four types of operator
orderings and we establish that two of them ($normal$ and
$particle-like$ $orderings$) are admissible $\forall p\in Z_{+}$,
but the other two ($string-like$ and $Weyl$ $orderings$) are
admissible only for $p=1$. The fact that the latter two orderings
lead to the appearance of critical dimension, and the former two
do not, is a consequence of the constraint algebra property to
have non-trivial central extension only for $p=1$. On the other
hand, the obtained nontrivial central extensions of the Virasoro
type for some of it subalgebras, provide an explanation why the
critical dimensions $D=25+p, p=1,2,...$ \cite{PS97}, re-derived
here, can emerge. However, our claim is that the critical
dimensions appearing in the subalgebras, must not be considered as
such for the given model as a whole. The model is represented by
the full constraint algebra, which does not possess non-trivial
central extension for $p\geq 2$. This leads us to the proposition
of the rule: the right operator orderings in the case of null
string ($p=1$) are those which are also admissible in the $p>1$
case.

\newpage
\vspace*{.5cm}
\section{\bf NULL BRANES IN CURVED\\ BACKGROUNDS}
\hspace{1cm} In this section, we consider null bosonic $p$-branes
moving in curved space-times. Some exact solutions of the
classical equations of motion and of the constraints for the null
membrane in general stationary, axially symmetric, four
dimensional gravity background are found. The results considered
here are obtained in \cite{B993}.

\subsection{\bf Classical formulation}
\hspace{1cm} The action for the bosonic null $p$-brane in a
$D$-dimensional curved space-time with metric tensor
$g_{\mu\nu}(x)$ can be written in the form: \ba\label{cba} S=\int
d^{p+1}\xi L, \h L=V^JV^K\pJ x^\mu\pK x^\nu g_{\mu\nu}(x),
\\ \nl
\pJ=\p/\p\xi^J, \h \xi^J=(\xi^0,\xi^j)=(\tau,\s^j),
\\ \nl
J,K=0,1,...,p \h,\h j,k=1,...,p \h,\h \mu,\nu=0,1,...,D-1. \ea It is
an obvious generalization of the flat space-time action given in
\cite{HLU94}.

To prove the invariance of the action under infinitesimal
diffeomorphisms on the world-volume (reparametrizations), we first
write down the corresponding transformation law for the (r,s)-type
tensor density of weight $a$ \ba\nl
\delta_{\varepsilon}T^{J_1...J_r}_{K_1...K_s}[a]&=&
L_{\varepsilon}T^{J_1...J_r}_{K_1...K_s}[a]= \varepsilon^L\p_L
T^{J_1...J_r}_{K_1...K_s}[a]\\ \label{cbdiff} &+&
T^{J_1...J_r}_{KK_2...K_s}[a]\p_{K_1}\varepsilon^K+...+
T^{J_1...J_r}_{K_1...K_{s-1}K}[a]\p_{K_s}\varepsilon^K \\ \nl &-&
T^{JJ_2...J_r}_{K_1...K_s}[a]\p_J\varepsilon^{J_1}-...-
T^{J_1...J_{r-1}J}_{K_1...K_s}[a]\p_J\varepsilon^{J_r} \\ \nl &+&
aT^{J_1...J_r}_{K_1...K_s}[a]\p_L\varepsilon^L , \ea where
$L_\varepsilon$ is the Lie derivative along the vector field
$\varepsilon$. Using (\ref{cbdiff}), one verifies that if
$x^\mu(\xi)$, $g_{\mu\nu}(\xi)$ are world-volume scalars ($a=0$)
and $V^J(\xi)$ is a world-volume (1,0)-type tensor density of
weight $a=1/2$, then $\pJ x^\nu$ is a (0,1)-type tensor, $\pJ
x^\mu \pK x^\nu g_{\mu \nu}$ is a (0,2)-type tensor and $L$ is a
scalar density of weight $a=1$. Therefore, \ba\nl
\delta_{\varepsilon}S=\int d^{p+1}\xi\p_J\bigl ( \varepsilon^J
L\bigr ) \ea and this variation vanishes under suitable boundary
conditions.

The equations of motion following from (\ref{cba}) are: \ba\nl
\pJ\Bigl (V^J V^K \pK x^{\lambda}\Bigr ) + \Gamma^{\lambda}_{\mu\nu}
V^J V^K \pJ x^{\mu}\pK x^{\nu} = 0 ,\\ \nl V^J \pJ x^{\mu}\pK x^{\nu}
g_{\mu\nu}(x) = 0 , \ea where $\Gamma^{\lambda}_{\mu\nu}$ is the
connection compatible with the metric $g_{\mu\nu}(x)$: \ba\nl
\Gamma^{\lambda}_{\mu\nu}=\frac{1}{2}g^{\lambda\rho}\bigl(
\p_{\mu}g_{\nu\rho}+\p_{\nu}g_{\mu\rho}-\p_{\rho}g_{\mu\nu}\bigr) .
\ea

For the transition to Hamiltonian picture it is convenient to
rewrite the Lagrangian density (\ref{cba}) in the form
($\pu=\p/\p\tau, \pj=\p/\p\s^j$): \ba\label{cbL}
L=\frac{1}{4\mu^0} g_{\mu\nu}(x)\bigl (\pu-\mu^j\pj\bigr )x^\mu
\bigl (\pu-\mu^k\pk\bigr )x^\nu , \ea where \ba\nl
V^J=\bigl(V^0,V^j\bigr)=\Biggl(-\frac{1}{2\sqrt{\mu^0}},
\frac{\mu^j}{2\sqrt{\mu^0}}\Biggr) . \ea Now the equation of
motion for $x^\nu$ takes the form: \ba\label{cbeqx} \pu\Bigl
[\frac{1}{2\mu^0}\bigl (\pu-\mu^k\pk\bigr )x^{\lambda}\Bigr ]
-\pj\Bigl [\frac{\mu^{j}}{2\mu^{0}}\bigl (\pu-\mu^k\pk\bigr
)x^{\lambda} \Bigr ] \\ \nl +
\frac{1}{2\mu^0}\Gamma^{\lambda}_{\mu\nu} \bigl (\pu-\mu^j\pj\bigr
)x^\mu \bigl (\pu-\mu^k\pk\bigr )x^\nu = 0 . \ea The equations of
motion for the Lagrange multipliers $\mu^{0}$ and $\mu^{j}$ which
follow from (\ref{cbL}) give the constraints : \ba\label{cbTx1}
g_{\mu\nu}(x)\bigl (\pu-\mu^j\pj\bigr )x^\mu \bigl
(\pu-\mu^k\pk\bigr )x^\nu = 0 , \\ \nl g_{\mu\nu}(x)\bigl
(\pu-\mu^k\pk\bigr )x^\mu \pj x^\nu = 0 . \ea In terms of $x^\nu$
and the conjugated momentum $p_\nu$ they read: \ba\label{cbTpx}
T_0=g^{\mu\nu}(x)p_\mu p_\nu = 0 \h,\h T_j=p_\nu\pj x^\nu = 0 .
\ea

The Hamiltonian which corresponds to the Lagrangian density
(\ref{cbL}) is a linear combination of the constraints
(\ref{cbTpx}) : \ba\nl H_0=\int d^p\sigma\bigl (\mu^0 T_0+\mu^j
T_j \bigr ) . \ea They satisfy the following (equal $\tau$)
Poisson bracket algebra \ba \nl \{T_0(\ul \sigma_1),T_0(\ul
\sigma_2)\}&=&0,
\\ \label {CA}
\{T_0(\ul \sigma_1),T_{j}(\ul \sigma_2)\}&=& [T_0(\ul \sigma_1) +
T_0(\ul \sigma_2)] \p_j \delta^p (\ul \sigma_1 - \ul \sigma_2) ,
\\ \nl
\{T_{j}(\ul \sigma_1),T_{k}(\ul \sigma_2)\}&=&
[\delta_{j}^{l}T_{k}(\ul \sigma_1) + \delta_{k}^{l}T_{j}(\ul
\sigma_2)]\p_l\delta^p(\ul \sigma_1-\ul \sigma_2) ,\\ \nl
\us&=&(\s^1,...,\s^p) . \ea The equalities (\ref{CA}) show that
the constraint algebra is the same for flat and for curved
backgrounds. Consequently, one can apply the BFV-BRST approach in
exactly the same way as in the previous section.

\subsection{\bf Null membranes in D=4}
\hspace{1cm} Here we confine ourselves to the case of membranes
moving in a four dimensional, stationary, axially symmetric gravity
background of the type \ba\label{gm}
ds^2&=&g_{00}(dx^0)^2+g_{11}(dx^1)^2+g_{22}(dx^2)^2+
g_{33}(dx^3)^2+2g_{03}dx^0 dx^3 , \\ \nl
g_{\mu\nu}&=&g_{\mu\nu}(x^1,x^2) . \ea We will work in the gauge
$\mu^0, \mu^j = constants $, in which the equations of motion
(\ref{cbeqx}) and constraints (\ref{cbTx1}) for the membrane
$(j,k=1,2)$ have the form: \ba\label{eqxf}
\bigl(\pu-\mu^j\pj\bigr)\bigl (\pu-\mu^k\pk\bigr )x^{\lambda} +
\Gamma^{\lambda}_{\mu\nu} \bigl (\pu-\mu^j\pj\bigr )x^\mu \bigl
(\pu-\mu^k\pk\bigr )x^\nu = 0 .
\\ \label{cf1}
g_{\mu\nu}(x)\bigl (\pu-\mu^j\pj\bigr )x^\mu \bigl
(\pu-\mu^k\pk\bigr )x^\nu = 0 , \\ \nl g_{\mu\nu}(x)\bigl
(\pu-\mu^k\pk\bigr )x^\mu \pj x^\nu = 0 . \ea To establish the
correspondence with the null geodesics we note that if we
introduce the quantities \ba\label{unu}
u^{\nu}(x)=\bigl(\pu-\mu^{j}\pj\bigr)x^{\nu} , \ea the equations
of motion (\ref{eqxf}) can be rewritten as \ba\nl
u^{\nu}\bigl(\p_{\nu}u^{\lambda}+\Gamma^{\lambda}_{\mu\nu}u^{\mu}\bigr)=0
. \ea Then it follows from here that $u^2$ does not depend on
$x^{\nu}$. In this notations, the constraints are: \ba\nl
g_{\mu\nu}u^{\mu}u^{\nu} = 0 \h,\h g_{\mu\nu}u^{\mu}\pj x^{\nu} =
0 . \ea

Taking into account the metric (\ref{gm}), one can write the
equations of motion (\ref{eqxf}) and the constraint (\ref{cf1}) in
the form: \ba\nl \Dj
u^0+2\Bigl(\Gamma^0_{01}u^0+\Gamma^0_{13}u^3\Bigr)\Dj x^1 +
\\ \nl
+2\Bigl(\Gamma^0_{02}u^0+\Gamma^0_{23}u^3\Bigr)\Dj x^2 = 0 ,\\ \nl
\Dj\Dk x^1+\Gamma^1_{11}\Dj x^1\Dk x^1+
\\ \nl
+2\Gamma^1_{12}\Dj x^1\Dk x^2 +\Gamma^1_{22}\Dj x^2\Dk x^2+
\\ \nl
+\Gamma^1_{00}(u^0)^2+ 2\Gamma^1_{03}u^0 u^3+\Gamma^1_{33}(u^3)^2 = 0
,
\\ \label{l2}
\Dj\Dk x^2+\Gamma^2_{11}\Dj x^1\Dk x^1+
\\ \nl
+2\Gamma^2_{12}\Dj x^1\Dk x^2+\Gamma^2_{22}\Dj x^2\Dk x^2+
\\ \nl
+\Gamma^2_{00}(u^0)^2+2\Gamma^2_{03}u^0 u^3+\Gamma^2_{33}(u^3)^2 = 0
,
\\ \nl
\Dj u^3+2\Bigl(\Gamma^3_{01}u^0+\Gamma^3_{13}u^3\Bigr)\Dj x^1 +
\\ \nl
+2\Bigl(\Gamma^3_{02}u^0+\Gamma^3_{23}u^3\Bigr)\Dj x^2 = 0 ,\\ \nl
g_{11}\Dj x^1\Dk x^1+g_{22}\Dj x^2\Dk x^2+
\\ \nl
+g_{00}(u^0)^2+2g_{03}u^0 u^3+g_{33}(u^3)^2 = 0 ,\\ \nl g_{11}\Dk
x^1\pj x^1+g_{22}\Dk x^2\pj x^2 +
\\ \nl
+\bigl(g_{00}\pj x^0+g_{03}\pj x^3\bigr)u^0 + \bigl(g_{03}\pj
x^0+g_{33}\pj x^3\bigr)u^3 = 0 , \ea where the notation introduced
in (\ref{unu}) is used. To simplify these equations, we use the
ansatz \ba\nl x^0(\tau,\us)&=&f^0(z^1,z^2)+t(\tau) ,
\\ \label{az}
x^1(\tau,\us)&=&r(\tau) \h,\h x^2(\tau,\us)=\theta(\tau) ,
\\ \nl
x^3(\tau,\us)&=&f^3(z^1,z^2)+\varphi(\tau) ,
\\ \nl
z^j&=&\mu^j \tau + \sigma^j , \ea where $f^0, f^3$ are arbitrary
functions of their arguments.

After substituting (\ref{az}) in (\ref{l2}), we receive (the dot is
used for differentiation with respect to the affine parameter
$\tau$): \ba\label{l01} \dot u^0+\Bigl[\left(g^{00}\p_1
g_{00}+g^{03}\p_1 g_{03}\right)u^0 +\left(g^{00}\p_1
g_{03}+g^{03}\p_1 g_{33}\right)u^3\Bigr]\dot r
\\ \nl
+\Bigl[\left(g^{00}\p_2 g_{00}+g^{03}\p_2 g_{03}\right)u^0
+\left(g^{00}\p_2 g_{03}+g^{03}\p_2 g_{33}\right)u^3\Bigr]\dot\theta
= 0 ,
\\ \label{l11}
2g_{11}\ddot r+\p_1 g_{11}\dot r^2+2\p_2 g_{11}\dot r\dot\theta -\p_1
g_{22}\dot\theta^2
\\ \nl
-\bigl[\p_1 g_{00}(u^0)^2+2\p_1 g_{03}u^0 u^3 +\p_1
g_{33}(u^3)^2\bigr]=0 ,
\\ \label{l21}
2g_{22}\ddot\theta+\p_2 g_{22}\dot\theta^2+2\p_1 g_{22}\dot
r\dot\theta-\p_2 g_{11}\dot r^2
\\ \nl
-\bigl[\p_2 g_{00}(u^0)^2+2\p_2 g_{03}u^0 u^3 +\p_2
g_{33}(u^3)^2\bigr]=0 ,
\\ \label{l31}
\dot u^3+\Bigl[\left(g^{33}\p_1 g_{03}+g^{03}\p_1 g_{00}\right)u^0
+\left(g^{33}\p_1 g_{33}+g^{03}\p_1 g_{03}\right)u^3\Bigr]\dot r
\\ \nl
+\Bigl[\left(g^{33}\p_2 g_{03}+g^{03}\p_2 g_{00}\right)u^0
+\left(g^{33}\p_2 g_{33}+g^{03}\p_2 g_{03}\right)u^3\Bigr]\dot\theta
= 0 ,
\\ \label{c11}
g_{11}\dot r^2+g_{22}\dot\theta^2+g_{00}(u^0)^2 +2g_{03}u^0
u^3+g_{33}(u^3)^2 = 0 ,
\\ \label{c21}
\bigl(g_{00}\pj f^0+g_{03}\pj f^3\bigr)u^0 +\bigl(g_{03}\pj
f^0+g_{33}\pj f^3\bigr)u^3 = 0 . \ea If we choose \ba\nl
f^0(z^1,z^2)=f^0(w) \h,\h f^3(z^1,z^2)=f^3(w) , \ea where
$w=w(z^1,z^2)$ is an arbitrary function of $z^1$ and $z^2$, then the
system of equations (\ref{c21}) reduces to the single equation
\ba\label{c22}
\Bigl(g_{00}\frac{df^0}{dw}+g_{03}\frac{df^3}{dw}\Bigr)u^0
+\Bigl(g_{03}\frac{df^0}{dw}+g_{33}\frac{df^3}{dw}\Bigr)u^3=0 . \ea

To be able to separate the variables $u^0, u^3$ in the system of
differential equations (\ref{l01}), (\ref{l31}) with the help of
(\ref{c22}), we impose the following condition on $f^0(w)$ and
$f^3(w)$ \ba\nl f^0(w) = C^0 f[w(z^1,z^2)] \h,\h f^3(w) = C^3
f[w(z^1,z^2)] , \ea where $C^0, C^3$ are constants, and $f(w)$ is an
arbitrary function of $w$. Then the solution of (\ref{l01}),
(\ref{l31}) and (\ref{c22}) is \cite{PP97} ($C_1 = const $): \ba\nl
u^0(\tau)=-C_1\left(C^0 g_{03}+C^3 g_{33}\right)\exp(-H) ,
\\ \label{s032}
u^3(\tau)=+C_1\left(C^0 g_{00}+C^3 g_{03}\right)\exp(-H) ,
\\ \nl
H = \int\Bigl(g^{00}dg_{00}+2g^{03}dg_{03}+g^{33}dg_{33}\Bigr) . \ea
The condition for the compatibility of (\ref{s032}) with (\ref{l11}),
(\ref{l21}) and (\ref{c11}) is: \ba\nl u^0(\tau)&=&-C_1\left(C^0
g_{03}+C^3 g_{33}\right)h^{-1}
\\ \nl
&=&-C_1\left(C^3 g^{00}-C^0 g^{03}\right)=\dot t(\tau) ,
\\ \label{su3}
u^3(\tau)&=&+C_1\left(C^0 g_{00}+C^3 g_{03}\right)h^{-1}
\\ \nl
&=&-C_1\left(C^3 g^{03}-C^0 g^{33}\right)=\dot\varphi (\tau) ,
\\ \nl
h&=&g_{00}g_{33}-g_{03}^2 . \ea

On the other hand, from (\ref{l21}) and (\ref{c11}) one has:
\ba\label{r.2} \dot
r^2&=&-g^{11}\Bigl[C_1^2\frac{G}{h}+g^{22}\left(g_{22}^2
\dot\theta^2\right)\Bigr]
\\ \nl
&=&g^{11}\Bigl\{C_1^2\bigl[2C^0 C^3 g^{03}-(C^3)^2 g^{00}-(C^0)^2
g^{33}\bigr]-g^{22}\bigl(g_{22}^2\dot\theta^2\bigr)\Bigr\} ,
\\ \label{t.2}
g_{22}^2\dot\theta^2&=&C_2+C_1^2\int\limits^{\theta}d\theta h^{-2}
\Bigl[g_{22}G\frac{\p h}{\p\theta}-h\frac{\p
g_{22}G}{\p\theta}\Bigr],
\\ \nl
G&=&(C^0)^2 g_{00}+2C^0 C^3 g_{03}+(C^3)^2 g_{33}. \ea In obtaining
(\ref{t.2}), we have used the gauge freedom in the metric (\ref{gm}),
to impose the condition \cite{C83}: \ba\nl
\p_2\Biggl(\frac{g_{22}}{g_{11}}\Biggr) = 0 . \ea

As a final result we have \ba\nl x^0 &=& C^0 f[w(z^1,z^2)] + t(\tau),
\\ \nl x^1 &=& r(\tau), \\ \nl x^2 &=& \theta (\tau), \\ \nl x^3 &=&
C^3 f[w(z^1,z^2)] + \varphi (\tau) , \ea where $\dot t(\tau), \dot
r(\tau), \dot\theta (\tau), \dot\varphi (\tau)$ are given by
(\ref{su3}), (\ref{r.2}), and (\ref{t.2}).

In the particular case when $x^2=\theta=\theta_0=const$, one can
integrate to obtain the following exact solution of the equations of
motion and constraints for the null membrane in the gravity
background (\ref{gm}): \ba\nl x^0 (\tau,\s^1,\s^2)&=&C^0
f[w(z^1,z^2)] + t_0 \\ \nl &\pm&\int\limits_{r_0}^{r}dr\left(C^3
g^{00}-C^0 g^{03}\right)W^{-1/2} ,
\\ \label{GSC}
x^3 (\tau,\s^1,\s^2)&=&C^3 f[w(z^1,z^2)] + \varphi_{0}
\\ \nl
&\pm&\int\limits_{r_0}^{r}dr\left(C^3 g^{03}-C^0
g^{33}\right)W^{-1/2} ,
\\ \nl
C_1 (\tau -\tau_{0})&=& \pm\int\limits_{r_0}^{r}drW^{-1/2} ,\\ \nl
W&=&g^{11}\left[2C^0 C^3 g^{03}-\left(C^3\right)^2
g^{00}-\left(C^0\right)^2 g^{33}\right] ,
\\ \nl
t_0, r_0, \varphi_{0}, \tau_{0} &-& constants. \ea

\subsection{\bf Examples}
\hspace{1cm} Here we give some examples of solutions of the type
received in the previous section. To begin with, let us start with
the simplest case of {\it Minkowski space-time}. The metric is \ba\nl
g_{00}=-1,\h g_{11}=1,\h g_{22}=r^2,\h g_{33}=r^2 \sin^2\theta , \ea
and equalities (\ref{su3}), (\ref{r.2}), (\ref{t.2}) take the form:
\ba\nl \dot t&=&C_1 C^3 ,\\ \nl \dot r^2&=&(C_1 C^3)^2 -
\frac{C_2}{r^2} ,\\ \nl r^4 \dot\theta^2&=&C_2-\frac{(C_1
C^0)^2}{\sin^2\theta} ,
\\ \nl
\dot\varphi&=&\frac{C_1 C^0}{r^2 \sin^2\theta} . \ea When
$\theta=\theta_0=const$, the solution (\ref{GSC}) is: \ba\nl x^0
(\tau,\s^1,\s^2)&=&C^0 f[w(z^1,z^2)] + t_0 \mp C^3
\int\limits_{r_0}^{r}\frac{dr} {\bigl [(C^3)^2-(C^0)^2
r^{-2}\sin^{-2}\theta_0\bigr ]^{1/2}} ,
\\ \nl
x^3 (\tau,\s^1,\s^2)&=&C^3 f[w(z^1,z^2)] + \varphi_{0}
\mp\frac{C^0}{\sin^2\theta_0}\int\limits_{r_0}^{r}\frac{dr} {r^2\bigl
[(C^3)^2-(C^0)^2 r^{-2}\sin^{-2}\theta_0\bigr ]^{1/2}} ,
\\ \nl
C_1(\tau - \tau_{0}) &=& \pm\int\limits_{r_0}^{r}\frac{dr}{\bigl
[(C^3)^2 -(C^0)^2 r^{-2}\sin^{-2}\theta_{0}\bigr ]^{1/2}} . \ea

Our next example is the {\it de Sitter space-time}. We take the
metric in the form \ba\nl g_{00}=-\left(1-kr^2\right),
g_{11}=\left(1-kr^2\right)^{-1}, g_{22}=r^2, g_{33}=r^2 \sin^2\theta
, \ea where $k$ is the constant curvature. Now we have \ba\nl \dot
t&=&\frac{C_1 C^3}{1-kr^2} ,\\ \nl \dot r^2&=&(C_1 C^3)^2 + C_2
(k-r^{-2}) ,
\\ \nl
r^4 \dot\theta^2&=&C_2-\frac{(C_1 C^0)^2}{\sin^2\theta} ,
\\ \nl
\dot\varphi&=&\frac{C_1 C^0}{r^2 \sin^2\theta} , \ea and the
corresponding solution (\ref{GSC}) is: \ba\nl x^0
(\tau,\s^1,\s^2)&=&C^0 f[w(z^1,z^2)] + t_0
\\ \nl
&\mp& C^3 \int\limits_{r_0}^{r}\frac{dr}{(1-kr^2) \bigl
[(C^3)^2+(C^0)^2 (k-r^{-2})\sin^{-2}\theta_0\bigr ]^{1/2}} ,
\\ \nl
x^3 (\tau,\s^1,\s^2)&=&C^3 f[w(z^1,z^2)] + \varphi_{0}
\\ \nl
&\mp&\frac{C^0}{\sin^2\theta_0}\int\limits_{r_0}^{r}\frac{dr}
{r^2\bigl[(C^3)^2+(C^0)^2 (k-r^{-2})\sin^{-2}\theta_0\bigr]^{1/2}} ,
\\ \nl
C_1(\tau - \tau_{0}) &=& \pm\int\limits_{r_0}^{r}\frac{dr}{\bigl
[(C^3)^2 +(C^0)^2 (k-r^{-2})\sin^{-2}\theta_0\bigr ]^{1/2}} . \ea

Now let us turn to the case of {\it Schwarzschild space-time}. The
corresponding metric may be written as \ba\nl g_{00}&=&-(1-2Mr^{-1})
\h,\h g_{11}=(1-2Mr^{-1})^{-1} ,
\\ \nl
g_{22}&=&r^2 \h,\h g_{33} = r^2 \sin^2\theta , \ea where $M$ is the
Schwarzschild mass. The equalities (\ref{su3}), (\ref{r.2}) and
(\ref{t.2}) read \ba\nl \dot t&=&\frac{C_1 C^3}{1-2Mr^{-1}} ,\\ \nl
\dot r^2&=&(C_1 C^3)^2 -\frac{C_2}{r^2}(1-2Mr^{-1}) ,
\\ \label{S}
r^4 \dot\theta^2&=&C_2-\frac{(C_1 C^0)^2}{\sin^2\theta} ,
\\ \nl
\dot\varphi&=&\frac{C_1 C^0}{r^2 \sin^2\theta} . \ea When
$\theta=\theta_0=const$, one obtains from (\ref{GSC}) \ba\nl x^0
(\tau,\s^1,\s^2)&=&C^0 f[w(z^1,z^2)] + t_0
\\ \nl
&\mp& C^3\int\limits_{r_0}^{r}\frac{dr}{(1-2Mr^{-1}) \bigl [(C^3)^2
-(C^0)^2 r^{-2}(1-2Mr^{-1})\sin^{-2}\theta_0]^{1/2}} ,
\\ \nl
x^3 (\tau,\s^1,\s^2)&=&C^3 f[w(z^1,z^2)] + \varphi_{0}
\\ \nl
&\mp&\frac{C^0}{\sin^2\theta_0}\int\limits_{r_0}^{r}\frac{dr}
{r^2\bigl[(C^3)^2 -(C^0)^2
r^{-2}(1-2Mr^{-1})\sin^{-2}\theta_0\bigr]^{1/2}} ,
\\ \nl
C_1(\tau - \tau_{0}) &=& \pm\int\limits_{r_0}^{r}\frac{dr}{\bigl
[(C^3)^2-(C^0)^2 r^{-2} (1-2Mr^{-1})\sin^{-2}\theta_0\bigr ]^{1/2}} .
\ea

For the {\it Taub-NUT space-time} we take the metric as \ba\nl
g_{00}=-\frac{\delta}{R^2}\h,\h g_{11}=\frac{R^2}{\delta}\h,\h
g_{22}=R^2 ,\\ \nl
g_{33}=R^2\sin^2\theta-4l^2\frac{\delta\cos^2\theta}{R^2}\h,\h
g_{03}=-2l\frac{\delta\cos\theta}{R^2} , \\ \nl
\delta(r)=r^2-2Mr-l^2\h,\h R^2(r)=r^2+l^2 , \ea where $M$ is the mass
and $l$ is the NUT-parameter. Now we have from (\ref{su3}),
(\ref{r.2}) and (\ref{t.2}): \ba\nl \dot t&=&\frac{C_1}{R^2}\Biggl[
C^3\Biggl(\frac{R^4}{\delta}+4l^2\Biggr)-\frac{2l}{\sin^2\theta}\left(
C^0\cos\theta+2C^3 l\right)\Biggr], \\ \nl R^4\dot
r^2&=&\left(C_1C^3\right)^2\left(R^4+4l^2\delta\right)-C_2\delta, \\
\nl R^4\dot\theta^2&=&C_2-\frac{C_1^2}{\sin^2\theta}\Bigl[\left(
C^0\right)^2+\left(2C^3 l\right)^2+4C^0 C^3 l\cos\theta\Bigr],
\\ \nl \dot
\varphi&=&\frac{C_1}{R^2\sin^2\theta}\left(C^0+2C^3
l\cos\theta\right). \ea In the Taub-NUT metric the solution
(\ref{GSC}) is \ba\nl x^0 (\tau,\s^1,\s^2)&=&C^0 f[w(z^1,z^2)] + t_0
\\ \nl
&\pm&\int\limits_{r_0}^{r}dr\Bigl[C^3\left(R^4
\delta^{-1}+4l^2\right)-2l
\sin^{-2}\theta_0\left(C^0\cos\theta_{0}+2C^3
l\right)\Bigr]U^{-1/2}(r) , \\ \nl x^3 (\tau,\s^1,\s^2)&=&C^3
f[w(z^1,z^2)] + \varphi_{0}
\\ \nl
&\pm&\frac{1}{\sin^2\theta_0}\int\limits_{r_0}^{r}dr\left(C^0+2C^3
l\cos\theta_{0}\right)U^{-1/2}(r) ,\\ \nl C_1(\tau - \tau_{0})&=&
\pm\int\limits_{r_0}^{r}drR^2 U^{-1/2}(r) ,\ea where \ba\nl
U(r)=\left(C^3\right)^2\left(R^4-4l^2\delta\cot^2\theta_{0}\right)
-\delta\sin^{-2}\theta_{0}\left[\left(C^0\right)^2+4C^0 C^3
l\cos\theta_{0}\right] . \ea

 Finally, we consider the {\it Kerr space-time} with metric taken
in the form \ba\nl g_{00}&=&-\Biggl(1-\frac{2Mr}{\rho^2}\Biggr)\h,\h
g_{11}=\frac{\rho^2}{\Delta},
\\ \nl
g_{22}&=&\rho^2 \h,\h g_{33}=\Biggl(r^2+a^2+\frac{2Ma^2
r\sin^2\theta}{\rho^2}\Biggr)\sin^2\theta ,
\\ \nl
g_{03}&=&-\frac{2Mar\sin^2\theta}{\rho^2} , \ea where \ba\nl
\rho^2=r^2+a^2\cos^2\theta \h,\h \Delta=r^2-2Mr+a^2 , \ea $M$ is the
mass and $a$ is the angular momentum per unit mass of the Kerr black
hole. With this input in equations (\ref{su3}), (\ref{r.2}) and
(\ref{t.2}), we have: \ba\nl \dot
t&=&\frac{C_1}{\Delta\rho^2}\Bigl[C^3 (r^2+a^2)^2 -C^3 a^2\Delta
\sin^2\theta - 2C^0 M a r\Bigr] ,
\\ \nl
\rho^4\dot r^2&=&C_1^2\Bigl[(C^3)^2 (r^2+a^2)^2 -4C^0 C^3 Mar+(C^0)^2
a^2\Bigr] - C_2 \Delta ,
\\ \nl
\rho^4\dot\theta^2&=&C_2-C_1^2\Biggl(\frac{(C^0)^2}{\sin^2\theta}
+(C^3)^2 a^2 \sin^2\theta\Biggr)
\\ \nl
\dot\varphi&=&\frac{C_1}{\Delta\rho^2}\Biggl( \frac{C^0
\Delta}{\sin^2\theta}+2C^3 Mar-C^0 a^2\Biggr) . \ea In this case, the
exact solution (\ref{GSC}) takes the form: \ba\nl x^0
(\tau,\s^1,\s^2)&=&C^0 f[w(z^1,z^2)] + t_0
\\ \nl
&\pm&\int\limits_{r_0}^{r}dr\Delta^{-1} \Bigl\{2C^0 Mar-C^3\left[
(r^2+a^2)\rho_{0}^2+2Ma^2 r\sin^2\theta_0\right]\Bigr\}V^{-1/2}(r)
,\\ \label{K} x^3 (\tau,\s^1,\s^2)&=&C^3 f[w(z^1,z^2)] + \varphi_{0}
\\ \nl
&\pm&\frac{1}{\sin^2\theta_0}\int\limits_{r_0}^{r}dr\Delta^{-1}
\left[C^0\left(2Mr-\rho^2_{0}\right)-2C^3
Mar\sin^2\theta_0\right]V^{-1/2}(r) ,\\ \nl C_1(\tau - \tau_{0})&=&
\pm\int\limits_{r_0}^{r}dr\rho_0^2 V^{-1/2}(r) ,\\ \nl
V(r)&=&\left(C^0\right)^2\left(a^2-\Delta\sin^{-2}\theta_{0}\right)
-4C^0 C^3 Mar\\ \nl &+&\left(C^3\right)^2\Bigl[\left(r^2+a^2\right)^2
-a^2\Delta\sin^2\theta_{0}\Bigr] ,\\ \nl
\rho^2_{0}&=&r^2+a^2\cos^2\theta_{0} . \ea

\subsection{\bf Comments}
\hspace{1cm} In the previous subsection we restrict ourselves to
some particular cases of the generic solution (\ref{GSC}) and did
not pay attention to the existing possibilities for obtaining
solutions in the case $\theta\neq const$ in the considered
examples.

Obviously, the examples given in the previous subsection do not
exhaust all possibilities contained in the metric (\ref{gm})
\cite{KSMH80}. On the other hand, in different particular cases of
this type of metric, there exist more general brane solutions.
They will be not considered here. We only mention that in the
gauge $\mu^{k}=const$, \ba\nl
x^{\nu}(\tau,\us)=x^{\nu}(\mu^{k}\tau+\s^k) \ea is an obvious
nontrivial solution of the equations of motion and of the
constraints (\ref{cbeqx}), (\ref{cbTx1}) depending on $D$
arbitrary functions of $p$ variables for the null $p$-brane in
arbitrary $D$-dimensional gravity background.

From the results of the previous subsection, it is easy to extract
the corresponding formulas for the null string case simply by
putting $\s^1=\s, \mu^1=\mu, \s^2=\mu^2=0$. For example, our
equalities (\ref{S}) coincide with the ones obtained in
\cite{DL96} for the null string moving in Schwarzschild space-time
after identification: \ba\nl E=C_1 C^3 ,\h L=C_1 C^0 ,\h L^2+K=C_2
. \ea Moreover, our solution (\ref{K}) in the case $p=1$,
generalizes the solution given in \cite{PP97}. The latter
corresponds to fixing the arbitrary function $f(w)$ to a linear
one and fixing the gauge to $\mu=0$, i.e. \ba\nl
f[w(\mu^1\tau+\s^1,\mu^2\tau+\s^2)]\mapsto
f(\mu\tau+\s)=\mu\tau+\s ,\h \mbox{with}\h \mu=0 . \ea

\newpage
\vspace*{.5cm}
\section{\bf D=10 CHIRAL NULL SUPER p-BRANES}
\hspace{1cm} Here we consider a model for tensionless super
$p$-branes with $N$ chiral supersymmetries in ten dimensional flat
space-time. After establishing the symmetries of the action, we
give the general solution of the classical equations of motion in
a particular gauge. In the case of a null superstring ($p$=1) we
find the general solution in an arbitrary gauge. Then, using a
harmonic superspace approach, the initial algebra of first and
second class constraints is converted into an algebra of
Lorentz-covariant, BFV-irreducible, first class constraints only.
The corresponding BRST charge is as for a first rank dynamical
system. This section is based on the papers \cite{B986,B991,B992}.

\subsection{\bf Lagrangian formulation}
\hspace{1cm} We define our model for $D=10$ $N$-extended chiral
tensionless super $p$-branes by the action: \ba\label{a} S=\int
d^{p+1}\xi L \h,\h L=V^JV^K\Pi_J^\mu\Pi_K^\nu\eta_{\mu\nu},
\\ \nl
\Pi_J^\mu=\pJ x^\mu+i\sum_{A=1}^N (\theta^A\s^\mu\pJ\theta^A)\h,\h
\pJ=\p/\p\xi^J,
\\ \nl
\xi^J=(\xi^0,\xi^j)=(\tau,\s^j),\h diag(\eta_{\mu\nu})=(-,+,...,+),
\\ \nl
J,K=0,1,...,p \h,\h j,k=1,...,p \h,\h \mu,\nu=0,1,...,9. \ea Here
$(x^\nu,\theta^{A\alpha})$ are the superspace coordinates,
$\theta^{A\alpha}$ are $N$ left Majorana-Weyl space-time spinors
($\alpha=1,...,16$ , $N$ being the number of the supersymmetries) and
$\s^\mu$ are the 10-dimensional Pauli matrices (our spinor
conventions are given in the Appendix B). Actions of this type are
first given in \cite{LST911} for the case of tensionless superstring
($p=1,N=1$) and in \cite{HLU94} for the bosonic case ($N=0$).

The action (\ref{a}) has an obvious global Poincar$\acute{e}$
invariance. Under global infinitesimal supersymmetry
transformations, the fields $\theta^{A\alpha}(\xi)$, $x^\nu(\xi)$
and $V^J(\xi)$ transform as follows: \ba\nl
\delta_\eta^{A\alpha}=\eta^{A\alpha}\h,\h \delta_\eta
x^\mu=i\sum_A(\theta^A\s^\mu\delta_\eta\theta^A)\h,\h \delta_\eta
V^J=0. \ea As a consequence $\delta_\eta\Pi_J^\mu=0$ and hence
also $\delta_\eta L =\delta_\eta S=0$.

Using (\ref{cbdiff}), one verifies that if $x^\mu(\xi)$,
$\theta^{A\alpha}(\xi)$ are world-volume scalars ($a=0$) and
$V^J(\xi)$ is a world-volume (1,0)-type tensor density of weight
$a=1/2$, then $\Pi_J^\nu$ is a (0,1)-type tensor, $\Pi_J^\nu
\Pi_{K\nu}$ is a (0,2)-type tensor and $L$ is a scalar density of
weight $a=1$. Therefore, \ba\nl \delta_{\varepsilon}S=\int
d^{p+1}\xi\p_J\bigl ( \varepsilon^J L\bigr ) \ea and this variation
vanishes under suitable boundary conditions.

Let us now check the $\kappa$-invariance of the action. We define
the $\kappa$-variations of $\theta^{A\alpha}(\xi)$, $x^\nu(\xi)$
and $V^J(\xi)$ as follows: \ba\label{k}
\delta_\kappa\theta^{A\alpha}=i\bigl(\Gamma\kappa^A\bigr)^\alpha=
iV^J\bigl(\not{\Pi_J}\kappa^A\bigr)^\alpha,\h \delta_\kappa
x^\nu=-i\sum_A(\theta^A\s^\nu\delta_\kappa\theta^A),
\\ \nl
\delta_\kappa V^K=2V^K V^L \sum_A(\p_L\theta^A\kappa^A). \ea
Therefore, $\kappa^{A\alpha}(\xi)$ are left Majorana-Weyl space-time
spinors and world-volume scalar densities of weight $a=-1/2$.

From (\ref{k}) we obtain: \ba\nl \delta_\kappa\bigl(\Pi_J^\nu
\Pi_{K\nu}\bigr)=-2i\sum_A\bigl[
\p_J\theta^A\not{\Pi_K}+\p_K\theta^A\not{\Pi_J}\bigr]
\delta_\kappa\theta^A \ea and \ba\nl \delta_\kappa L=2V^J \Pi_J^\nu
\Pi_{K\nu}\bigl[\delta_\kappa V^K- 2V^K
V^L\sum_A(\p_L\theta^A\kappa^A)\bigr] = 0 . \ea

The algebra of kappa-transformations closes only on the equations
of motion, which can be written in the form: \ba\label{eqm}
\p_J\bigl(V^JV^K\Pi_{K\nu}\bigr)=0,\h
V^JV^K\bigl(\p_J\theta^A\not{\Pi_K}\bigr)_\alpha=0,\h V^J
\Pi_J^\nu \Pi_{K\nu}=0 . \ea As usual, an additional local bosonic
world-volume symmetry is needed for its closure. In our case, the
Lagrangian, and therefore the action, are invariant under the
following transformations of the fields: \ba\nl
\delta_\lambda\theta^A(\xi)=\lambda V^J\p_J\theta^A,\h
\delta_\lambda x^\nu(\xi)=-i\sum_A(\theta^A\s^\nu\delta_\lambda
\theta^A),\h \delta_\lambda V^J(\xi)=0 . \ea Now, checking the
commutator of two $\kappa$-transformations, we find: \ba\nl
[\delta_{\kappa_1},\delta_{\kappa_2}]\theta^{A\alpha}(\xi)&=&
\delta_\kappa\theta^{A\alpha}(\xi)+ \mbox{terms $\propto$ eqs. of
motion} , \\ \nl
[\delta_{\kappa_1},\delta_{\kappa_2}]x^\nu(\xi)&=&
(\delta_\kappa+\delta_\varepsilon+\delta_\lambda)x^\nu(\xi)+
\mbox{terms $\propto$ eqs. of motion} , \\ \nl
[\delta_{\kappa_1},\delta_{\kappa_2}]V^J(\xi)&=&
\delta_\varepsilon V^J(\xi)+ \mbox{terms $\propto$ eqs. of motion}
. \ea Here $\kappa^{A\alpha}(\xi)$, $\lambda(\xi)$ and
$\varepsilon(\xi)$ are given by the expressions: \ba\nl
\kappa^{A\alpha}=-2V^K\sum_B[(\p_K\theta^B\kappa_1^B)\kappa_2^{A\alpha}-
(\p_K\theta^B\kappa_2^B)\kappa_1^{A\alpha}],\\ \nl
\lambda=4iV^K\sum_A(\kappa_1^A\not{\Pi_K}\kappa_2^A)\h,\h
\varepsilon^J=-V^J\lambda . \ea

We note that $\Gamma_{\alpha\beta}=\bigl(V^J\not{\Pi_J}\bigr)_
{\alpha\beta}$ in (\ref{k}) has the following property on the
equations of motion \ba\nl \Gamma^2 = 0 . \ea This means, that the
local $\kappa$-invariance of the action indeed eliminates half of
the components of $\theta^A$ as is needed.

For transition to Hamiltonian picture, it is convenient to rewrite
the Lagrangian density (\ref{a}) in the form ($\pu=\p/\p\tau,
\p_j=\p/\p\s^j$): \ba\label{L} L=\frac{1}{4\mu^0}\Bigl [\bigl
(\pu-\mu^j\pj\bigr )x+ i\sum_A\theta^A\s\bigl (\pu-\mu^j\pj\bigr
)\theta^A\Bigr ]^2 , \ea where \ba\nl
V^J=\bigl(V^0,V^j\bigr)=\Biggl(-\frac{1}{2\sqrt{\mu^0}},
\frac{\mu^j}{2\sqrt{\mu^0}}\Biggr). \ea The equations of motion
for the Lagrange multipliers $\mu^{0}$ and $\mu^{j}$ which follow
from (\ref{L}) give the constraints ($p_\nu$ and
$p^A_{\theta\alpha}$ are the momenta conjugated to $x^\nu$ and
$\theta^{A\alpha}$): \ba\label{T0,Tj} T_0=p^2\h,\h T_j=p_\nu\p_j
x^\nu+\sum_A p^A_{\theta\alpha}\p_j\theta^{A\alpha}. \ea The
remaining constraints follow from the definition of the momenta
$p^A_{\theta\alpha}$: \ba\label{D}
D_\alpha^A=-ip_{\theta\alpha}^A-(\not{p}\theta^A)_\alpha. \ea

\subsection{\bf Hamiltonian formulation}
\hspace{1cm} The Hamiltonian which corresponds to the Lagrangian
density (\ref{L}) is a linear combination of the constraint
(\ref{T0,Tj}) and (\ref{D}): \ba\label{H0} H_0=\int d^p\sigma\bigl
[\mu^0 T_0+\mu^j T_j+ \sum_A\mu^{A\alpha} D^A_{\alpha}\bigr ] \ea It
is a generalization of the Hamiltonians for the bosonic null
$p$-brane and for the $N$-extended Green-Schwarz superparticle.

The equations of motion which follow from the Hamiltonian (\ref{H0})
are: \ba\nl
(\pu-\mu^j\pj)x^\nu&=&2\mu^0p^\nu-\sum_A(\mu^A\s^\nu\theta^A),\h
(\pu-\mu^j\pj)p_\nu=(\pj\mu^j)p_\nu,\\ \label{em}
(\pu-\mu^j\pj)\theta^{A\alpha}&=&i\mu^{A\alpha},\h
(\pu-\mu^j\pj)p^A_{\theta\alpha}=(\pj\mu^j)p^A_{\theta\alpha}
+(\mu^A\not{p})_\alpha . \ea In (\ref{em}), one can consider $\mu^0$,
$\mu^j$ and $\mu^{A\alpha}$ as depending only on
$\us=(\s^1,...,\s^p)$, but not on $\tau$ as a consequence from their
equations of motion.

In the gauge when $\mu^0$, $\mu^j$ and $\mu^{A\alpha}$ are constants,
the general solution of (\ref{em}) is \ba \nl
x^\nu(\tau,\us)&=&x^\nu(\uz)+\tau\bigl [2\mu^0p^\nu(\uz)-
\sum_A(\mu^A\s^\nu\theta^A(\tau,\us))\bigr ], \\ \nl
&=&x^\nu(\uz)+\tau\bigl [2\mu^0p^\nu(\uz)-
\sum_A(\mu^A\s^\nu\theta^A(\uz))\bigr ] \\ \label{gsp}
p_\nu(\tau,\us)&=&p_\nu(\uz) ,\h \theta^{A\alpha}(\tau,\us)=
\theta^{A\alpha}(\uz)+i\tau\mu^{A\alpha} ,\\ \nl
p^A_{\theta\alpha}(\tau,\us)&=&p^A_{\theta\alpha}(\uz)+ \tau
(\mu^A\s^\nu)_{\alpha} p_{\nu}(\uz) , \ea where $x^\nu(\uz)$,
$p_\nu(\uz)$, $\theta^{A\alpha}(\uz)$ and $p^A_{\theta\alpha}(\uz)$
are arbitrary functions of their arguments \ba\nl z^j =
\mu^j\tau+\s^j . \ea In the case of null strings ($p=1$), one can
write explicitly the general solution of the equations of motion in
an arbitrary gauge:  $\mu^0=\mu^0(\s)$, $\mu^1\equiv\mu=\mu(\s)$,
$\mu^{A\alpha}=\mu^{A\alpha}(\s)$. This solution is given by \ba \nl
x^\nu(\tau,\s)&=&g^\nu(w)-2\int\limits_{}^\s\frac{\mu^0(s)}{\mu^2(s)}ds
f^\nu(w)+\sum_A\int\limits^\s\frac{\mu^{A\alpha}(s)}{\mu(s)}ds \bigl
[\s^\nu\zeta^A(w)\bigr ]_\alpha \\ \nl &-&i\sum_A\int\limits^\s
ds_1\frac{(\mu^A\s^\nu)_\alpha(s_1)}{\mu(s_1)}
\int\limits^{s_1}\frac{\mu^{A\alpha}(s)}{\mu(s)}ds ,\\ \label{gs1}
p_\nu(\tau,\s)&=&\mu^{-1}(\s)f_\nu(w) ,\\ \nl
\theta^{A\alpha}(\tau,\s)&=&\zeta^{A\alpha}(w)-i\int\limits^\s
\frac{\mu^{A\alpha}(s)}{\mu(s)}ds ,\\ \nl
p^A_{\theta\alpha}(\tau,\s)&=&\mu^{-1}(\s)\Biggl [h^A_\alpha(w)-
\int\limits^\s\frac{(\mu^A\s^\nu)_\alpha(s)}{\mu(s)}ds f_\nu(w)
\Biggr ] . \ea Here $g^\nu(w)$, $f_\nu(w)$, $\zeta^{A\alpha}(w)$ and
$h^A_\alpha(w)$ are arbitrary functions of the variable \ba\nl w =
\tau + \int\limits^\s \frac{ds}{\mu(s)} \ea The solution (\ref{gsp})
at $p=1$ differs from (\ref{gs1}) by the choice of the particular
solutions of the inhomogenious equations.  As for $z$ and $w$, one
can write for example ($\mu^0$, $\mu$, $\mu^{A\alpha}$ are now
constants) \ba\nl
p_\nu(\tau,\s)=\mu^{-1}f_\nu(\tau+\s/\mu)=\mu^{-1}f_\nu[\mu^{-1}
(\mu\tau+\s)]=p_\nu(z) \ea and analogously for the other arbitrary
functions in the general solution of the equations of motion.

Let us now consider the properties of the constraints (\ref{T0,Tj}),
(\ref{D}). They satisfy the following (equal $\tau$) Poisson bracket
algebra \ba \nl \{T_0(\ul \sigma_1),T_0(\ul \sigma_2)\}&=&0,\h
\{T_0(\ul \sigma_1),D^A_{\alpha}(\ul \sigma_2)\}=0 , \\ \nl \{T_0(\ul
\sigma_1),T_{j}(\ul \sigma_2)\}&=& [T_0(\ul \sigma_1) + T_0(\ul
\sigma_2)] \p_j \delta^p (\ul \sigma_1 - \ul \sigma_2) , \\ \nl
\{T_{j}(\ul \sigma_1),T_{k}(\ul \sigma_2)\}&=&
[\delta_{j}^{l}T_{k}(\ul \sigma_1) + \delta_{k}^{l}T_{j}(\ul
\sigma_2)]\p_l\delta^p(\ul \sigma_1-\ul \sigma_2) , \\ \nl
\{T_{j}(\ul \sigma_1),D^A_{\alpha}(\ul \sigma_2)\}&=&
D^A_{\alpha}(\ul \sigma_1) \p_j \delta^p (\ul \sigma_1-\ul \sigma_2)
, \\ \nl \{D^A_{\alpha}(\ul \sigma_1),D^B_{\beta}(\ul \sigma_2)\}&=&
2i\delta^{AB}\not{p}_{\alpha\beta} \delta^p (\ul \sigma_1-\ul
\sigma_2) . \ea From the condition that the constraints must be
maintained in time, i.e. \cite{D64} \ba\label{Dc} \bigl
\{T_0,H_0\bigr \}\approx 0,\h \bigl \{T_j,H_0\bigr \}\approx 0, \h
\bigl \{D^A_\alpha,H_0\bigr \}\approx 0, \ea it follows that in the
Hamiltonian $H_0$ one has to include the constraints \ba \nl
T^A_{\alpha}=\not p_{\alpha\beta} D^{A\beta} \ea instead of
$D^A_{\alpha}$. This is because the Hamiltonian has to be first class
quantity, but $D^A_{\alpha}$ are a mixture of first and second class
constraints. $T^A_{\alpha}$ has the following non-zero Poisson
brackets \ba \nl \{T_{j}(\ul \sigma_1),T^A_{\alpha}(\ul
\sigma_2)\}&=& [T^A_{\alpha}(\ul \sigma_1)+T^A_{\alpha}(\ul
\sigma_2)] \p_j \delta^p (\ul \sigma_1-\ul \sigma_2) , \\ \nl
\{T^A_{\alpha}(\ul \sigma_1),T^B_{\beta}(\ul \sigma_2)\}&=&
2i\delta^{AB}\not{p}_{\alpha\beta}T_0 \delta^p (\ul \sigma_1-\ul
\sigma_2) . \ea In this form, our constraints are first class and the
Dirac consistency conditions (\ref{Dc}) (with $D^A_\alpha$ replaced
by $T^A_\alpha$) are satisfied identically.  However, one now
encounters a new problem. The constraints $T_0$, $T_j$ and
$T^A_{\alpha}$ are not BFV-irreducible, i.e. they are functionally
dependent: \ba \nl (\not p T^A)^{\alpha} - D^{A\alpha} T_0 = 0 . \ea
It is known, that in this case after BRST-BFV quantization an
infinite number of ghosts for ghosts appear, if one wants to preserve
the manifest Lorentz invariance.  The way out consists in the
introduction of auxiliary variables, so that the mixture of first and
second class constraints $D^{A\alpha}$ can be appropriately
covariantly decomposed into first class constraints and second class
ones. To this end, here we will use the auxiliary harmonic variables
introduced in \cite{S867} and \cite{NP88}.  These are $u_{\mu}^a$ and
$v_{\alpha}^{\pm}$, where superscripts $a=1,...,8$ and $\pm$
transform under the 'internal' groups $SO(8)$ and $SO(1,1)$
respectively. The just introduced variables are constrained by the
following orthogonality conditions \ba\nl u_{\mu}^a u^{b\mu}=C^{ab},
\h u_{\mu}^{\pm}u^{a\mu}=0, \h u_{\mu}^{+} u^{-\mu}=-1, \ea where
\ba\nl u_{\mu}^{\pm}=v_{\alpha}^\pm
\sigma_{\mu}^{\alpha\beta}v_{\beta}^\pm , \ea $C^{ab}$ is the
invariant metric tensor in the relevant representation space of
$SO(8)$ and $(u^\pm)^2=0$ as a consequence of the Fierz identity for
the 10-dimensional $\sigma$-matrices. We note that $u^a_\nu$ and
$v_{\alpha}^\pm$ do not depend on $\us$.

Now we have to ensure that our dynamical system does not depend on
arbitrary rotations of the auxiliary variables $(u_{\mu}^a$,
$u_{\mu}^\pm)$. It can be done by introduction of first class
constraints, which generate these transformations \ba\nl
I^{ab}&=&-(u_{\nu}^a p^{b\nu}_u - u_{\nu}^b p^{a\nu}_u +
\frac{1}{2}v^+\sigma^{ab}p_v^+ +\frac{1}{2}v^-\sigma^{ab}p_v^-), \h
\sigma^{ab}=u_{\mu}^a u_{\nu}^b \sigma^{\mu\nu},
\\
\label{ncon} I^{-+}&=&-\frac{1}{2}(v_{\alpha}^+ p_v^{+\alpha} -
v_{\alpha}^- p_v^{-\alpha}),
\\ \nl
I^{\pm a}&=&-(u_{\mu}^\pm p_u^{a\mu} + \frac{1}{2}v^\mp \sigma^\pm
\sigma^a p_v^\mp) , \h \sigma^\pm =u^\pm_\nu \sigma^\nu, \h \sigma^a
= u^a_\nu \sigma^\nu . \ea In the above equalities, $p_u^{a\nu}$ and
$p_v^{\pm \alpha}$ are the momenta canonically conjugated to
$u^a_\nu$ and $v^\pm_\alpha$.

The newly introduced constraints (\ref{ncon}) obey the following
Poisson bracket algebra \ba\nl
\{I^{ab},I^{cd}\}&=&C^{bc}I^{ad}-C^{ac}I^{bd}+C^{ad}I^{bc}-C^{bd}I^{ac},
\\ \nl
\{I^{-+},I^{\pm a}\}&=&\pm I^{\pm a},
\\ \nl
\{I^{ab},I^{\pm c}\}&=&C^{bc}I^{\pm a}-C^{ac}I^{\pm b},
\\ \nl
\{I^{+a},I^{-b}\}&=&C^{ab}I^{-+} + I^{ab} . \ea This algebra is isomorphic
to the $SO(1,9)$ algebra: $I^{ab}$ generate $SO(8)$ rotations, $I^{-+}$ is
the generator of the subgroup $SO(1,1)$ and $I^{\pm a}$ generate the
transformations from the coset $SO(1,9)/\left(SO(1,1)\times SO(8)\right)$.

Now we are ready to separate $D^{A\alpha}$ into first and second
class constraints in a Lorentz-covariant form. This separation is
given by the equalities \cite{NPS89}: \ba\label{rev} D^{A\alpha}
&=&\frac{1}{p^+}\bigl [(\sigma^a v^+)^\alpha D^A_a + (\not p \sigma^+
\sigma^a v^-)^\alpha G^A_a \bigr ], \h p^+ = p^\nu u^+_\nu,
\\ \nl
D^{Aa} &=&(v^+ \sigma^a \not p )_\beta D^{A\beta}, \h G^{Aa} =
\frac{1}{2}(v^- \sigma^a \sigma^+)_\beta D^{A\beta}. \ea Here
$D^{Aa}$ are first class constraints and $G^{Aa}$ are second class
ones: \ba\nl \{D^{Aa}(\us_1),D^{Bb}(\us_2)\}&=&
-2i\delta^{AB}C^{ab}p^+ T_0 \delta^p(\us_1 - \us_2) \\ \nl
\{G^{Aa}(\us_1),G^{Bb}(\us_2)\}&=&i\delta^{AB}C^{ab}p^+ \da . \ea It
is convenient to pass from second class constraints $G^{Aa}$ to first
class constraints $\hat G^{Aa}$, without changing the actual degrees
of freedom \cite{NPS89}, \cite{EM93} : \ba\nl G^{Aa} \rightarrow \hat
G^{Aa} = G^{Aa} + (p^+)^{1/2} \Psi^{Aa} \h \Rightarrow \h \{\hat
G^{Aa}(\us_1),\hat G^{Bb}(\us_2)\} = 0 , \ea where $\Psi^{Aa}(\us)$
are fermionic ghosts which abelianize our second class constraints as
a consequence of the Poisson bracket relation \ba\nl
\{\Psi^{Aa}(\us_1),\Psi^{Bb}(\us_2)\} = -i\delta^{AB}C^{ab}\da . \ea

It turns out that the constraint algebra is much more simple, if
we work not with $D^{Aa}$ and $\hat G^{Aa}$ but with $\hat
T^{A\alpha}$ given by \ba\nl \hat T^{A\alpha} &=&(p^+)^{-1/2}\bigl
[(\s^a v^+)^\alpha D^A_a + (\not p \s^+ \s^a v^-)^\alpha \hat
G^A_a \bigr ]
\\ \nl &=&(p^+)^{1/2}D^{A\alpha} + (\not p \s^+ \s^a v^-)^\alpha
\Psi^A_a . \ea After the introduction of the auxiliary fermionic
variables $\Psi^{Aa}$, we have to modify some of the constraints, to
preserve their first class property. Namely $T_j$, $I^{ab}$ and
$I^{-a}$ change as follows \ba\nl \hat T_j &=& T_j +
\frac{i}{2}C^{ab}\sum_A\Psi^A_a \p_j \Psi^A_b , \\ \nl \hat I^{ab}
&=& I^{ab} + J^{ab}, \hspace{.3cm} J^{ab}=\int d^p \s j^{ab}(\us),
\hspace{.3cm} j^{ab}=\frac{i}{4}(v^-\s_c\s^{ab}\s^+\s_d v^-)
\sum_A\Psi^{Ac}\Psi^{Ad},
\\ \nl
\hat I^{-a}&=&I^{-a} + J^{-a}, \h J^{-a}=\int d^p\s j^{-a}(\us), \h
j^{-a}=-(p^+)^{-1}j^{ab}p_b . \ea As a consequence, we can write down
the Hamiltonian for the considered model in the form: \ba\nl H=\int
d^p\s\bigl [\la^0 T_0(\us)+\la^j \hat T_j(\us)+
\sum_A\la^{A\alpha}\hat T^A_\alpha(\us)\bigr ] +
\\ \nl
\la_{ab}\hat I^{ab}+\la_{-+}I^{-+}+\la_{+a}I^{+a}+\la_{-a}\hat I^{-a}
. \ea The constraints entering $H$ are all first class, irreducible
and Lorentz- covariant. Their algebra reads (only the non-zero
Poisson brackets are written): \ba\nl \{T_0(\us_1),\hat
T_j(\us_2)\}&=&\bigl (T_0(\us_1)+T_0(\us_2)\bigr )\p_j \da ,
\\ \nl
\{\hat T_j(\us_1),\hat T_k(\us_2)\}&=& \bigl (\delta_j^l \hat
T_k(\us_1)+\delta_k^l \hat T_j(\us_2)\bigr )\p_l \da ,
\\ \nl
\{\hat T_j(\us_1),\hat T^A_\alpha(\us_2)\}&=& \bigl (\hat
T^A_\alpha(\us_1)+ \frac{1}{2}\hat T^A_\alpha(\us_2)\bigr )\p_j \da ,
\\ \nl
\{\hat T^A_\alpha(\us_1),\hat T^B_\beta(\us_2)\}&=&
i\delta^{AB}\s^+_{\alpha\beta}T_0 \da ,
\\ \nl
\{I^{-+},\hat T^A_\alpha \}&=&\frac{1}{2}\hat T^A_\alpha , \h \{\hat
I^{-a},\hat T^A_\alpha \}=(2p^+)^{-1}\bigl [ p^a\hat T^A_\alpha +
(\s^+ \s^{ab}v^-)_\alpha\Psi^A_b T_0 \bigr ] ,
\\ \nl
\{\hat I^{ab},\hat I^{cd}\}&=&C^{bc}\hat I^{ad}-C^{ac}\hat I^{bd}+
C^{ad}\hat I^{bc}-C^{bd}\hat I^{ac} ,
\\ \nl
\{I^{-+},I^{+a}\}&=&I^{+a} , \h \{I^{-+},\hat I^{-a}\}=-\hat I^{-a} ,
\\ \nl
\{\hat I^{ab},I^{+c}\}&=&C^{bc}I^{+a}-C^{ac}I^{+b} , \h \{\hat
I^{ab},\hat I^{-c}\}=C^{bc}\hat I^{-a}-C^{ac}\hat I^{-b} ,
\\ \nl
\{I^{+a},\hat I^{-b}\}&=&C^{ab}I^{-+} + \hat I^{ab} ,
\\ \nl
\{\hat I^{-a},\hat I^{-b}\}&=&-\int d^p\s(p^+)^{-2}j^{ab}T_0 . \ea

Having in mind the above algebra, one can construct the corresponding
BRST charge $\Omega$ \cite{FF78} ($*$=complex conjugation) \ba
\label{O} \Omega = \Omega^{min}+\pi_M \bar {\cal P}^M , \h
\{\Omega,\Omega\} = 0 , \h \Omega^* = \Omega , \ea where
$M=0,j,A\alpha,ab,-+,+a,-a$.  $\Omega^{min}$ in (\ref{O}) can be
written as \ba\nl \Omega^{min}&=&\Omega^{brane}+\Omega^{aux} ,
\\ \nl
\Omega^{brane}&=&\int d^p\s\{T_0\eta^0+\hat T_j\eta^j+ \sum_A\hat
T^A_\alpha \eta^{A\alpha} + {\cal P}_0 [(\p_j\eta^j)\eta^0 +
(\p_j\eta^0)\eta^j ] +
\\ \nl &+&{\cal P}_k(\p_j\eta^k)\eta^j +
\sum_A{\cal P}^A_\alpha [\eta^j\p_j\eta^{A\alpha} -
\frac{1}{2}\eta^{A\alpha}\p_j\eta^j ] - \frac{i}{2}{\cal P}_0
\sum_A\eta^{A\alpha}\s^+_{\alpha\beta}\eta^{A\beta}\} ,
\\ \nl
\Omega^{aux}&=& \hat
I^{ab}\eta_{ab}+I^{-+}\eta_{-+}+I^{+a}\eta_{+a}+\hat I^{-a}\eta_{-a}
\\ \nl
&+&({\cal P}^{ac}\eta^{b.}_{.c}-{\cal P}^{bc}\eta^{a.}_{.c} + 2{\cal
P}^{+a}\eta^b_+ + 2{\cal P}^{-a}\eta^b_-)\eta_{ab}
\\ \nl
&+&({\cal P}^{+a}\eta_{+a}-{\cal P}^{-a}\eta_{-a})\eta_{-+} + ({\cal
P}^{-+}\eta^a_- + {\cal P}^{ab}\eta_{-b})\eta_{+a}
\\ \nl
&+&\frac{1}{2}\int d^p\s\{ \sum_A{\cal
P}^A_\alpha\eta^{A\alpha}\eta_{-+} + (p^+)^{-1}\sum_A[p^a{\cal
P}^A_\alpha - (\s^+\s^{ab}v^-)_\alpha\Psi^A_b {\cal P}_0]
\eta^{A\alpha}\eta_{-a}
\\ \nl
&-&(p^+)^{-2}j^{ab}{\cal P}_0\eta_{-b}\eta_{-a}\} . \ea These
expressions for $\Omega^{brane}$ and $\Omega^{aux}$ show that we have
found a set of constraints which ensure the first rank property of
the model.

$\Omega^{min}$ can be represented also in the form \ba\nl
\Omega^{min}=\int d^p\s\bigl [\bigl (T_0+\frac{1}{2}T_0^{gh}\bigr
)\eta^0 +\bigl (\hat T_j+\frac{1}{2}T_j^{gh}\bigr )\eta^j+ \sum_A
\bigl (\hat T^A_\alpha +\frac{1}{2}T_\alpha^{A gh}\bigr )
\eta^{A\alpha}\bigr ]
\\ \nl
+\bigl (\hat I^{ab}+\frac{1}{2}I^{ab}_{gh}\bigr )\eta_{ab} +\bigl
(I^{-+}+\frac{1}{2}I^{-+}_{gh}\bigr )\eta_{-+} +\bigl
(I^{+a}+\frac{1}{2}I^{+a}_{gh}\bigr )\eta_{+a} +\bigl (\hat
I^{-a}+\frac{1}{2}I^{-a}_{gh}\bigr )\eta_{-a}
\\ \nl
+\int d^p\s\p_j\Bigl (\frac{1}{2}{\cal P}_k\eta^k\eta^j +\frac{1}{4}
\sum_A{\cal P}^A_\alpha\eta^{A\alpha}\eta^j\Bigr ) . \ea Here a
super(sub)script $gh$ is used for the ghost part of the total gauge
generators \ba\nl \nl G^{tot}=\{\Omega,{\cal
P}\}=\{\Omega^{min},{\cal P}\}=G+G^{gh} . \ea We recall that the
Poisson bracket algebras of $G^{tot}$ and $G$ coincide for first rank
systems. The manifest expressions for $G^{gh}$ are: \ba\nl
T_0^{gh}&=&2{\cal P}_0\p_j\eta^j+\bigl (\p_j{\cal P}\bigr )\eta^j ,
\\ \nl
T_j^{gh}&=&2{\cal P}_0\p_j\eta^0+\bigl (\p_j{\cal P}_0\bigr )\eta^0+
{\cal P}_j\p_k\eta^k+{\cal P}_k\p_j\eta^k+\bigl (\p_k{\cal P}_j\bigr
)\eta^k
\\ \nl
&+&\frac{3}{2}\sum_A{\cal P}^A_\alpha\p_j\eta^{A\alpha}
+\frac{1}{2}\sum_A\bigl (\p_j{\cal P}^A_\alpha\bigr )\eta^{A\alpha},
\\ \nl
T_\alpha^{A gh}&=&-\frac{3}{2}{\cal P}^A_\alpha\p_j\eta^j -\bigl
(\p_j{\cal P}^A_\alpha\bigr )\eta^j -i{\cal
P}_0\s^+_{\alpha\beta}\eta^{A\beta} +
\\ \nl
&+&\frac{1}{2}{\cal P}^A_\alpha\eta_{-+} +(2p^+)^{-1}\Bigl [p^a{\cal
P}^A_\alpha -(\s^+\s^{ab}v^-)_\alpha\Psi^A_b{\cal P}_0\Bigr
]\eta_{-a} ,
\\ \nl
I^{ab}_{gh}&=&2\bigl ({\cal P}^{ac}\eta^{b.}_{.c}-{\cal
P}^{bc}\eta^{a.}_{.c} \bigr )+\bigl ({\cal P}^{+a}\eta^b_+-{\cal
P}^{+b}\eta^a_+\bigr )+ \bigl ({\cal P}^{-a}\eta^b_--{\cal
P}^{-b}\eta^a_-\bigr ) ,
\\ \nl
I^{-+}_{gh}&=&{\cal P}^{+a}\eta_{+a}-{\cal P}^{-a}\eta_{-a}+
\frac{1}{2}\int d^p\s\sum_A{\cal P}^A_\alpha\eta^{A\alpha} ,
\\ \nl
I^{+a}_{gh}&=&2{\cal P}^{+b}\eta^{a.}_{.b}-{\cal P}^{+a}\eta_{-+}+
{\cal P}^{-+}\eta^a_-+{\cal P}^{ab}\eta_{-b} ,
\\ \nl
I^{-a}_{gh}&=&2{\cal P}^{-b}\eta^{a.}_{.b}+{\cal P}^{-a}\eta_{-+}-
{\cal P}^{-+}\eta^a_++{\cal P}^{ab}\eta_{+b}+
\\ \nl
&+&\int d^p\s\Bigl \{(2p^+)^{-1} \sum_A\Bigl [p^a{\cal P}^A_\alpha-
(\s^+\s^{ab}v^-)_\alpha\Psi^A_b{\cal P}_0\Bigr ]\eta^{A\alpha}-
(p^+)^{-2}j^{ab}{\cal P}_0 \eta_{-b}\Bigr \} . \ea Up to now, we
introduced canonically conjugated ghosts $\bigl (\eta^M,{\cal
P}_M\bigr )$, $\bigl (\bar \eta_M,\bar {\cal P}^M\bigr)$ and momenta
$\pi_M$ for the Lagrange multipliers $\lambda^M$ in the Hamiltonian.
They have Poisson brackets and Grassmann parity as follows
($\epsilon_M$ is the Grassmann parity of the corresponding
constraint): \ba\nl \bigl \{\eta^M,{\cal P}_N\bigr \}&=&\delta^M_N ,
\h \epsilon (\eta^M)=\epsilon ({\cal P}_M)=\epsilon_M + 1 , \\ \nl
\bigl \{\bar \eta_M,\bar {\cal P}^N \bigr
\}&=&-(-1)^{\epsilon_M\epsilon_N} \delta^N_M , \h \epsilon
(\bar\eta_M)=\epsilon (\bar {\cal P}^M)=\epsilon_M + 1 ,
\\ \nl
\bigl \{\lambda^M,\pi_N\bigr \}&=&\delta^M_N , \h \epsilon
(\lambda^M)=\epsilon (\pi_M)=\epsilon_M . \ea

The BRST-invariant Hamiltonian is \ba\label{H} H_{\tilde
\chi}=H^{min}+\bigl \{\tilde \chi,\Omega\bigr \}= \bigl \{\tilde
\chi,\Omega\bigr \} , \ea because from $H_{canonical}=0$ it
follows that $H^{min}=0$. In this formula $\tilde \chi$ stands for
the gauge fixing fermion $(\tilde \chi^* = -\tilde \chi)$. We use
the following representation for the latter \ba\nl \tilde
\chi=\chi^{min}+\bar\eta_M(\chi^M+\frac{1}{2}\rho_{(M)}\pi^M) , \h
\chi^{min}=\la^M{\cal P}_M , \ea where $\rho_{(M)}$ are scalar
parameters and we have separated the $\pi^M$-dependence from
$\chi^M$. If we adopt that $\chi^M$ does not depend on the ghosts
$(\eta^M,{\cal P}_M)$ and $(\bar\eta_M,\bar {\cal P}^M)$, the
Hamiltonian $H_{\tilde\chi}$ from (\ref{H}) takes the form \ba
\label{r} H_{\tilde\chi}&=&H_{\chi}^{min}+{\cal P}_M \bar {\cal
P}^M - \pi_M(\chi^M+\frac{1}{2}\rho_{(M)}\pi^M)+ \\ \nl
&+&\bar\eta_M\Bigl [\bigl \{\chi^M,G_N\bigr \}\eta^N
+\frac{1}{2}(-1)^{\epsilon_N} {\cal P}_Q\bigl
\{\chi^M,U^Q_{NP}\bigr \}\eta^P\eta^N \Bigr ], \ea where \ba\nl
H_{\chi}^{min}=\bigl \{\chi^{min},\Omega^{min}\bigr \} , \ea and
generally $\bigl \{\chi^M,U^Q_{NP}\bigr \}\not=0$ as far as the
structure coefficients of the constraint algebra $U^M_{NP}$ depend
on the phase-space variables.

One can use the representation (\ref{r}) for $H_{\tilde\chi}$ to
obtain the corresponding BRST invariant Lagrangian \ba\nl
L_{\tilde\chi}=L+L_{GH}+L_{GF} . \ea Here $L_{GH}$ stands for the
ghost part and $L_{GF}$ for the gauge fixing part of the
Lagrangian. If one does not intend to pass to the Lagrangian
formalism, one may restrict oneself to the minimal sector $\bigl
(\Omega^{min},\chi^{min},H_\chi^{min}\bigr )$. In particular, this
means that Lagrange multipliers are not considered as dynamical
variables anymore. With this particular gauge choice,
$H_\chi^{min}$ is a linear combination of the total constraints
\ba\nl H_\chi^{min}&=&H_{brane}^{min}+H_{aux}^{min}=
\\ \nl
&=&\int d^p\s\Bigl [\La^0 T_0^{tot}(\us)+\La^j T_j^{tot}(\us)+
\sum_A\La^{A\alpha} T_\alpha^{A tot}(\us)\Bigr ] +
\\ \nl
&+&\La_{ab}I^{ab}_{tot}+\La_{-+}I^{-+}_{tot}+\La_{+a}I^{+a}_{tot}+
\La_{-a}I^{-a}_{tot} , \ea and we can treat here the Lagrange
multipliers $\La^0,...,\La_{-a}$ as constants. Of course, this does
not fix the gauge completely.

\subsection{\bf Comments}
\hspace{1cm} To ensure that the harmonics and their conjugate momenta
are pure gauge degrees of freedom, we have to consider as physical
observables only such functions on the phase space which do not carry
any $SO(1,1)\times SO(8)$ indices.  More precisely, these functions
are defined by the following expansion \ba\nl
F(y,u,v;p_y,p_u,p_v)=\sum_{}\bigl [u_{\nu_1}^{a_1}...u_{\nu_k}^{a_k}
p_{u\nu_{k+1}}^{a_{k+1}}...p_{u\nu_{k+l}}^{a_{k+l}}\bigr ]_{SO(8)
singlet}\\ \nl
v_{\alpha_1}^+...v_{\alpha_m}^+v_{\alpha_{m+1}}^-...v_{\alpha_{m+n}}^-
p_v^{+\beta_1}...p_v^{+\beta_r}p_v^{-\beta_{r+1}}...p_v^{-\beta_{m-n+r}}\\
\nl
F_{\beta_1...\beta_{m-n+r}}^{\alpha_1...\alpha_{m+n},\nu_1...\nu_{k+l}}
(y,p_y) , \ea where $(y,p_y)$ are the non-harmonic phase space
conjugated pairs.

\newpage
\section{\bf CONCLUSIONS}
\hspace{1cm}The dissertation is devoted to the description and
further investigation of the properties of {\it null} $p$-branes.
The necessary preliminaries are given in the Introduction. Then we
explain how the above extended objects arise in the context of
string theory. The following section consider the known classical
and quantum properties of the tensionless branes. The original
results are described in sections 4-6.

In the forth section we perform BRST quantization of the null
bosonic $p$-branes using four different types of operator
ordering. It is shown that one can or can not obtain critical
dimension for the null string $(p=1)$, depending on the choice of
the operator ordering and corresponding vacuum states. When $p>1$,
operator orderings leading to critical dimension in the $p=1$ case
are forbidden by the Jacobi identities. Admissible orderings give
no restrictions on the dimension of the embedding space-time. This
is connected with the fact that the full constraint algebra has no
nontrivial central extension, but there are $p$ subalgebras which
possess non-trivial central extensions. When $p=1$, there is one
such subalgebra and it coincides with the full algebra. This
section is based on the paper \cite{B9711}.

In the fifth section we perform some investigation on the
classical dynamics of the null bosonic branes in curved
background. We write down the action, prove its reparametrization
invariance and give the equations of motion and constraints in an
arbitrary gauge. Then we construct the corresponding Hamiltonian
and compute the constraint algebra. In the following subsection we
consider the dynamics of null membranes ($p=2$) in a four
dimensional, stationary, axially symmetric gravity background.
Some exact solutions of the equations of motion and of the
constraints are found there. The next subsection is devoted to
examples of such solutions in Minkowski, de Sitter, Schwarzschild,
Taub-NUT and Kerr space-times. This section is based on the paper
\cite{B993}.

In the sixth section we consider a model for tensionless super
$p$-branes with N global chiral supersymmetries in 10-dimensional
Minkowski space-time. We show that the action is reparametrization
and $\kappa$-invariant. After establishing the symmetries of the
action, we give the general solution of the classical equations of
motion in a particular gauge. In the case of null superstrings
($p$=1) we find the general solution in an arbitrary gauge.
Starting with a Hamiltonian which is a linear combination of first
and mixed (first and second) class constraints, we succeed to
obtain a new one, which is a linear combination of first class,
BFV-irreducible and Lorentz-covariant constraints only. This is
done with the help of the introduced auxiliary harmonic variables.
Then we give manifest expressions for the classical BRST charge,
the corresponding total constraints and BRST-invariant
Hamiltonian. It turns out that in the given formulation our model
is a first rank dynamical system. This section is based on the
papers \cite{B986,B991,B992}.

\newpage
\section{\bf ACKNOWLEDGEMENTS}
\hspace{1cm}First of all, I would like to thank Prof. D. Stoyanov
for all I have learned from him in the years. This dissertation is
a result of the scientific experience I have obtained working with
him.

I am grateful to my family and many colleagues for the support I
have obtained from them.

The author acknowledges the hospitality of the {\it Bogoliubov
Laboratory of Theoretical Physics}, JINR, Dubna, Russia, where
this work has been done.

The papers on which the dissertation is based are supported in
part by the National Science Foundation of Bulgaria under contract
$\Phi-620/1996$.

\newpage
\section{\bf Appendix A}
\hspace{1cm} Here we briefly comment on the possible central
extensions of the algebra, given by the commutators: \ba \nl
[A_{\ul n},A_{\ul m}]&=&0 ,
\\
\nl [A_{\ul n},B_{j,\ul m}]&=&(n_j - m_j)A_{\ul n +\ul m} ,
\\
\nl [B_{j,\ul n},B_{k,\ul m}]&=&(\delta_{j}^{l} n_k -
\delta_{k}^{l} m_j) B_{l,\ul n +\ul m} \h , \h (j,k=1,2,...,p) .
\ea To begin with, we modify the right hand sides of the above
equalities as follows: \ba \nl [A_{\ul n},A_{\ul m}]&=&d(\ul n,\ul
m)
\\
\nl [A_{\ul n},B_{j,\ul m}]&=&(n_j - m_j)A_{\ul n +\ul m}+d_j (\ul
n,\ul m) ,
\\
\nl [B_{j,\ul n},B_{k,\ul m}]&=&(\delta_{j}^{l} n_k - \delta_{k}^{l}
m_j) B_{l,\ul n +\ul m}+d_{jk}(\ul n,\ul m) . \ea Checking the Jacobi
identities, involving the triplets  $(A,A,B)$ ,
 $(A,B,B)$  and  $(B,B,B)$ , one shows that there are only trivial solutions
for $d(\ul n,\ul m)$, $d_{j}(\ul n,\ul m)$ and  $d_{jk}(\ul n,\ul
m)$. Namely, \ba \nl d(\ul n,\ul m)&=&0 \h , \h d_{j}(\ul n,\ul
m)=(n_j - m_j)f(\ul n +\ul m) ,
\\
\nl d_{jk}(\ul n,\ul m)&=&(\delta_{j}^{l}n_k - \delta_{k}^{l}m_j)
g_{l}(\ul n + \ul m) , \ea where $f, g_j$ are arbitrary functions of
their arguments. In particular, there exist the solutions \ba \nl
d_{j}(\ul n,\ul m)&=&2\alpha n_j \delta_{\ul n +\ul m,\ul 0} \h , \h
\alpha=const ,
\\
\nl d_{jk}(\ul n,\ul m)&=&(\beta_{j}n_k + \beta_{k}n_j)\delta_{\ul n
+\ul m,\ul 0} \h , \h \beta_j = const , \ea which might appear
because of the operator ordering in $A_{\ul n}$ and $B_{j,\ul n}$.
However, there are $p$ subalgebras with non-trivial central
extensions (no summation over $j$): \ba \nl [A_{\ul n},A_{\ul m}]&=&0
,
\\
\nl [A_{\ul n},B_{j,\ul m}]&=&(n_j - m_j)A_{\ul n +\ul m}+
(q_{j}n_{j}^2+r_{j})n_{j}\delta_{\ul n +\ul m,\ul 0} ,
\\
\nl [B_{j,\ul n},B_{j,\ul m}]&=&(n_j - m_j) B_{j,\ul n +\ul
m}+(s_{j}n_{j}^2 + t_{j})n_{j}\delta_{\ul n +\ul m,\ul 0} ,
\hspace{1cm} q_{j}, r_{j}, s_{j}, t_{j} - const . \ea When $p=1$,
there is one such subalgebra and it coincides with the full algebra.

\section{\bf Appendix B}
\hspace{1cm} We briefly describe here our 10-dimensional conventions.
Dirac $\gamma$- matrices obey \ba\nl
\Gamma_\mu\Gamma_\nu+\Gamma_\nu\Gamma_\mu=2\eta_{\mu\nu} \ea and are
taken in the representation \ba\nl \Gamma^\mu =
\left(\begin{array}{cc}0&(\s^\mu)_\alpha^{\dot\beta}\\
(\tilde\s^\mu)^\beta_{\dot\alpha}&0\end{array}\right) \h . \ea
$\Gamma^{11}$ and charge conjugation matrix $C_{10}$ are given by
\ba\nl \Gamma^{11}=\Gamma^0\Gamma^1 ... \Gamma^9 =
\left(\begin{array}{cc}\delta_\alpha^\beta &0\\
0&-\delta_{\dot\alpha}^{\dot\beta}\end{array}\right) \h , \ea \ba \nl
C_{10}=\left(\begin{array}{cc}0&C^{\alpha\dot\beta}\\
(-C)^{\dot\alpha\beta}&0\end{array}\right) \h , \ea and the indices
of right and left Majorana-Weyl fermions are raised as \ba\nl
\psi^\alpha=C^{\alpha\dot\beta}\psi_{\dot\beta} \h ,\h
\phi^{\dot\alpha}=(-C)^{\dot\alpha\beta}\phi_\beta . \ea

We use $D=10$ $\s$-matrices with undotted indices \ba\nl
(\s^\mu)^{\alpha\beta}=C^{\alpha\dot\alpha}(\tilde\s^\mu)_{\dot\alpha}^\beta
\h ,\h
(\s^\mu)_{\alpha\beta}=(-C)^{-1}_{\beta\dot\beta}(\s^\mu)_\alpha^{\dot\beta}
\h , \ea and the notation \ba\nl
\s^{\mu_1...\mu_n}\equiv\s^{[\mu_1}...\s^{\mu_n]} \ea for their
antisymmetrized products.

From the corresponding properties of $D=10$ $\gamma$-matrices, it
follows that \ba\nl (\s^\mu)_{\alpha\gamma}(\s^\nu)^{\gamma\beta}+
(\s^\nu)_{\alpha\gamma}(\s^\mu)^{\gamma\beta}=
-2\delta_\alpha^\beta\eta^{\mu\nu}\h ,
\\ \nl
(\s_{\mu_1 ... \mu_{2s+1}})^{\alpha\beta}= (-1)^s(\s_{\mu_1 ...
\mu_{2s+1}})^{\beta\alpha}\h ,
\\ \nl
\s^\mu\s^{\nu_1 ... \nu_n}=\s^{\mu\nu_1 ... \nu_n}+
\sum_{k=1}^{n}(-1)^k\eta^{\mu\nu_k} \s^{\nu_1 ... \nu_{k-1}\nu_{k+1}
... \nu_n}. \ea

The Fierz identity for the $\s$-matrices reads: \ba\nl
(\s_\mu)^{\alpha\beta}(\s^\mu)^{\gamma\delta}+
(\s_\mu)^{\beta\gamma}(\s^\mu)^{\alpha\delta}+
(\s_\mu)^{\gamma\alpha}(\s^\mu)^{\beta\delta} = 0 . \ea

\newpage


\end{document}